\pdfoutput=1
\documentclass[preprint,12pt]{elsarticle}

\usepackage[margin=0.9in]{geometry} % change top margin

\usepackage{color, colortbl}  % allow use color in table
\usepackage{xcolor} % allow use color in text
\definecolor{Mycolor1}{HTML}{f90b9b}   % define a pink color
\definecolor{Mycolor2}{HTML}{0042FF}   % define a blue color
\definecolor{LightCyan}{rgb}{0.88,1,1}

\usepackage{url}
\usepackage[utf8]{inputenc}
\usepackage{graphicx}

\usepackage{arydshln}  % allow to use dashed lines in the table

\usepackage{amssymb}

\usepackage{amsthm} % allow begin proof

\usepackage{amsmath}   % use math command

\usepackage{times}
\usepackage{sectsty}
\usepackage{amsfonts}   % allow use \mathbb

\usepackage{stackengine}
\usepackage{caption}

\biboptions{numbers,sort&compress}
\allsectionsfont{\normalsize} % font size for sections, such as introduction.

\makeatletter
\def\ps@pprintTitle{%
 \let\@oddhead\@empty
 \let\@evenhead\@empty
 \def\@oddfoot{}%
 \let\@evenfoot\@oddfoot}
\makeatother

\begin{document}

\begin{frontmatter}


\title{Analysis of node2vec random walks on networks}

\author[1]{Lingqi Meng}
\author[1,2]{Naoki Masuda\corref{cor1}%
}
\ead{naokimas@buffalo.edu}

\cortext[cor1]{Corresponding author}
\address[1]{Department of Mathematics, University at Buffalo, State University of New York, Buffalo, NY 14260-2900, USA}
\address[2]{Computational and Data-Enabled Science and Engineering Program, University at Buffalo, State University of New York, Buffalo, NY 14260-5030, USA}

\begin{abstract}
Random walks have been proven to be useful for constructing various algorithms to gain information on networks. Algorithm node2vec employs biased random walks to realize embeddings of nodes into low-dimensional spaces, which can then be used for tasks such as multi-label classification and link prediction. The performance of the node2vec algorithm in these applications is considered to depend on properties of random walks that the algorithm uses. In the present study, we theoretically and numerically analyze random walks used by the node2vec. Those random walks are second-order Markov chains. We exploit the mapping of its transition rule to a transition probability matrix among directed edges to analyze the stationary probability, relaxation times in terms of the spectral gap of the transition probability matrix, and coalescence time. In particular, we show that node2vec random walk accelerates diffusion when walkers are designed to avoid both back-tracking and visiting a neighbor of the previously visited node but do not avoid them completely.
\end{abstract}


\end{frontmatter}

\section{Introduction}
\label{S:1}

Random walks on finite networks have been a favorite research topic for decades \cite{doyle1984random, aldous1995reversible, noh2004random, masuda2017random}. Perhaps more importantly, random walks are a core technique for building algorithms to extract useful information from network data. Such applications of random walks include community detection, ranking of nodes and edges, dimension reduction of data, sampling, to name a few \cite{masuda2017random, xia2019random}. Many theoretical, computational, and algorithmic studies have employed simple random walks on unweighted networks, which by definition dictates that a walker moves to one of its neighbors with equal probability in each time step. However, there are also various other types of random walks, many of which have been fed to random walk algorithms \cite{masuda2017random, xia2019random}. 

The random walks developed for the algorithmic framework called the node2vec are one such random walk \cite{grover2016node2vec}. Unlike simple random walks, transitions of node2vec random walkers not only depend on the degree of the currently visited node or its variant with edge weights, but also on the structure of the local network and last visited node. Grover and Leskovec proposed node2vec for scalable feature learning on networks, which can be used in tasks such as community detection, multi-label classification, and link prediction. In node2vec, one can tune the weight of local versus global search of the network by modulating parameter values \cite{grover2016node2vec}. The node2vec has found applications in, for example, predicting genes associated with Parkinson’s disease \cite{peng2019predicting} and movie recommendation \cite{palumbo2018knowledge}.

To date, not much is known about behavior of node2vec random walks. Note that, among various properties of random walks, the stationary probability plays a key role in ranking the nodes \cite{masuda2017random, langville2006google, newman2018networks}, and the relaxation time affects, for example, the rate of the convergence of random-walk algorithms and quality of community structure \cite{masuda2017random}. In the present study, we theoretically and numerically examine the node2vec random walks on finite networks. In particular, we provide multiple lines of evidence supporting that diffusion (i.e., approaching to the stationary probability and coalescence of random walkers) is accelerated when the parameters of node2vec random walks are tuned such that back-tracking and visiting the neighbors of the last visited node are suppressed and exploration of the rest of the network, similar to depth-first sampling, is explicitly promoted. This is the case unless the avoidance of local sampling including back-tracking is not excessive.

\section{Model}
\label{S:2}

\begin{figure}[!h]
    \centering
    \includegraphics[width=0.5\textwidth]{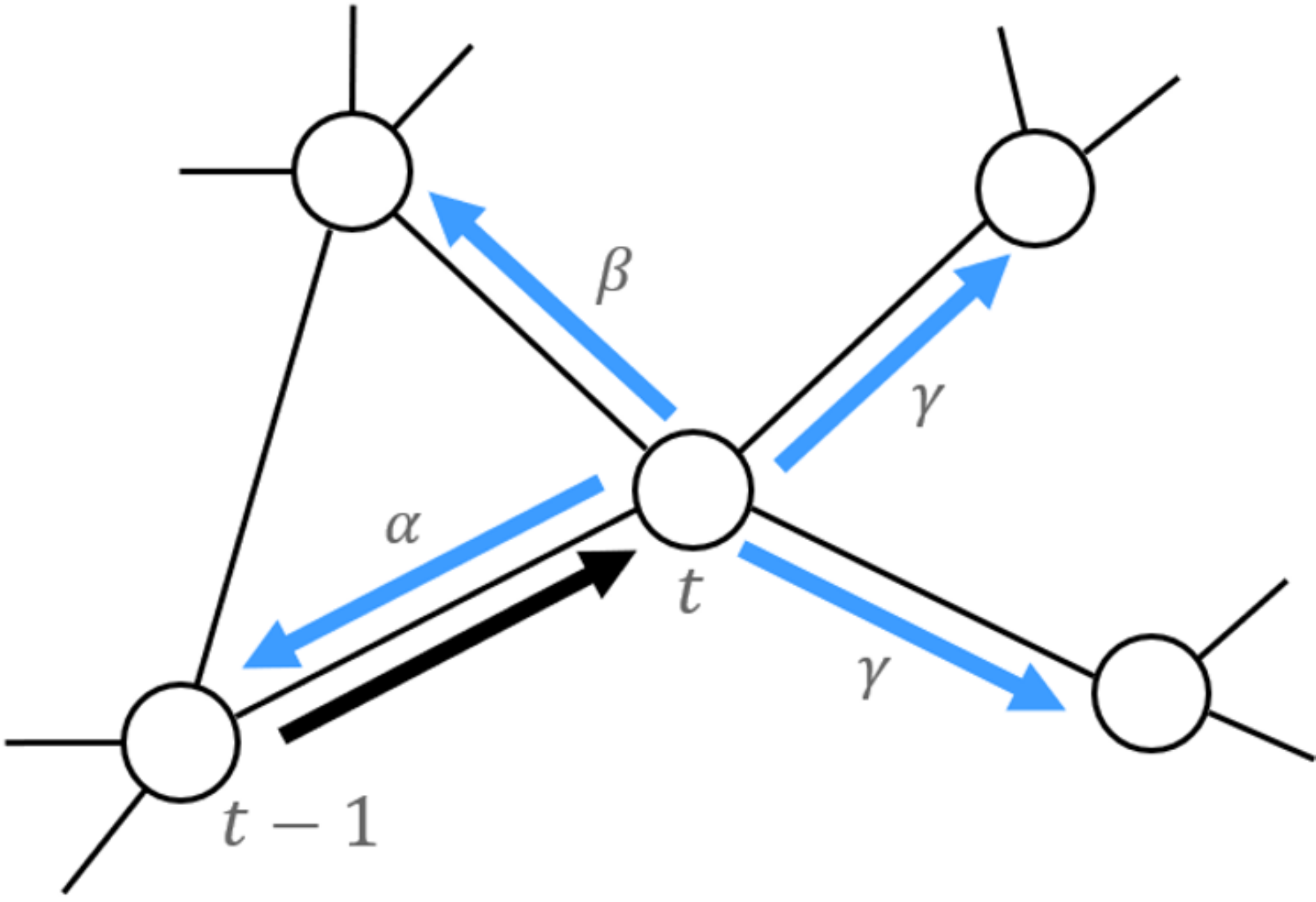}
    \caption{Schematic of the node2vec random walk. We assume that the network is unweighted. The transition probability to one of the four neighbors at time $t$ in this example is given by $\alpha/(\alpha+\beta+2\gamma), \beta/(\alpha+\beta+2\gamma$), or $\gamma/(\alpha+\beta+2\gamma)$.}
    \label{node2vec}
\end{figure}

Consider a finite network $G(V, E)$, where $V=\{1, \ldots, N\}$ is a finite set of nodes, $N$ is the number of nodes, and $E=\{(i, j)\ |\ (i,j)\in V\times V\  \mathrm{and}\ i\neq j \}$ is a set of edges. In the present study, we assume undirected and possible weighted networks that are free of self-loops and multiple edges, although the node2vec random walks and the formalism developed below are also valid for directed networks. Denote by $v_t$ ($t=0, 1, \ldots$) the position of a random walker at discrete time $t$. We say that a discrete-time random walk is node2vec if its transition probability $p_{i\to j}(t)$ at time $t$, where $(i,j)\in E$, is given by
\begin{align}
	p_{i\to j}(t) \propto
	\begin{cases}
        \alpha w_{ij} & \mathrm{if} \ v_{t-1}=v_{t+1}, \\
        \beta w_{ij} & \mathrm{if} \ (v_{t-1},v_{t+1})\in E, \\
        \gamma w_{ij} & \mathrm{if} \ (v_{t-1},v_{t+1})\not \in E,
	\end{cases}
	\label{eq2}
\end{align}
where $w_{ij}$ is the weight of edge $(i, j)$, and the symbol $\propto$ means ``proportional to'' \cite{grover2016node2vec}. The normalization is given by $\displaystyle \sum_{j=1}^N p_{i\to j}(t)=1$ for all $i\in V$ and $t>0$. Variable $\alpha$ represents the propensity for the random walk to backtrack, $\beta$ the weight of reaching a common neighbour of the currently visited node and the node visited in the last step, and $\gamma$ the weight of exploring any of the other neighbors. A large $\beta$ value implies an approximate breadth-first sampling, and a large $\gamma$ value implies an approximate depth-first sampling \cite{grover2016node2vec}. If $\alpha=\beta=\gamma\neq 0$, the node2vec random walk is reduced to a simple random walk. If $\alpha=0$ and $\beta=\gamma\neq 0$, the node2vec random walk is a non-backtracking random walk \cite{alon2007non, fitzner2013non}. Possible one-step transitions of the node2vec random walk are schematically shown in Fig.~\ref{node2vec}.

Equation (\ref{eq2}) implies that a node2vec random walk is a second-order Markov chain \cite{grover2016node2vec}. In other words, the transition probability $p_{i\to j}(t)$ depends on the currently visited node $i$ and the node visited in the previous time step (i.e., $t-1$), but not on the further history of the walk. To transform the node2vec random walk into a first-order Markov chain, we change the state space from the nodes of the network to the directed edges of the network, similar to the formation of memory networks \cite{rosvall2014memory, scholtes2014causality}. Let $M$ denote the number of undirected edges. Let $\overline{E}=\{e_1, \ldots , e_{2M}\}$ be the set of directed edges, which consists of each undirected edge $(u, v) \in E$ duplicated as directed edges $(u,v)$ and $(v, u)$. For notational convenience, we use $(\cdot, \cdot)$ to represent the cases of both an undirected and directed edge. For $e=(u,v)\in \overline{E}$, we denote $e(0)=u$ and $e(1)=v$. Under this transformation, the $2M \times 2M$ transition probability matrix $\overline{T}$ is given by
\begin{align}
	\overline{T}_{i,j}\propto
    \begin{cases} 
      \alpha w_{e_i(1), e_j(1)} & \mathrm{if} \ e_i(1)=e_j(0)\ \mathrm{and}\ e_i(0)=e_j(1), \\
      \beta w_{e_i(1), e_j(1)} & \mathrm{if} \ e_i(1)=e_j(0)\ \mathrm{and}\ (e_i(0),e_j(1))\in \overline{E}, \\
      \gamma w_{e_i(1), e_j(1)} & \mathrm{if} \ e_i(1)=e_j(0)\ \mathrm{and}\ (e_i(0),e_j(1))\not \in \overline{E}, \\
      0 & \mathrm{otherwise}.
   \end{cases}
   \label{eq edge2}
\end{align}
The normalization is given by $\displaystyle \sum_{j=1}^{2M} \overline{T}_{i,j}=1$ for $i=1, 2, \ldots, 2M$.

\section{Results}

\subsection{Stationary probability in special cases}

We start by briefly reviewing some definitions. A directed network is strongly connected if there exists a directed path from $u$ to $v$ and from $v$ to $u$ for any nodes $u$ and $v$. We say that a network is aperiodic if the greatest common divisor of the length of all the closed directed paths is equal to $1$. Most empirical networks are aperiodic although there are important exceptions such as bipartite networks including trees. Therefore, we assume aperiodicity throughout this paper.

A node2vec random walk on a strongly connected aperiodic finite network with state space $\overline{E}$ induces a unique positive probability vector $\boldsymbol{q}^*=(q_1^*, \ldots , q_{2M}^*)$, where $q_j^*$ is the stationary probability on directed edge $e_j$ ($j=1,2,\ldots,2M$), such that 
\begin{align}
\boldsymbol{q^*} \overline{T}= \boldsymbol{q^*}.
\end{align}
Denote $\boldsymbol{p}^*=(p_1^*, \ldots , p_N^*)$, where $p_i^*$ is the stationary probability at node $i$ ($i=1,\ldots, N$). Probability vectors $\boldsymbol{p}^*$ and $\boldsymbol{q}^*$ are related by
\begin{align}
    p_i^*=\sum_{\substack{e_j\in \overline{E}\\e_j(1)=i }} q_j^*.
    \label{eq 6}
\end{align}
In particular, if the network is undirected and the random walk is simple (i.e., $\alpha = \beta = \gamma$), one obtains
\begin{align}
\boldsymbol{q^*} = \frac{1}{2M}(1,...,1).
\label{eq6}
\end{align}
Therefore, for a simple random walk on undirected networks, we recover the well-known result given by
\begin{align}
    \boldsymbol{p}_i^*=\frac{d_i}{2M},
\label{eq7}
\end{align}
where $d_i$ is the weighted degree, which is called the node strength, of node $i$.

We say that a network is simple if it is unweighted, undirected, and free of self-loops and multiple edges. Non-backtracking random walks on a simple finite network with degree $d_i \geq 2$ $(i=1, \ldots, N)$ have the same stationary distribution as the simple random walk \cite{alon2007non}. Here we present a slight generalization of this result stated as follows:

\bigskip

\textbf{Theorem 1.} \textit{For a node2vec random walk on a simple finite network, the stationary distribution is the same as that for the simple random walk if $\beta=\gamma$, $\alpha>0$. In other words, it is given by Eq. (\ref{eq6}). Therefore, the stationary distribution for nodes is given by Eq. (\ref{eq7}). }

\begin{proof}
Let $\beta=\gamma$. In this case, we do not have to distinguish whether or not edges $(v_{t-1}, v_t)$, $(v_{t-1}, v_{t+1})$, and $(v_t, v_{t+1})$ form a triangle. Therefore, the transition probability matrix is given by
\begin{align}
	\overline{T}_{i,j}=
    \begin{cases} 
      \frac{\alpha}{\alpha+(d_{e_j(0)}-1)\beta} & \mathrm{if} \ e_i(1)=e_j(0)\ \mathrm{and}\ e_i(0)=e_j(1), \\
      \frac{\beta}{\alpha+(d_{e_j(0)}-1)\beta} & \mathrm{if} \ e_i(1)=e_j(0)\ \mathrm{and}\ e_i(0)\neq e_j(1), \\
      0 & \mathrm{otherwise}.
   \end{cases}
\end{align}
It is straightforward to verify that $\overline{T}$ has a left eigenvector $\mathbf{1}=(1,\ldots,1)$, such that $\mathbf{1} \overline{T}=\mathbf{1}$. Because of the uniqueness of the Perron-Frobenius vector, the stationary distribution is given by Eq. (\ref{eq6}).
\end{proof}

We remark that Theorem 1 allows nodes with degree $1$. If a node2vec random walker arrives at a node with degree 1, it always backtracks in the next time step because backtracking is the only possible move. This is consistent with the assumption $\alpha > 0$ in the theorem.

We now examine how symmetry in the network constrains the stationary distribution of the node2vec random walk. Consider a network $G(V,E)$ and its corresponding adjacency matrix $A$, where $G$ can be directed or undirected, and weighted or unweighted. An automorphism $\pi$ of network $G$ is a permutation of the nodes that preserves the adjacency of the nodes \cite{everett1988calculating, everett1990ego, biggs1993algebraic,barrett2017equitable}. In other words, automorphism $\pi\colon V\to V$ is a bijection that satisfies $A_{ij}=A_{\pi(i)\pi(j)}$, for any $i, j = 1, \ldots, N$. Two nodes, denoted by $v$ and $v^\prime$, are said to be automorphically equivalent if there is an automorphism that maps one node to the other, i.e., $\pi(v) = v^\prime$ \cite{everett1988calculating, everett1990ego}. A vertex-transitive network is an undirected network in which any pair of nodes is automorphically equivalent \cite{biggs1993algebraic, godsil2013algebraic}. 

\bigskip

\textbf{Theorem 2.} \textit{If nodes $u$ and $v$ are automorphically equivalent in undirected network $G(V,E)$, then they have the same stationary probability of being visited by a node2vec random walker, i.e., $p_u^*=p_v^*$.}
\begin{proof}
Let $\pi$ be an automorphism of $G$. Let $\overline{E}=\{e_1,e_2,\ldots,e_{2M}\}$ be an ordered set of the directed edges in the undirected network $G$, in which each undirected edge $(u, v) \in E$ is duplicated as directed edges $(u,v) \in \overline{E}$ and $(v,u)\in \overline{E}$. Define a permutation of $\overline{E}$ by $\phi(\overline{E})=\{\phi(e_1),\phi(e_2),\ldots,\phi(e_{2M})\}$, where a directed edge $\phi(e_i):=\left(\pi(e_i(0)),\pi(e_i(1))\right)$ for $i=1,\ldots, 2M$. Because $\phi(e_i)\in \overline{E}$ and $\phi(e_i) \neq \phi(e_j)$ if $i\neq j$, set $\phi(\overline{E})$ is also an ordered set of the directed edges in $G$. Therefore, $\phi$ is a permutation of $\overline{E}$.

First, we show that $\phi$ is an automorphism of a directed weighted network $\overline{G}$ derived from $G$. In $\overline{G}$, the set of nodes is given by $\overline{E}$, and the set of edges is specified by the weighted adjacency matrix, $\overline{T}$, given by Eq. (\ref{eq edge2}). Therefore, the two directed edges of $G$ (i.e., nodes of $\overline{G}$), denoted by $e_i$ and $e_j$, are connected by a directed edge of $\overline{G}$ if and only if random walkers that have traversed $e_i$ may traverse $e_j$ in the next time step. Formally, for arbitrary $e_i, e_j\in \overline{E}$, ordered pair $(\phi(e_i), \phi(e_j))$ is an edge of $\overline{G}$ if and only if $(e_i, e_j)$ is an edge of $\overline{G}$, because $\pi(e_i(1))=\pi(e_j(0))$ if and only if $e_i(1)=e_j(0)$. We also obtain
\begin{align}
    \overline{T}_{e_i, e_j} = \overline{T}_{\phi(e_i), \phi(e_j)}\propto
    \begin{cases} 
      \alpha w_{e_i(1) e_j(1)} & \mathrm{if} \ e_i(1)=e_j(0)\ \mathrm{and}\ e_i(0)=e_j(1),\ \mathrm{equivalently}, \\
      &\mathrm{if}\ \pi(e_i(1))=\pi(e_j(0))\ \mathrm{and}\ \pi(e_i(0))=\pi(e_j(1)),\\
      \beta w_{e_i(1) e_j(1)} & \mathrm{if} \ e_i(1)=e_j(0)\ \mathrm{and}\ (e_i(0),e_j(1))\in \overline{E},\ \mathrm{equivalently}, \\
      &\mathrm{if}\ \pi(e_i(1))=\pi(e_j(0))\ \mathrm{and}\ (\pi(e_i(0)),\pi(e_j(1)))\in \overline{E},\\
      \gamma w_{e_i(1) e_j(1)} & \mathrm{if} \ e_i(1)=e_j(0)\ \mathrm{and}\ (e_i(0),e_j(1))\not \in \overline{E},\ \mathrm{equivalently}, \\
      & \mathrm{if}\ \pi(e_i(1))=\pi(e_j(0))\ \mathrm{and}\ (\pi(e_i(0)),\pi(e_j(1)))\not \in \overline{E},\\
      0 & \mathrm{otherwise}.
   \end{cases}
   \label{eq8}
\end{align}
Therefore, $\phi$ is an automorphism of $\overline{G}$. Note that, in Eq. (\ref{eq8}), we used, for example, $e_i$ rather than $i$ to refer to the row and column of $\overline{T}$ to avoid an abuse of notation.

Second, we show that automorphically equivalent nodes in $\overline{G}$ have the same stationary probability of the random walk whose transition probability matrix is given by $\overline{T}$. To show this, let $\overline{T}^\prime$ be the weighted adjacency matrix of $\overline{G}$ when the rows and columns are reordered as $\phi(\overline{E}) = \{ \phi(e_1), \ldots, \phi(e_{2M}) \}$. Because $\phi$ is an automorphism, we obtain
\begin{align}
    \overline{T}_{e_i, e_j}=\overline{T}_{\phi(e_i), \phi(e_j)}= \overline{T}^\prime_{e_i, e_j},
\end{align}
for any $i, j=1,\ldots, 2M$. Let $\boldsymbol{q}^*$ and $\widetilde{\boldsymbol{q}}^*$ be the stationary probability of the random walk whose transition probability matrix is given by $\overline{T}$ and $\overline{T}^\prime$, respectively. Because $\overline{T}=\overline{T}^\prime$, we obtain $\boldsymbol{q}^*=\widetilde{\boldsymbol{q}}^*$, i.e., $q^*_{e_i} = q^*_{\phi(e_i)}$, $i=1, \ldots, 2M$.

Finally, assume that $u\in V$ and $v\in V$ are automorphically equivalent in $G$ and connected by an automorphism $\pi$, i.e., $v=\pi(u)$. For any directed edge $e_i$ incoming to $u$, i.e., $e_i(1)=u$, directed edge $\phi(e_i)$ is incoming to $v$ because $\phi(e_i) = (\pi(e_i(0)), \pi(e_i(1))) = (\pi(e_i(0)), v)$. Because $\phi$ is an automorphism of $\overline{G}$, we obtain $q^*_{e_i} = q^*_{\phi(e_i)}$. Because this argument holds true for any pair of $e_i \in \overline{E}$ incoming to $u$ and the corresponding edge incoming to $v$, we use Eq. (\ref{eq 6}) to conclude that $p_u^* = p_v^*$.
\end{proof}

\textbf{Corollary 1.} \textit{If network $G$ is vertex-transitive, $p_i^* = 1/N$ for all nodes.}

\subsection{Relaxation time}

The relaxation speed of the random walk is governed by the second largest eigenvalue of $T$ in modulus \cite{aldous1995reversible, masuda2017random, lovasz1993random}. The spectral gap defined by $1-|\lambda_2|$, where $\lambda_2$ is the second largest eigenvalue of $T$ in modulus, quantifies the relaxation speed (see SM for numerical examples). A large spectral gap implies a fast convergence.

A node2vec random walk is specified by three parameters $\alpha$, $\beta$, and $\gamma$. Because only the ratio among $\alpha$, $\beta$, and $\gamma$ specifies the transition probabilities, we set $\gamma=1$. Note that we are not interested in the case $\gamma=0$ because it implies that the walker always backtracks or visits the neighbor of the previously visited node without exploring a node different from $v_{t-1}$ or its neighbor. In this section, we examine relaxation time of node2vec random walks on empirical and synthetic networks.

\subsubsection{Empirical networks}
\label{section:Empirical networks}

\begin{table}[!t]
    \centering
    \caption{Properties of the empirical networks.}
    \begin{tabular}{ p{3cm} p{3cm} p{3cm} }
    \hline
    Network & $N$ & $M$\\
    \hline
    Vole & 51 & 105\\
    Dolphin & 62 & 159\\
    Enron & 143  & 623\\
    Jazz & 198 & 2742\\
    Coauthorship & 379 & 914\\
    Email & 1133 & 5451\\
    \hline
    \end{tabular}
    \label{tab1}
\end{table}

We study node2vec random walks on six empirical networks. Basic properties of the data sets are shown in Table \ref{tab1}. All the networks are treated as unweighted and undirected networks. The data sets can be downloaded at \cite{vole, dolphin, enron, jazz, coauthorship, email}.

\begin{figure}[!h]
  \includegraphics[width=0.47\textwidth]{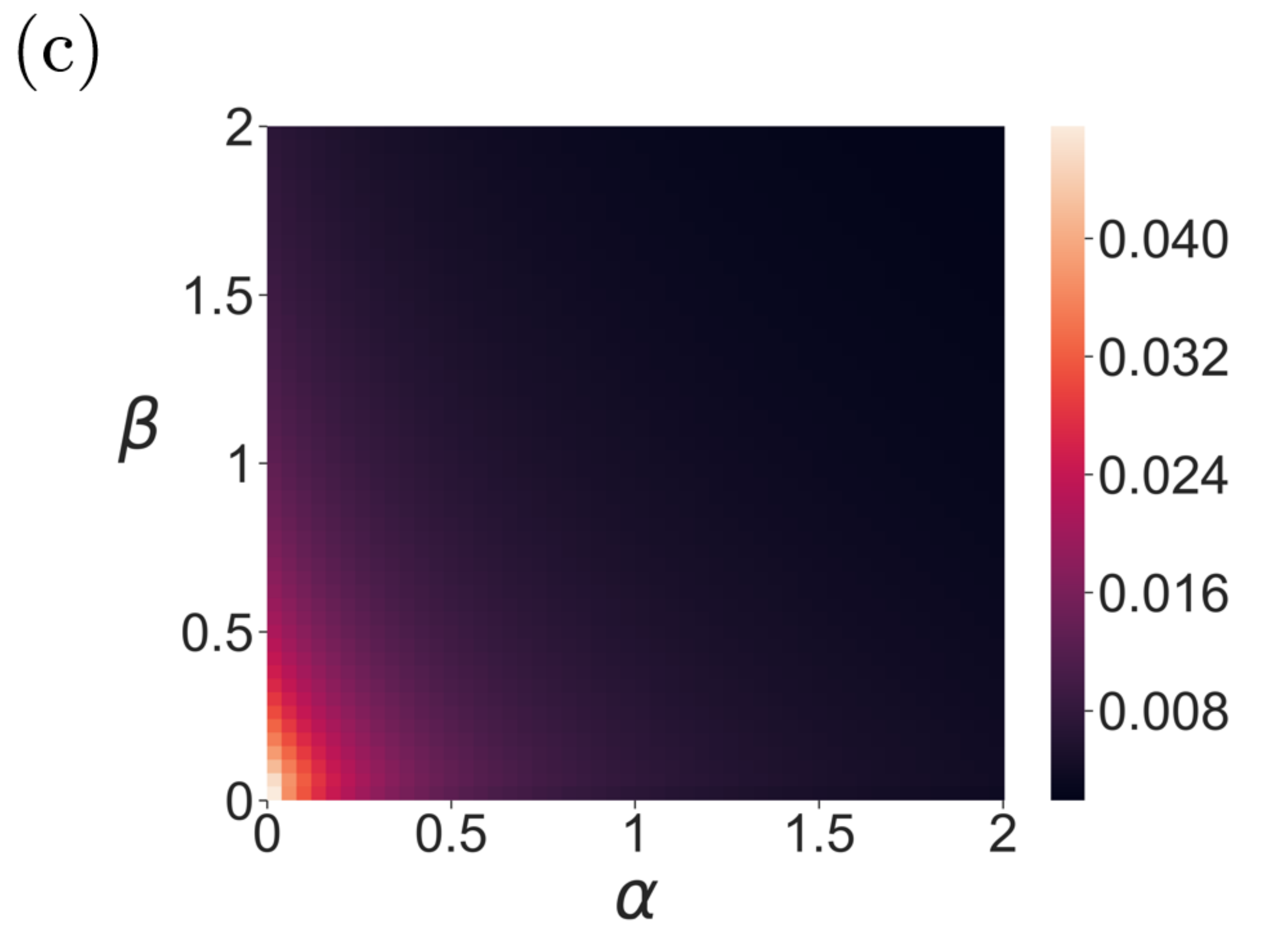}\label{fig2:sub3}\quad%
  \includegraphics[width=0.47\textwidth]{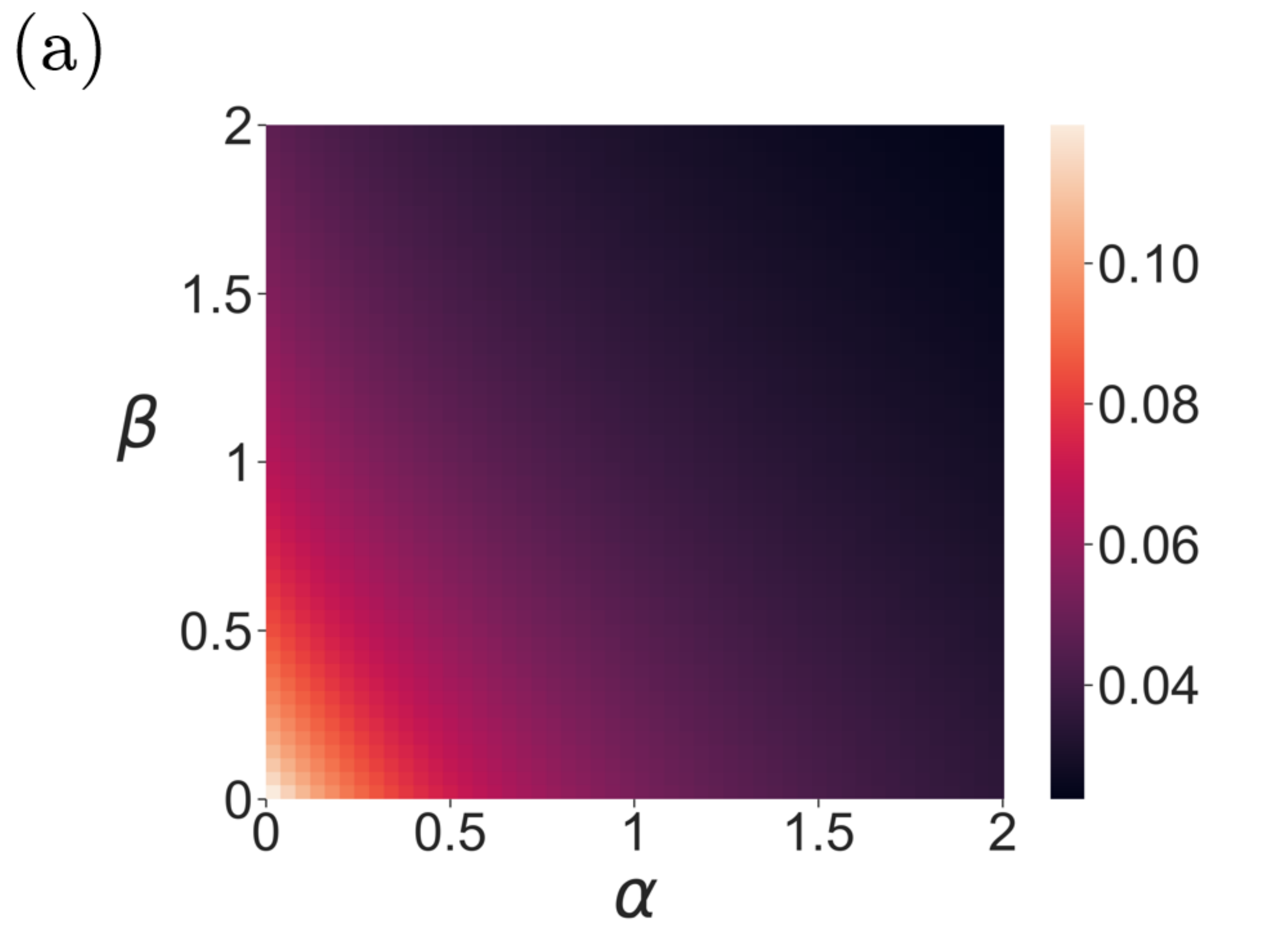}\label{fig2:sub1}\quad%
  \includegraphics[width=0.47\textwidth]{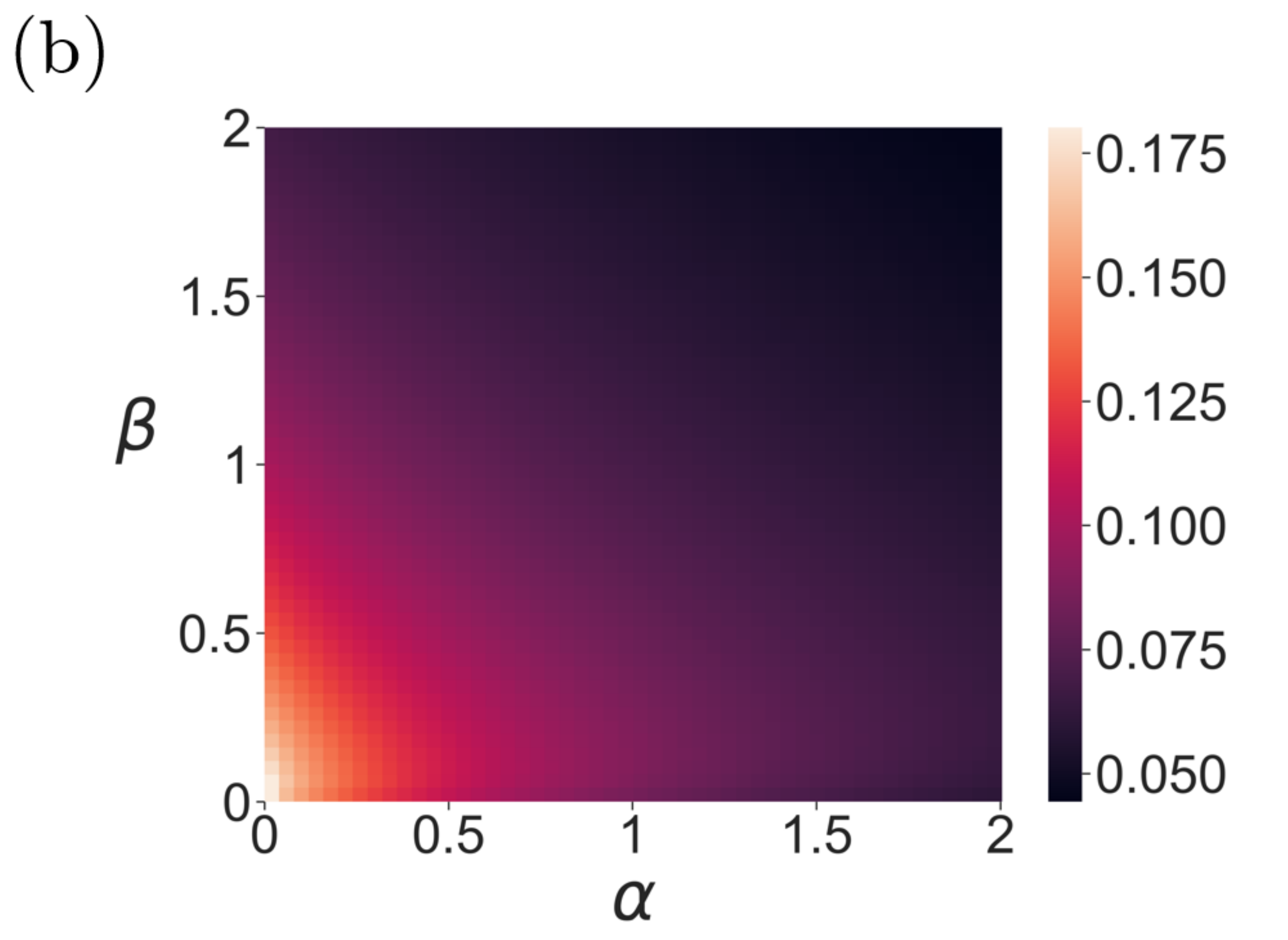}\label{fig2:sub2}\quad%
  \includegraphics[width=0.47\textwidth]{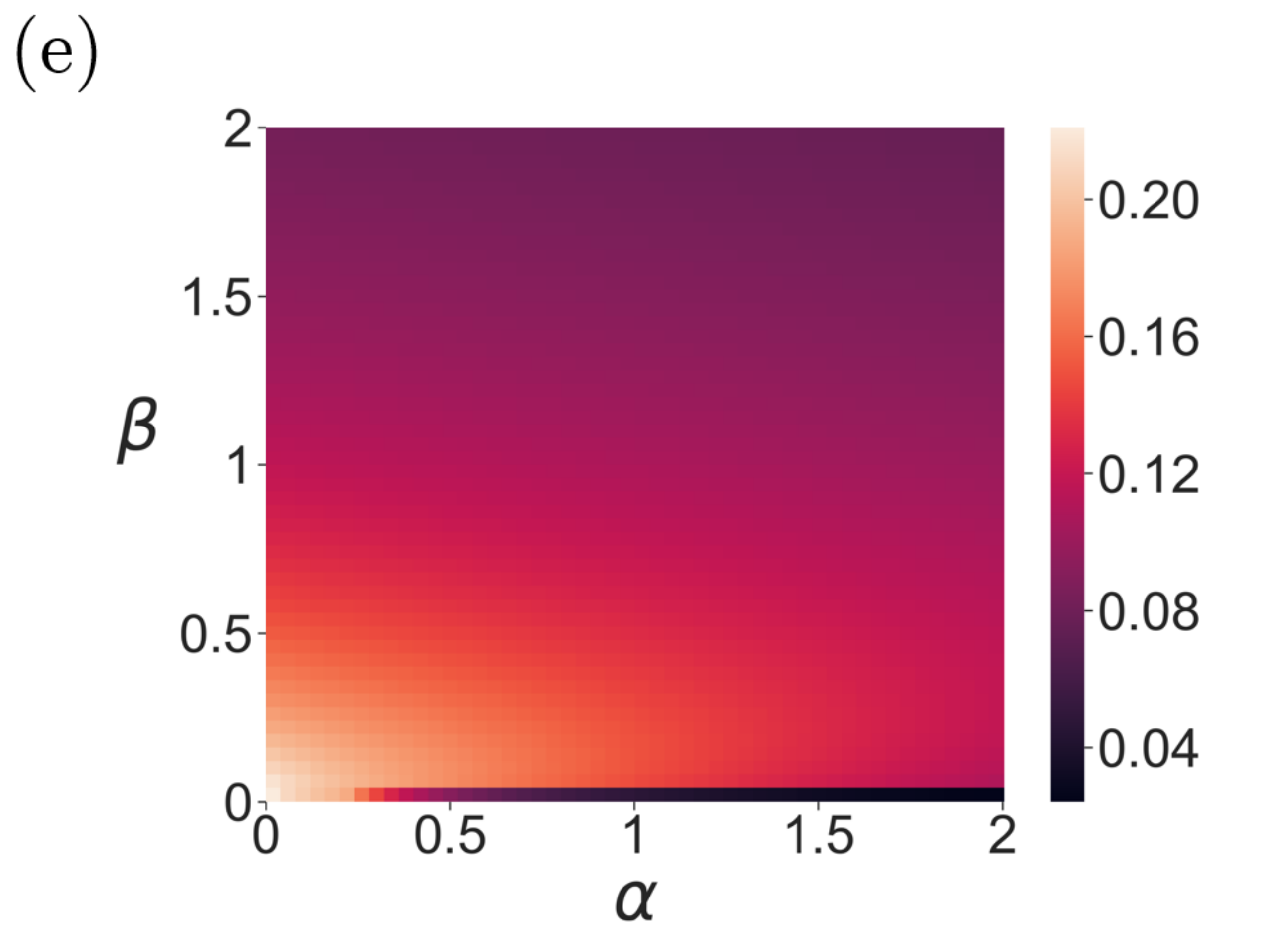}\label{fig2:sub5}\quad%
  \includegraphics[width=0.47\textwidth]{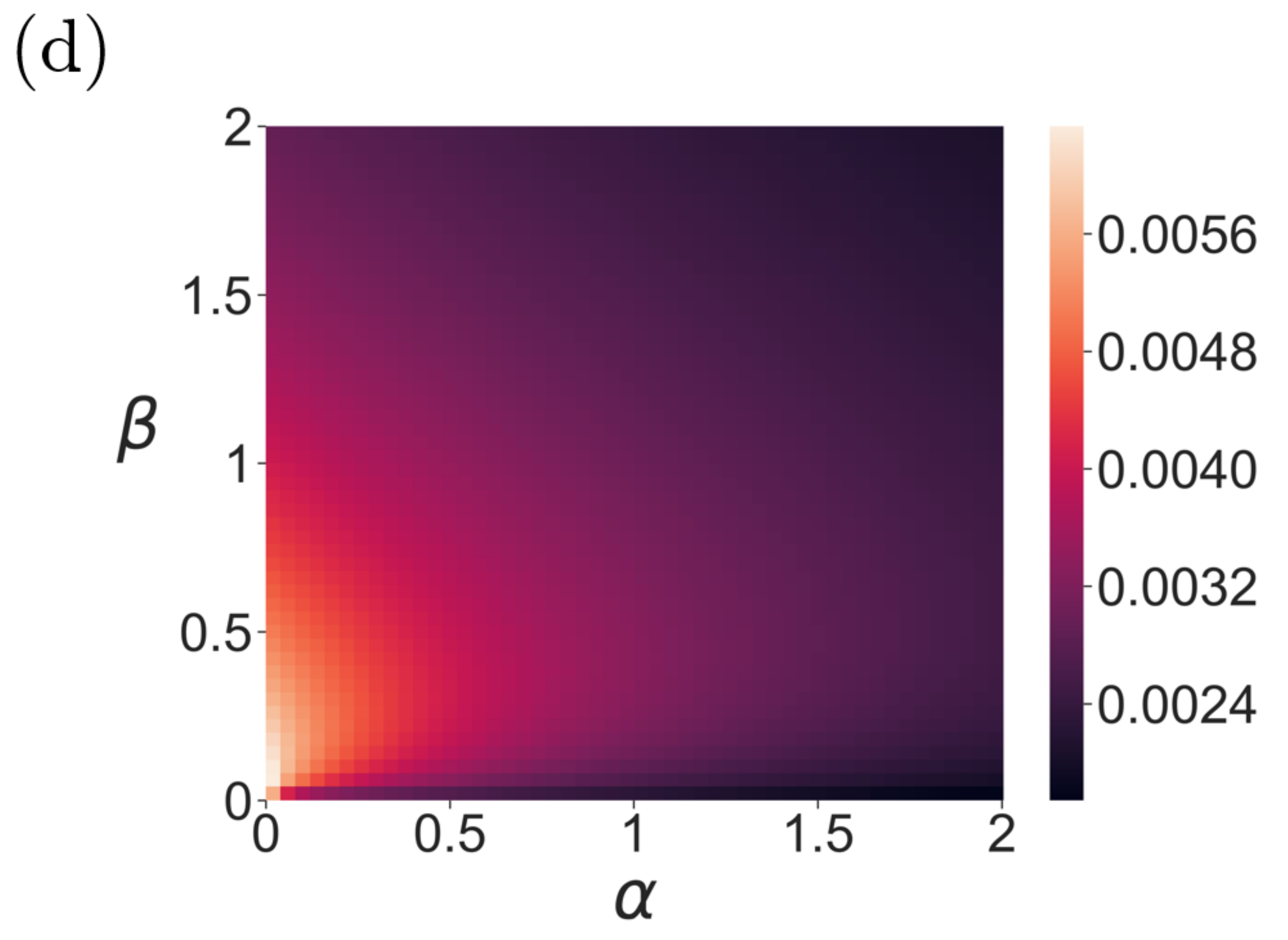}\label{fig2:sub4}\quad%
  \includegraphics[width=0.47\textwidth]{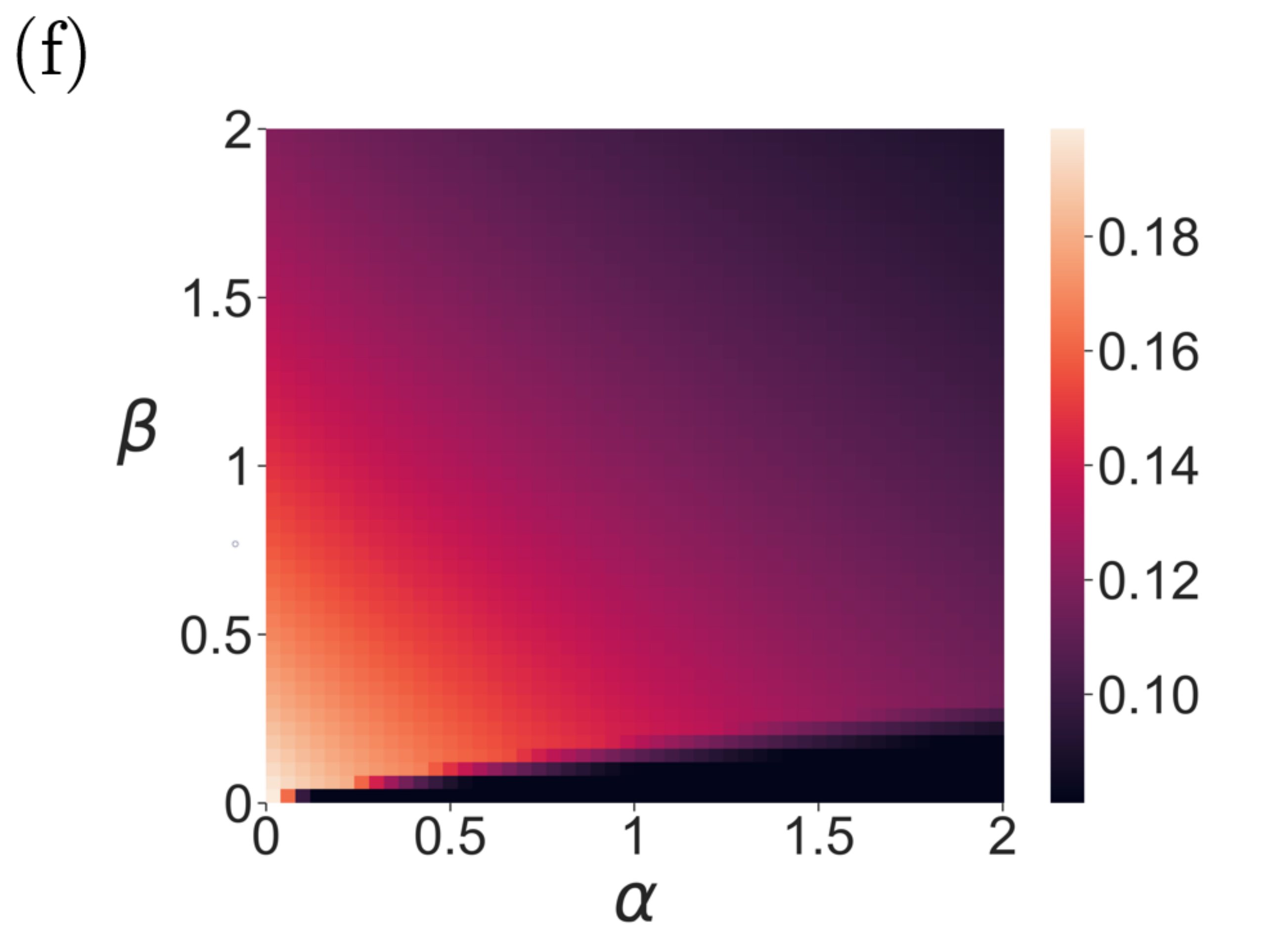}\label{fig2:sub6}\quad%
  \caption{Spectral gap for node2vec random walks on empirical networks. (a) Dolphin. (b) Enron. (c) Vole. (d) Coauthorship. (e) Jazz. (f) Email.}\label{fig2}
  \vspace*{-7pt}
\end{figure}

The voles network is one of the $128$ wild vole networks gathered in Kielder Forest on the English–Scottish border around $2001$ \cite{davis2015spatial}. Each node denotes a vole. An edge is present if two voles were caught in at least one common trap. The dolphin network is a social network, in which nodes are the bottlenose dolphins, and an edge occurs if there is a frequent association between two bottlenose dolphins \cite{lusseau2003bottlenose}. Enron Email Data set was collected and prepared by the CALO (A Cognitive Assistant that Learns and Organizes) project \cite{cohen2005enron}. Each node represents a manager or an employee of the Enron Corporation. There is an edge between two nodes if there is at least one email exchanged between the two individuals. The jazz network is constructed based on collaboration between jazz musician bands \cite{gleiser2003community}. Each node denotes a band. Two nodes are adjacent if they have a musician in common anytime between 1912 and 1940. The coauthorship network represents coauthor relationships between authors who published papers on network science up to 2006 \cite{newman2006finding}. The original data set has $1589$ nodes, and we only use the largest connected component. The email network is gathered from University at Rovira i Virgili in Tarragona, Spain, and contains 1669 users \cite{guimera2003self}. Each node represents an email address. An edges occurs between two nodes if there is an email communication between them at least once. Among the 1669 nodes, 1133 of them belongs to the largest connected component, which we use in the following analysis.

Figure 2 shows the numerically calculated spectral gap for the different empirical networks when we vary the $\alpha$ and $\beta$ values while keeping $\gamma=1$. The figure suggests that spectral gap largely decreases as $\alpha$ or $\beta$ increases for all the networks. The global maximum value of the spectral gap is obtained near $(\alpha,\beta)=(0,0)$. Therefore, smaller $\alpha$ and $\beta$ values, which imply a larger probability of exploring the network without backtracking or visiting common neighbors of the presently visited node and the last visited node, accelerate relaxation. In Figs. \ref{fig2}(d), \ref{fig2}(e), and \ref{fig2}(f), the spectral gap is small for excessively small $\beta$ even when $\alpha$ is relatively large. It is probably because a tiny $\beta$ value compels the random walker to leave local neighbors of a node, such as a community, before it sufficiently explores the neighborhood with a breadth-first sampling mechanism.

\subsubsection{Extended ring network with triangles}
\label{section:watts}

\begin{figure}[!h]
  \includegraphics[width=0.47\textwidth]{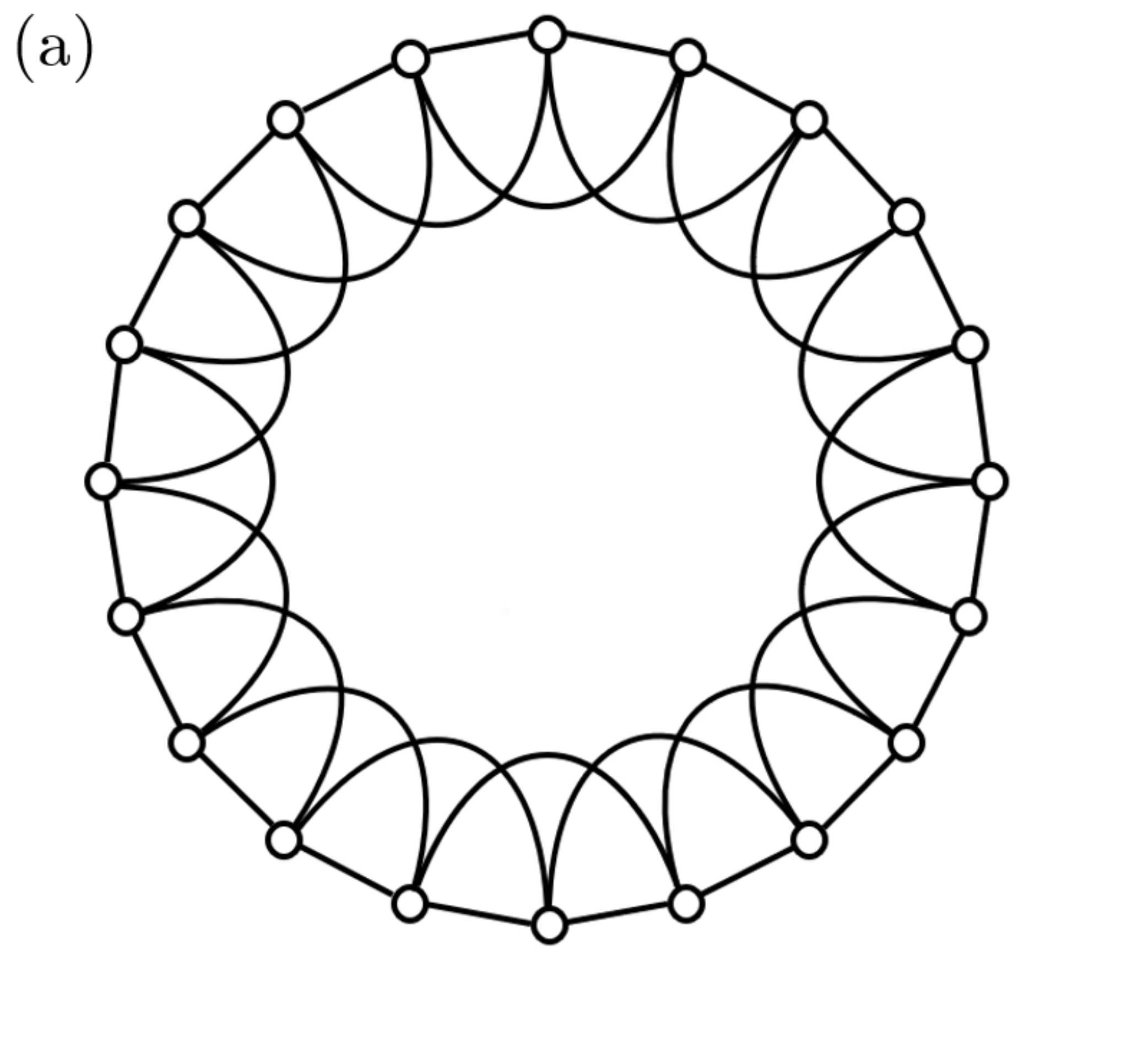}\label{fig3a}\quad%
  \includegraphics[width=0.47\textwidth]{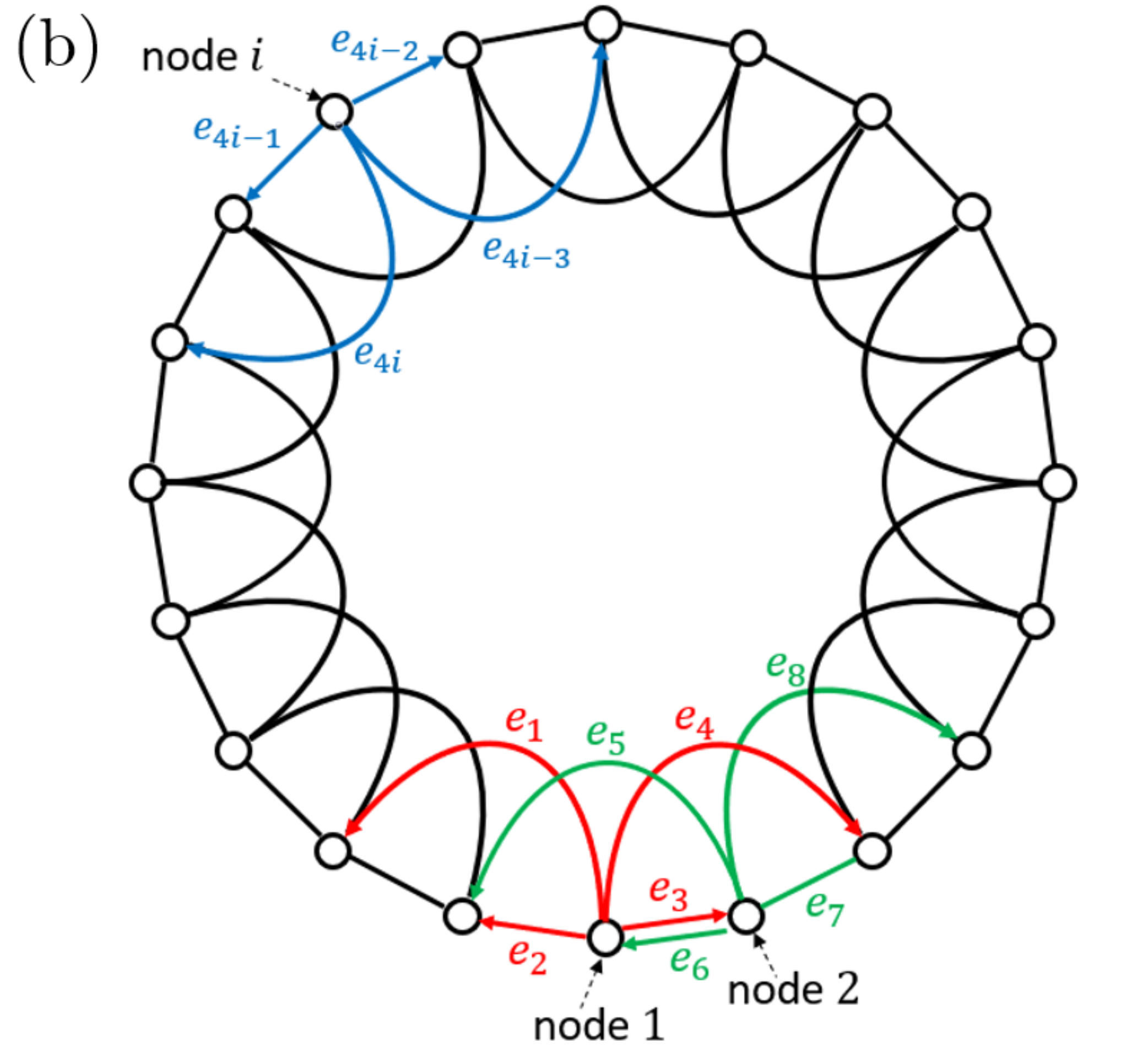}\label{fig3b}\quad%
  \caption{Schematic of the extended ring network and the method to label its directed edges. (a) Extended ring network with $N=20$. (b) The corresponding labeling method for its directed edges. The nodes and the corresponding directed edges are labeled counterclockwise. }\label{fig3}
  \vspace*{-7pt}
\end{figure}

Empirical networks are heterogeneous in terms of the node's degree and local abundance in triangles. Therefore, the stationary probability depends on the $\alpha$ and $\beta$ values given $\gamma = 1$, unless $\beta = 1$. Therefore, the result that a small $\alpha$ and $\beta$ largely accelerates the exploration of node2vec random walkers may partly rely on the change in the stationary probability as $\alpha$ or $\beta$ changes. To exclude this possibility, in this section and Section 3(b)(\ref{section:wattstwolayers}), we consider model networks whose stationary probability does not depend on $\alpha$ or $\beta$. Our choice of the model networks is based on analytical tractability rather than on sufficient similarity to empirical networks. Specifically, in this section we consider an extended ring network shown in Fig. \ref{fig3}(a). As the figure indicates, each node has degree $k=4$, and all the nodes are automorphically equivalent to each other. Therefore, Theorem 2 implies that owing to symmetry induced by the vertex-transitivity of the network, the stationary probability of the node2vec random walk is given by $\boldsymbol{p}^*=\mathbf{1}/N$ regardless of the values of $\alpha$, $\beta$, and $\gamma$.

To analyze the spectral gap, given $k\times k$ matrices $B_i$, where $i=1,2,\ldots ,n$, we define the $kn\times kn$ block circulant matrix $\mathrm{bcirc}(B_1,B_2,\ldots, B_n)$ by
\begin{align}
    \mathrm{bcirc}(B_1,B_2,\ldots,B_n):=
    \begin{pmatrix}
    B_1 & B_2 & \cdots & B_{n-1} & B_n \\
    B_n & B_1 & B_2 & \cdots & B_{n-1} \\
    \vdots  & B_n & B_1 & \ddots & \vdots  \\
    B_3 & \ & \ddots & \ddots & B_2\\
    B_2 & B_3 & \cdots & B_n & B_1
    \end{pmatrix}.
\end{align}

Consider the extended ring network and the set of directed edges $\overline{E}$. Note that there are $2M=4kN$ directed edges in $\overline{E}$. We order the directed edges in $\overline{E}$ as illustrated in Fig. \ref{fig3}(b). Then, the transition probability matrix $\overline{T}$ is block circulant and is given by
\begin{align}
    \overline{T}=\mathrm{bcirc}(0,A,B,0,\ldots,0,C,D),
\end{align}
where
\begin{align}
    A=\frac{1}{\alpha+2\beta+1}\begin{pmatrix}0&0&0&0 \\ 0\:&0&0&0\\ \beta&\alpha&\beta&1\\ 0\:&0&0&0\end{pmatrix},
\end{align}
\begin{align}
    B=\frac{1}{\alpha+\beta+2}\begin{pmatrix}0&0&0&0\\ 0&0&0&0\\ 0&0&0&0\\ \alpha&\beta&1&1\end{pmatrix},
\end{align}
\begin{align}
    C=\frac{1}{\alpha+\beta+2}\begin{pmatrix}1&1&\beta&\alpha\\ 0&0&0&0\\ 0&0&0&0\\ 0&0&0&0\end{pmatrix},
\end{align}
and
\begin{align}
    D=\frac{1}{\alpha+2\beta+1}\begin{pmatrix}0&0&0&0\\ 1&\beta&\alpha&\beta\\ 0&0&0&0\\ 0&0&0&0\end{pmatrix}.
\end{align}
We let
\begin{align}
    \rho_j=e^{i\frac{2\pi j}{N}}
    \label{eq20}
\end{align}
denote the $N$th roots of $1$, where $i$ is the imaginary unit and $j=0,1,2,...,N-1$. Then, we define $4\times 4$ matrices
\begin{align}
    H_j=A\rho_j+B\rho_j^2+C\rho_j^{N-2}+D\rho_j^{N-1},
    \label{eq21}
\end{align}
where $j=0, 1, \ldots, N-1$. In particular, 
\begin{align}
    H_0=A+B+C+D
\end{align}
has a right eigenvector $\boldsymbol 1^\top=(1,1,1,1)^\top$ corresponding to eigenvalue $1$. Theorem 3 in Ref. \cite{tee2007eigenvectors} guarantees that
\begin{align}
    \mathrm{spec}(\overline{T})=\bigcup_{j=0}^{N-1} \mathrm{spec}(H_j), 
    \label{eq23}
\end{align}
where $\mathrm{spec}(\cdot)$ denotes the spectrum of the matrix, i.e. the set of all its eigenvalues (also see Ref. \cite{barrett2017equitable}).

Equation (\ref{eq23}) allows us to calculate $\mathrm{spec}(\overline{T})$, and therefore the spectral gap of $\overline{T}$, by calculating the spectrum of $N$ matrices of size $4$. This method reduces the time for computing the spectral gap from $O(N^3)$ to $O(N)$. The method can be generalized to the $k$-regular extended ring without difficulty, where $k$ is an even number larger than $4$.

\begin{figure}[!h]
    \centering
    \includegraphics[width=0.325\textwidth]{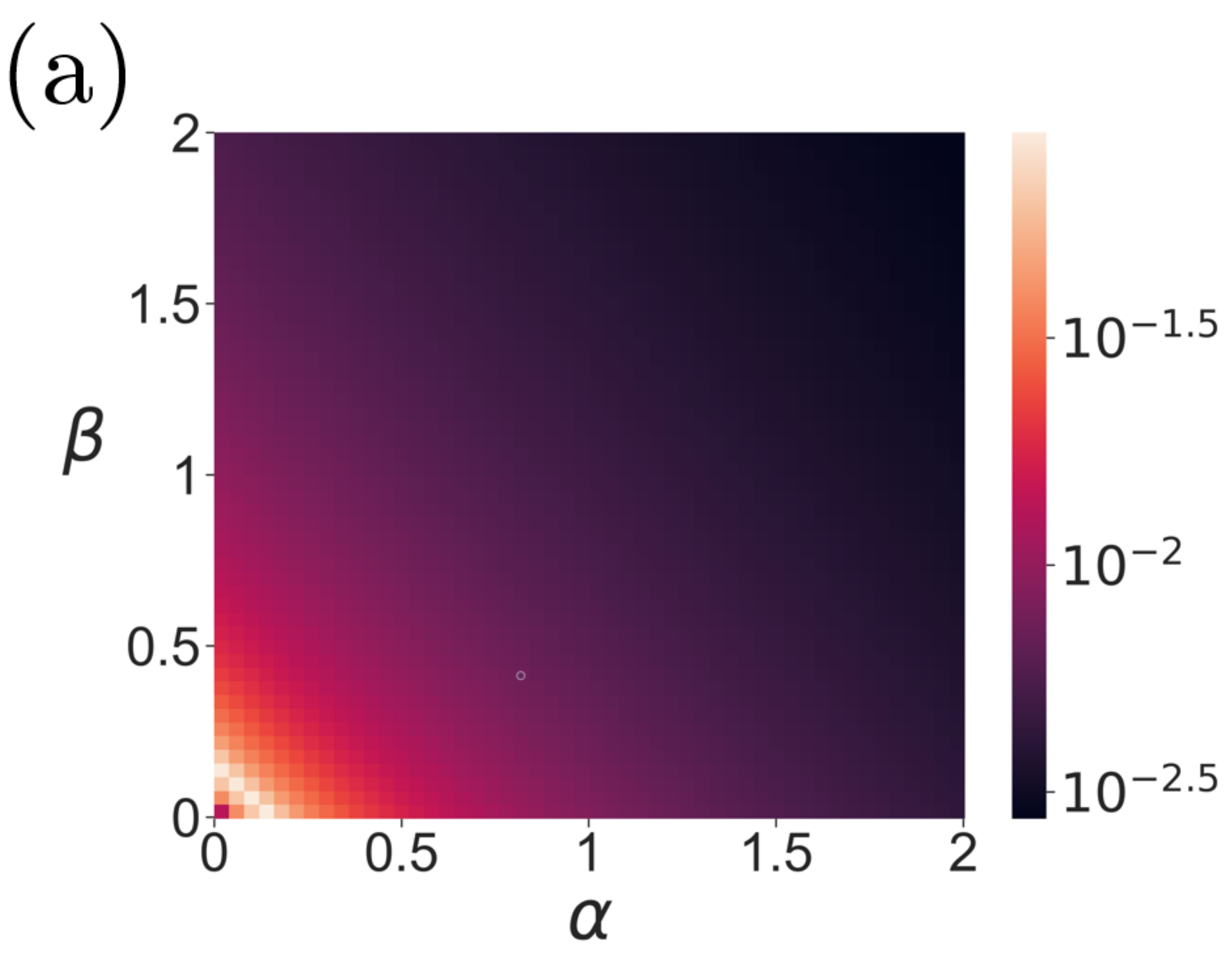}
    \includegraphics[width=0.325\textwidth]{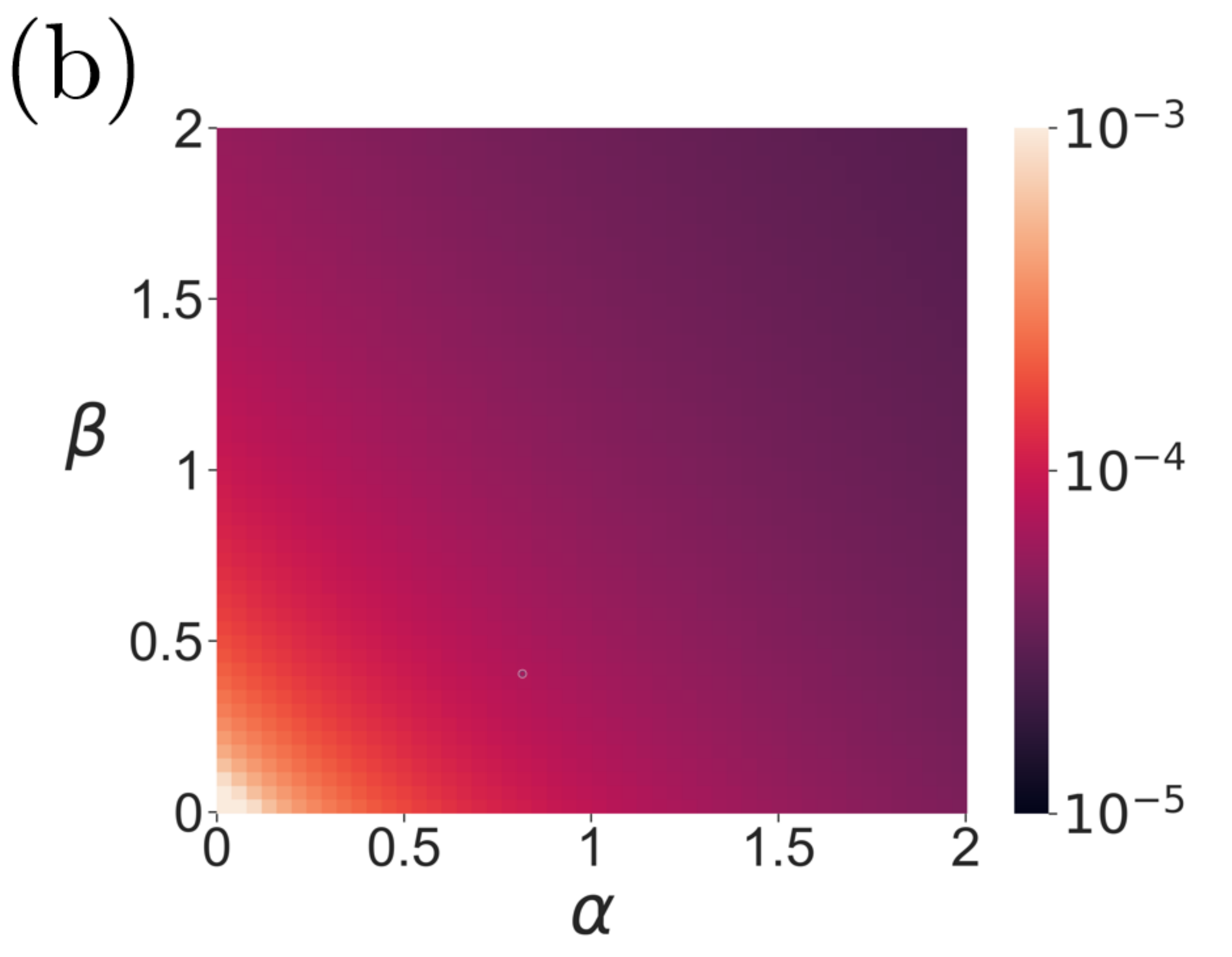}
    \includegraphics[width=0.325\textwidth]{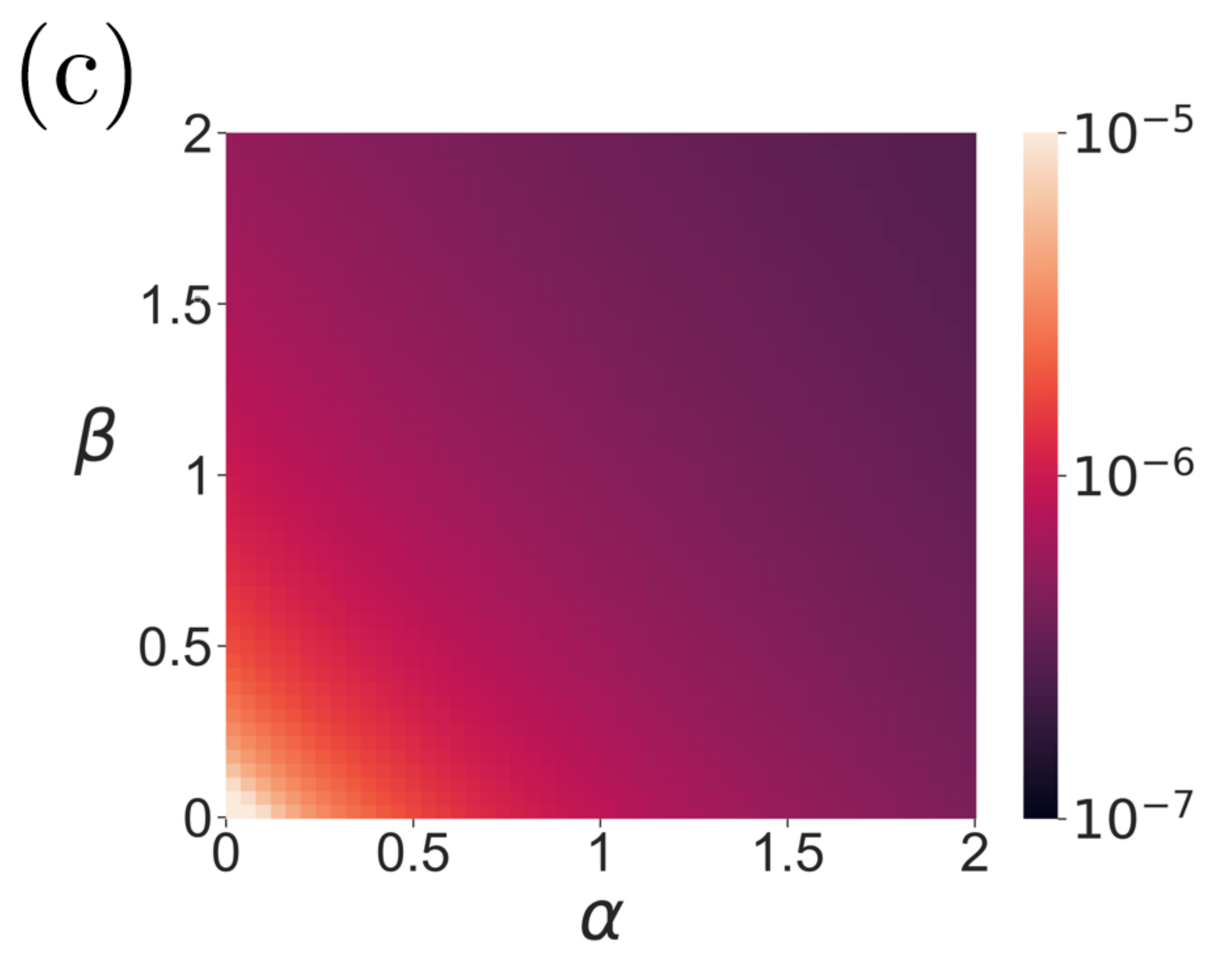}
    \caption{Spectral gap for node2vec random walks on the extended ring network with (a) $N=100$, (b) $N=1000$, and (c) $N=10000$.}
    \label{fig4}
\end{figure}

The spectral gap of $\overline{T}$ for the $4$-regular extended ring networks with $N=100, 1000$, and $10000$ nodes is shown in Fig. \ref{fig4}. The spectral gap is smaller when $N$ is larger for any $\alpha$ and $\beta$. This result is reasonable because the average path length between node pairs is proportional to $N$ for this network. Furthermore, Fig. \ref{fig4} indicates that the spectral gap is large when $\alpha$ and $\beta$ are small for any $N$. However, the spectral gap is not the largest when $\alpha$ and $\beta$ are the smallest when $N=100$ (see Fig. \ref{fig4}(a)). These results are roughly consistent with the results for the empirical networks shown in Section 3(b)(\ref{section:Empirical networks}). When $\alpha$ and $\beta$ are both extremely small, the random walker has to go clockwise or counterclockwise for a long time before changing the direction. We consider that the spectral gap is small when $\alpha$ and $\beta$ are both tiny because the walker skips to visit some nodes when unidirectionally sweeping the ring.

\subsubsection{Two-layer extended ring network}
\label{section:wattstwolayers}

\begin{figure}[!h]
  \includegraphics[width=0.47\textwidth]{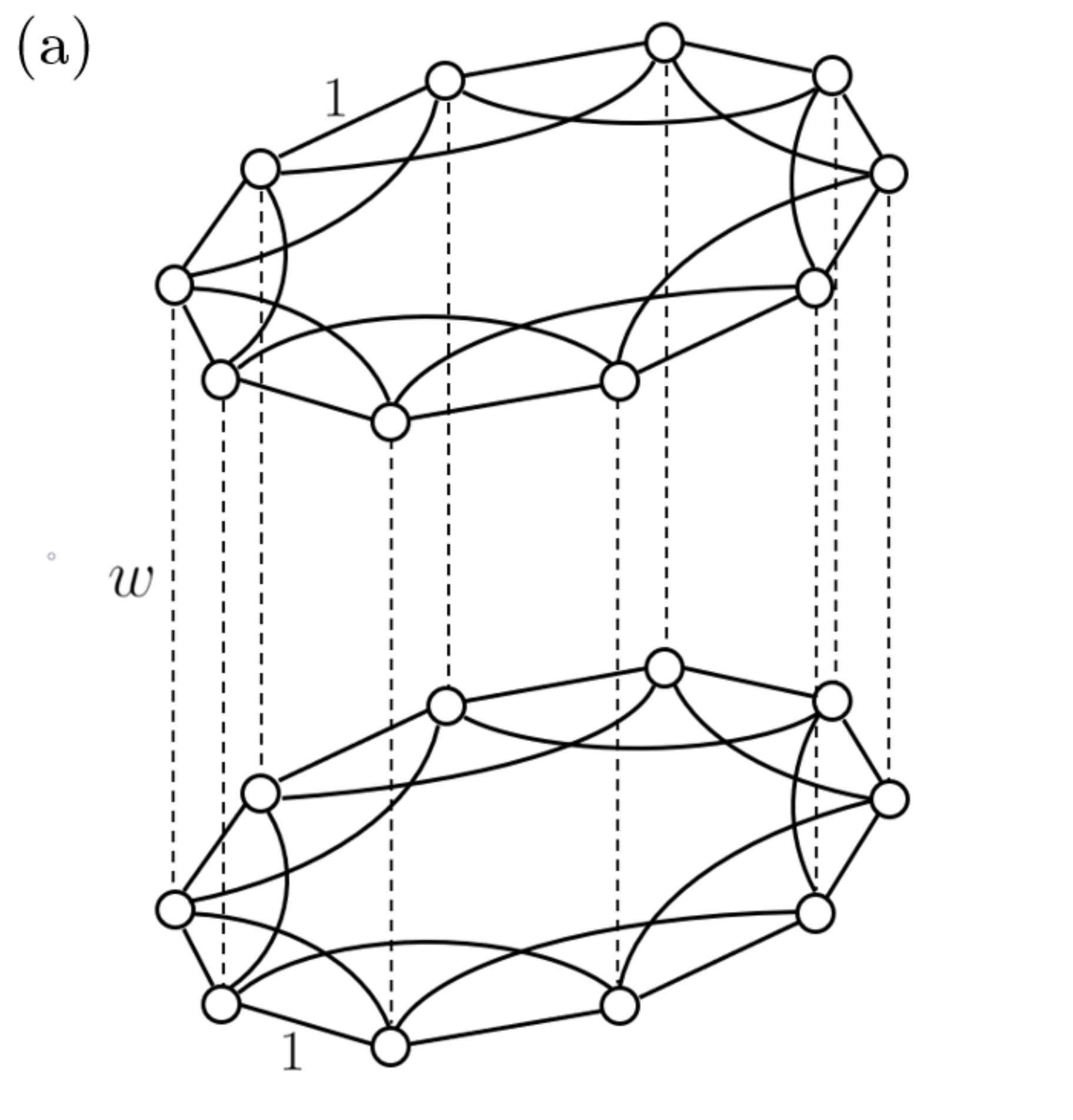}\label{fig5a}\quad%
  \includegraphics[width=0.47\textwidth]{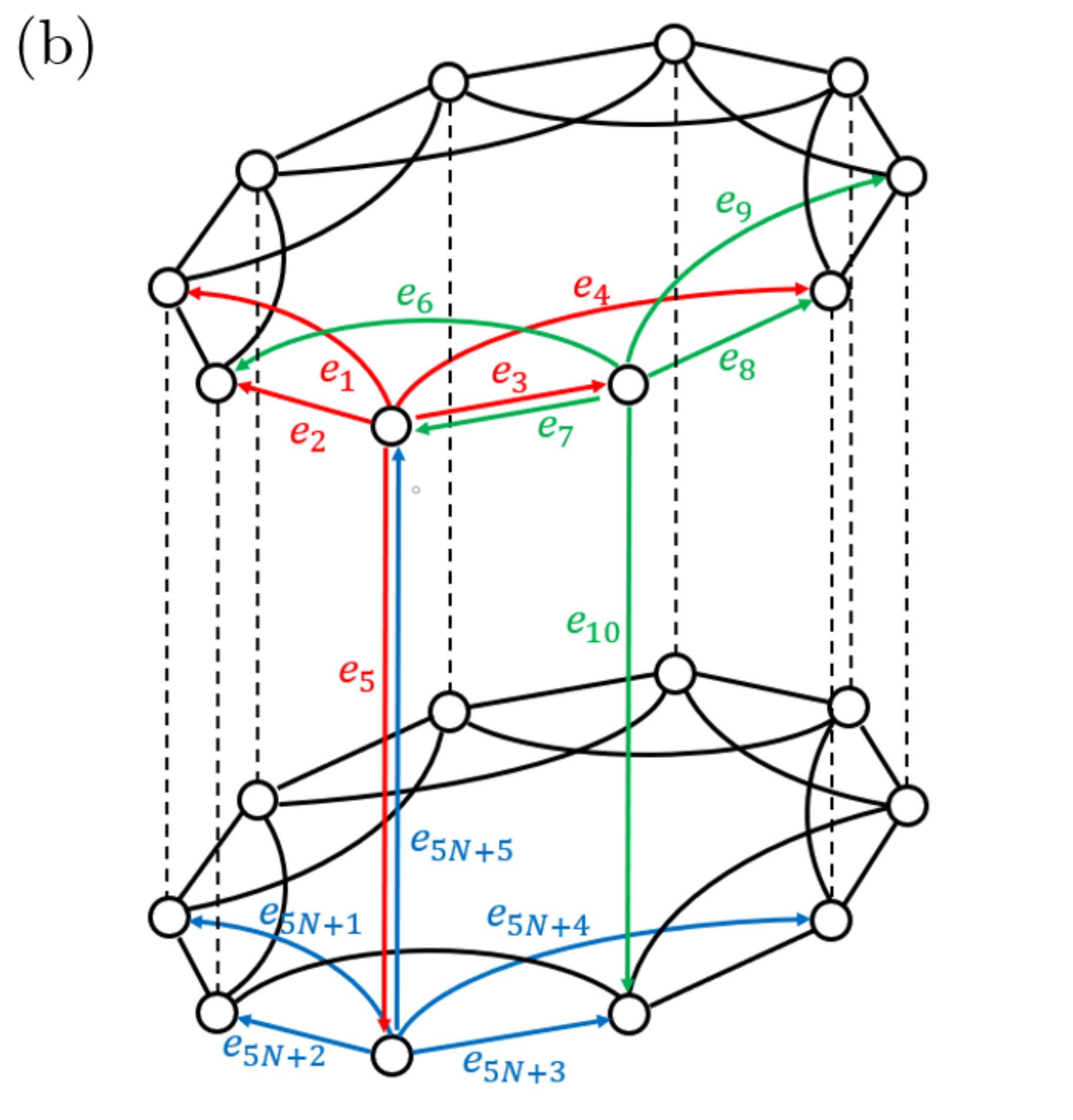}\label{fig5b}\quad%
  \caption{Schematic of the two-layer extended ring network. (a) Two-layer extended ring network with $N^\prime=10$. The solid and dashed lines represent the edges with weight $1$ and $w$, respectively. (b) Labeling convention for its directed edges. }\label{fig5}
  \vspace*{-7pt}
\end{figure}

Similarly, one can also semi-analytically calculate the spectral gap of the transition probability matrix of node2vec random walks on two-layer extended ring networks defined as follows. Consider a pair of extended ring network each of which has $N^{\prime}$ nodes labeled $1, 2, \ldots, N'$ in the same manner, e.g., counterclockwise. Then, we connect the nodes with the same label in the different layers by an edge with weight $w$ (Fig. \ref{fig5}(a)). We assume that the edges within each extended ring have weight $1$. The obtained network is an undirected weighted network with $N=2N^\prime$ nodes. Note that each node $v$ has degree $5$; four edges in the same layer as $v$ have weight $1$, and the other edge connecting the two layers has weight $w$. The network is composed of two communities when $w$ is small. Furthermore, it can be regarded as a multilayer network with two layers under the so-called ordinal coupling \cite{kivela2014multilayer, boccaletti2014structure, bianconi2018multilayer}. 

Consider the node2vec random walk on this network. For example, as the first-order random walk on the $2M$ directed edges, the transition probability from $e_5$ to $e_{5N+4}$ in Fig. \ref{fig5}(b) is given by $\overline{T}_{5,5N+4}=\gamma/(\alpha w +4\gamma)$, and that from $e_6$ to $e_5$ is given by $\overline{T}_{6,5}=\gamma w/(\alpha +2\beta+\gamma+\gamma w)$.

Because the network is vertex-transitive, Theorem 2 implies that the stationary probability $\boldsymbol{p}^*=\mathbf{1}/N$. To analyze the spectral gap of this network, we label the $5N$ directed edges as shown in Fig. \ref{fig5}(b).

The transition probability matrix $\overline{T}$ is a block circulant matrix given by
\begin{align}
    \overline{T} = \mathrm{bcirc}(M_1, M_2) = \begin{pmatrix}M_1&M_2\\ M_2&M_1\end{pmatrix},
\end{align}
where $5N^\prime \times 5N^\prime$ matrices $M_1$ and $M_2$ are themselves block circulant matrices. Matrices $M_1$ and $M_2$ are given by
\begin{align}
    M_1=\mathrm{bcirc}(0,A,B,0,\ldots,0,C,D),
\end{align}
and
\begin{align}
    M_2=\mathrm{bcirc}(E,0,\ldots,0),
\end{align}
where
\begin{align}
    A=\frac{1}{\alpha+2\beta+w+1}\begin{pmatrix}0&0&0&0&0\\ 0&0&0&0&0\\ \beta&\alpha&\beta&1&w\\ 0&0&0&0&0\\ 0&0&0&0&0\end{pmatrix},
\end{align}
\begin{align}
    B=\frac{1}{\alpha+\beta+w+2}\begin{pmatrix}0&0&0&0&0\\ 0&0&0&0&0\\ 0&0&0&0&0\\ \alpha&\beta&1&1&w\\ 0&0&0&0&0\end{pmatrix},
\end{align}
\begin{align}
    C=\frac{1}{\alpha+\beta+w+2}\begin{pmatrix}1&1&\beta&\alpha&w\\ 0&0&0&0&0\\ 0&0&0&0&0\\ 0&0&0&0&0\\ 0&0&0&0&0\end{pmatrix},
\end{align}  
\begin{align}
    D=\frac{1}{\alpha+2\beta+w+1}\begin{pmatrix}0&0&0&0&0\\ 1&\beta&\alpha&\beta&w\\ 0&0&0&0&0\\ 0&0&0&0&0\\ 0&0&0&0&0\end{pmatrix},
\end{align}
\begin{align}
    E=\frac{1}{\alpha w+4}\begin{pmatrix}0&0&0&0&0\\ 0&0&0&0&0\\ 0&0&0&0&0\\ 0&0&0&0&0\\ 1&1&1&1&\alpha w\end{pmatrix}.
\end{align}
Theorem 3 in Ref. \cite{tee2007eigenvectors} yields
\begin{align}
    \mathrm{spec}(\overline{T})=\mathrm{spec}(M_1+M_2)\cup \mathrm{spec}(M_1-M_2).
\end{align}
We define
\begin{align}
    H_j=E+A\rho_j+B\rho_j^2+C\rho_j^{N-2}+D\rho_j^{N-1}
\end{align}
and
\begin{align}
    G_j=-E+A\rho_j+B\rho_j^2+C\rho_j^{N-2}+D\rho_j^{N-1},
\end{align}
where $\rho_j$ is given by Eq. (\ref{eq20}). Because $M_1+M_2$ and $M_1-M_2$ are block circulant, one obtains
\begin{align}
    \mathrm{spec}(M_1+M_2)=\bigcup_{j=0}^{N-1} \mathrm{spec}(H_j),
\end{align}
and 
\begin{align}
    \mathrm{spec}(M_1-M_2)=\bigcup_{j=0}^{N-1} \mathrm{spec}(G_j).
\end{align}
Therefore, the spectrum of $\overline{T}$ is given by
\begin{align}
    \mathrm{spec}(\overline{T})=\bigcup_{j=0}^{N-1} \left[\mathrm{spec}(H_j)\cup \mathrm{spec}(G_j)\right].
    \label{eq39}
\end{align}

Similar to the case of mono-layer extended ring networks, this method enables practical computation of the spectrum and the spectral gap for two-layer extended ring networks of various sizes and can be easily generalized to two-layer $k$-regular extended ring networks. Equation (\ref{eq39}) implies that one can reduce the computation time from $O(N^3)$ to $O(N)$.

\begin{figure}[!h]
  \flushleft
  \includegraphics[width=0.47\textwidth]{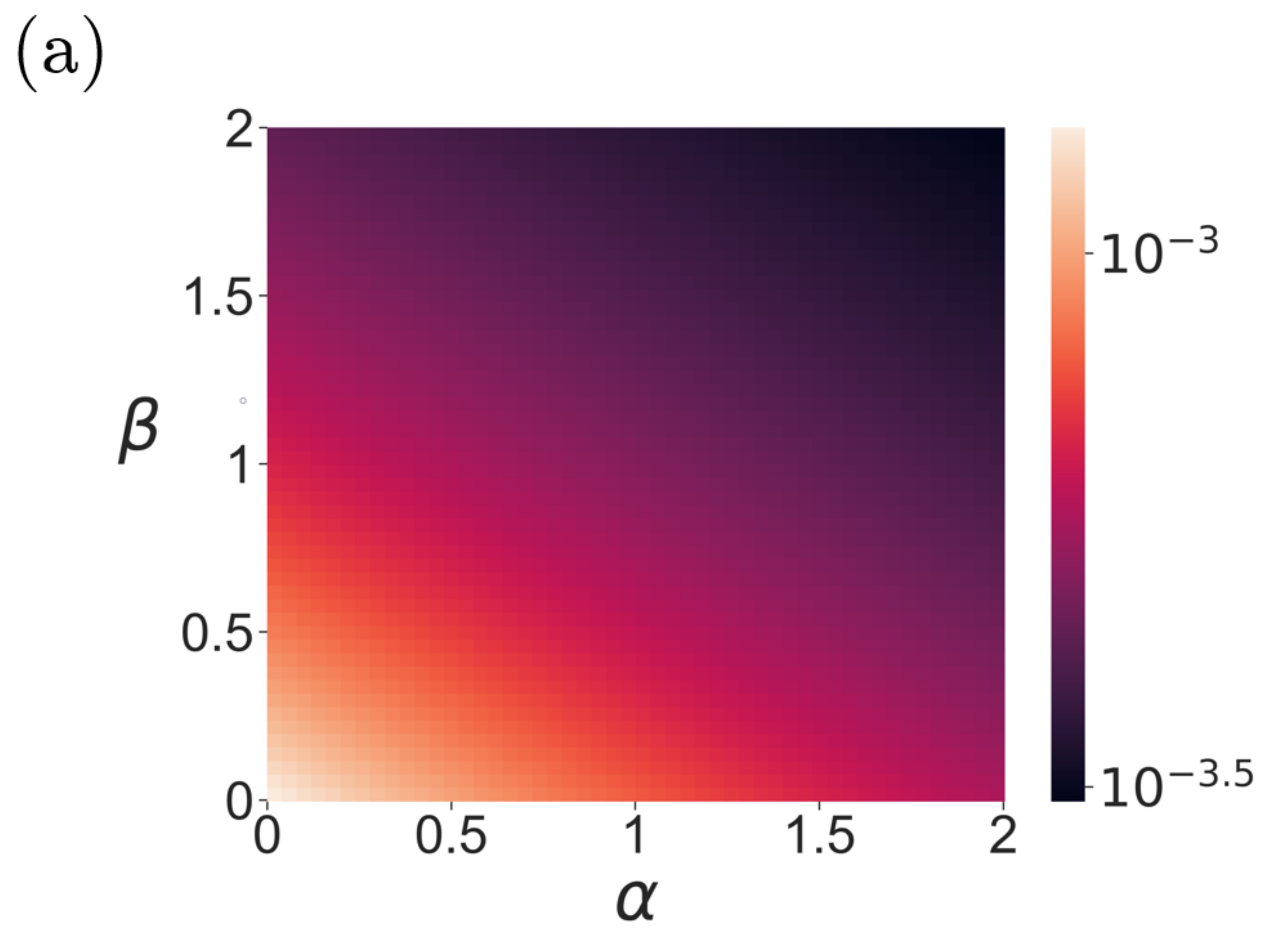}\label{fig6a0}\quad%
  \includegraphics[width=0.47\textwidth]{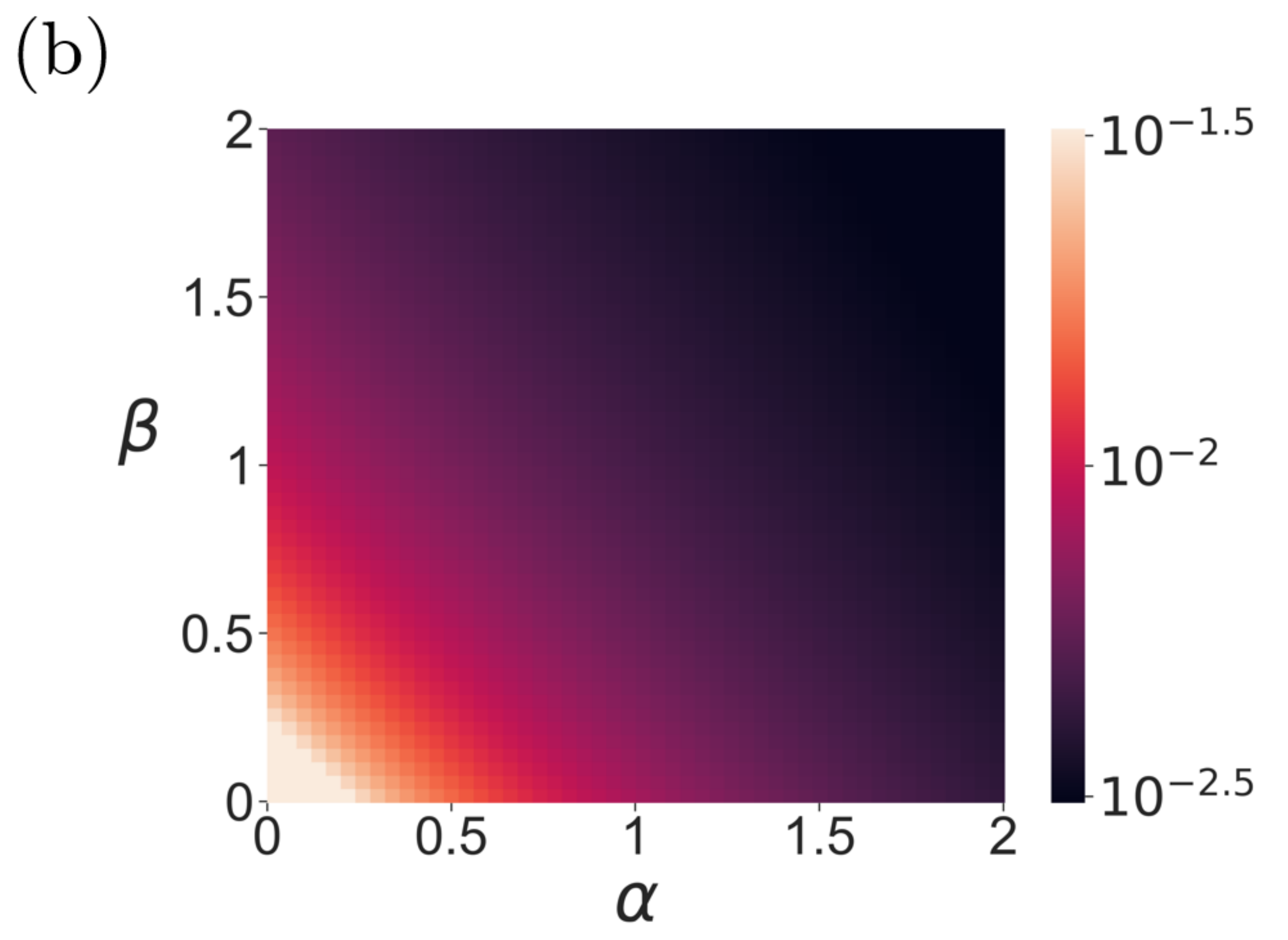}\label{fig6a}\quad%
  \includegraphics[width=0.47\textwidth]{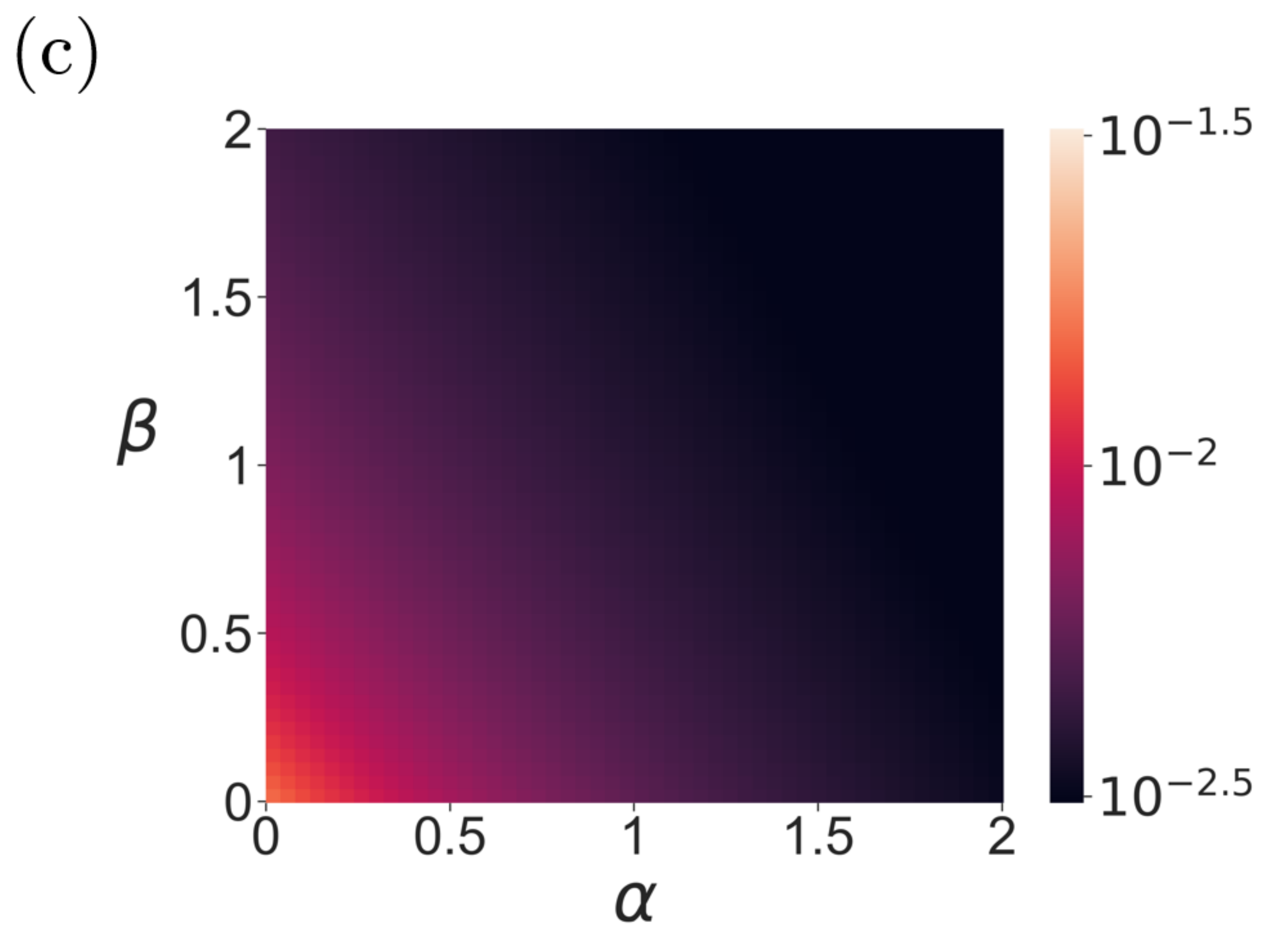}\label{fig6b}\quad%
  \includegraphics[width=0.47\textwidth]{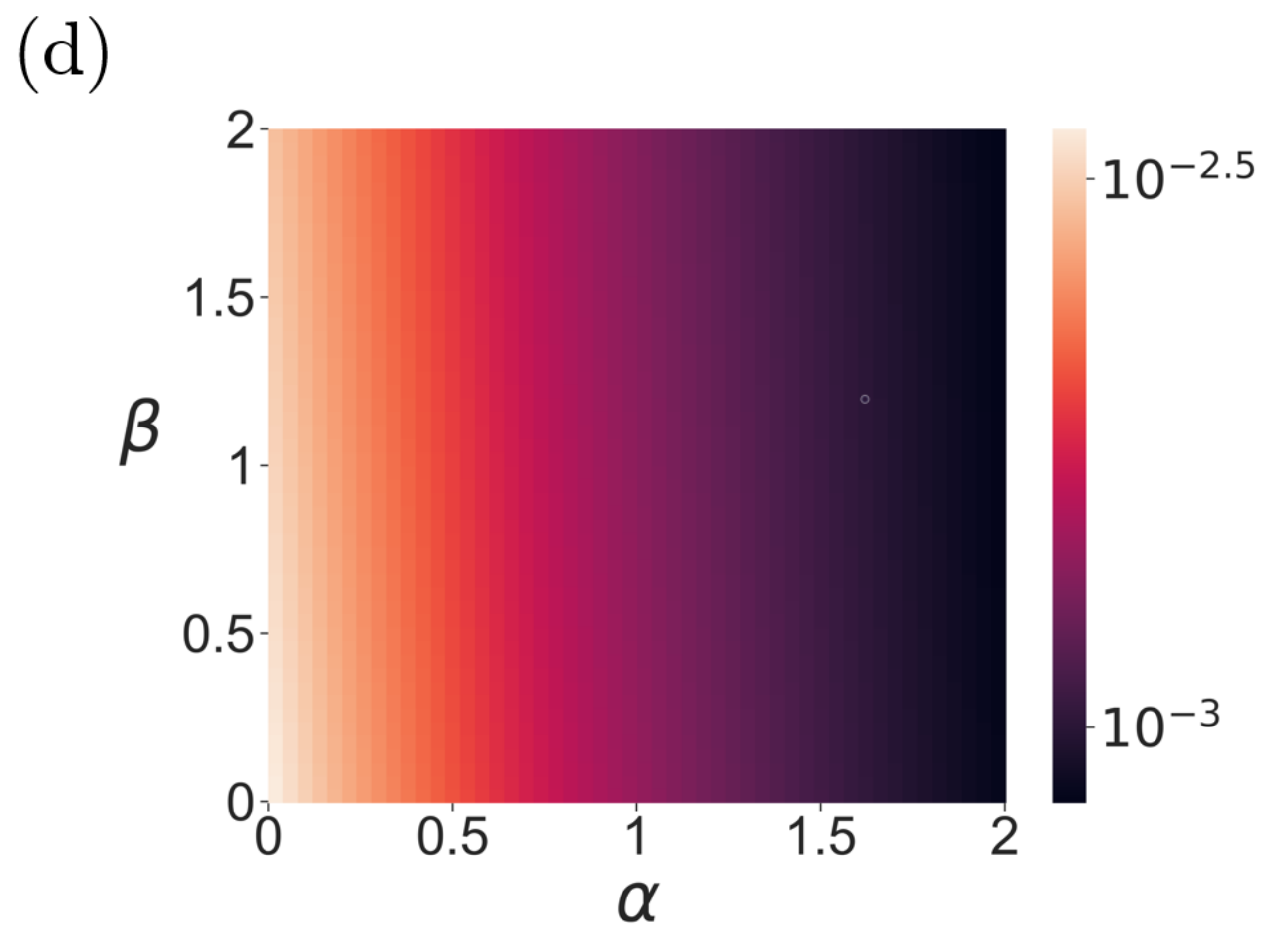}\label{fig6c}\quad%
  \caption{Spectral gap for node2vec random walks on the  two-layer extended ring network with $N^\prime= 100$. (a) $w=0.001$. (b) $w=0.1$. (c) $w=1$. (d) $w=10$.}\label{fig6}
  \vspace*{-7pt}
\end{figure}

Numerically calculated spectral gaps for the two-layer extended ring networks with $N=200$ nodes are shown in Fig. \ref{fig6} for various $\alpha$ and $\beta$ values and four values of $w$. We find that backtracking (i.e., large $\alpha$) slows down mixing for all the $w$ values. When $w$ is small, the spectral gap increases as $\alpha$ or $\beta$ decreases (Figs. \ref{fig6}(a), \ref{fig6}(b) and \ref{fig6}(c)). These results are consistent with the results for the empirical networks and the mono-layer extended ring network. When $w$ is large, movements between the two layers are frequent. In this case, the spectral gap decreases as $\alpha$ increases, whereas it is relatively insensitive to $\beta$ within the range of $\beta$ values that we have explored (Fig. \ref{fig6}(d)). In this situation, a random walker that visits more neighbors within the same layer by the breadth-first sampling mechanism (i.e., large $\beta$) mixes roughly as fast as a walker that frequently switches the layer (i.e., small $\beta$). The dependence of the spectral gap on $N$ is examined in the SM.

Last, Fig. \ref{fig6} indicates that the spectral gap is not monotonic in terms of $w$ for any given $\alpha$ and $\beta$ values. When $w$ is small (Fig. \ref{fig6}(a)), walkers find it difficult to transit from one layer to the other, which poses a bottleneck of diffusion. The spectral gap is the largest (i.e., relaxation is the fastest) for an intermediate value of $w$ ($w=0.1$ among the four values of $w$; Fig. \ref{fig6}(b)). When $w$ is larger (Figs. \ref{fig6}(c) and \ref{fig6}(d)), the diffusion is decelerated presumably because exploration within the individual layers is not enough relative to inter-layer moves. This deceleration result is opposite to the previous result that strong inter-layer coupling makes the spectral gap larger than for random walks confined to the individual layers for simple random walks \cite{PhysRevLett.110.028701}. The difference may be ascribed to the different types of random walks employed in these studies, i.e., simple random walks in Ref. \cite{PhysRevLett.110.028701} and node2vec random walks in the present study.

\subsection{Mean coalescence time on two-clique networks}

In this section, we provide an analysis that is different from the spectral gap with the aim of supporting our main claim that diffusion accelerates with small $\alpha$ and $\beta$ values. The voter model is a linear stochastic model of collective opinion formation, where each node in the network has one of the two opinions, denoted by $A$ and $B$ \cite{liggett2012interacting}. At least in finite networks, the consensus of opinion $A$ and that of $B$ are the only absorbing states. The duality relationship guarantees that the mean time to consensus is given by the mean time to coalescence of $N$ coalescing random walkers deployed on each node of the edge-reversed network into one walker \cite{aldous1995reversible, masuda2017random, liggett2012interacting, donnelly1983finite}. There are two random walkers just before all the $N$ walkers coalesce into one walker. Therefore, in this section, we evaluate the mean time to coalescence of two node2vec random walkers as an alternative measure of speed of diffusion.

\begin{figure}[!h]
  \flushleft
  \includegraphics[width=0.31\textwidth]{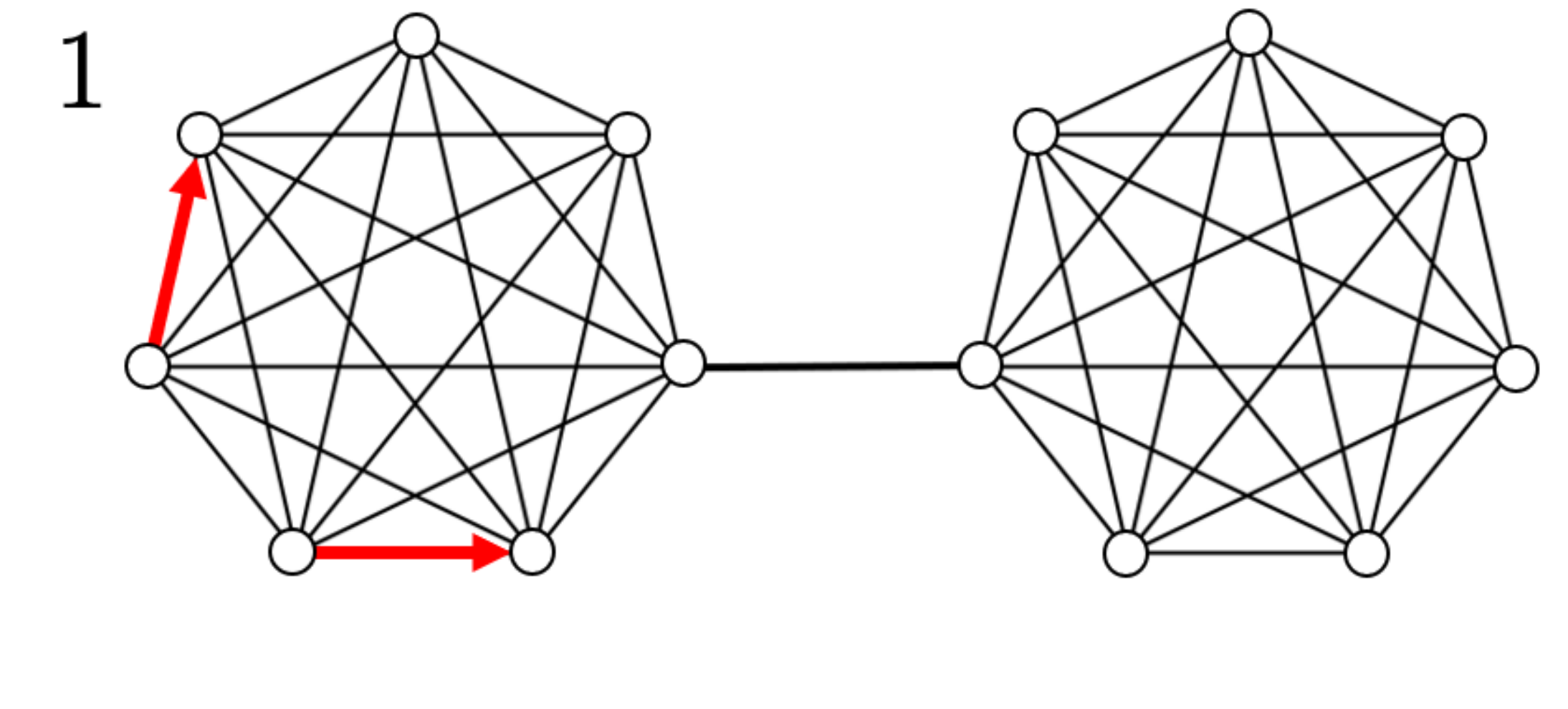}\label{fig:sub1}\quad%
  \includegraphics[width=0.31\textwidth]{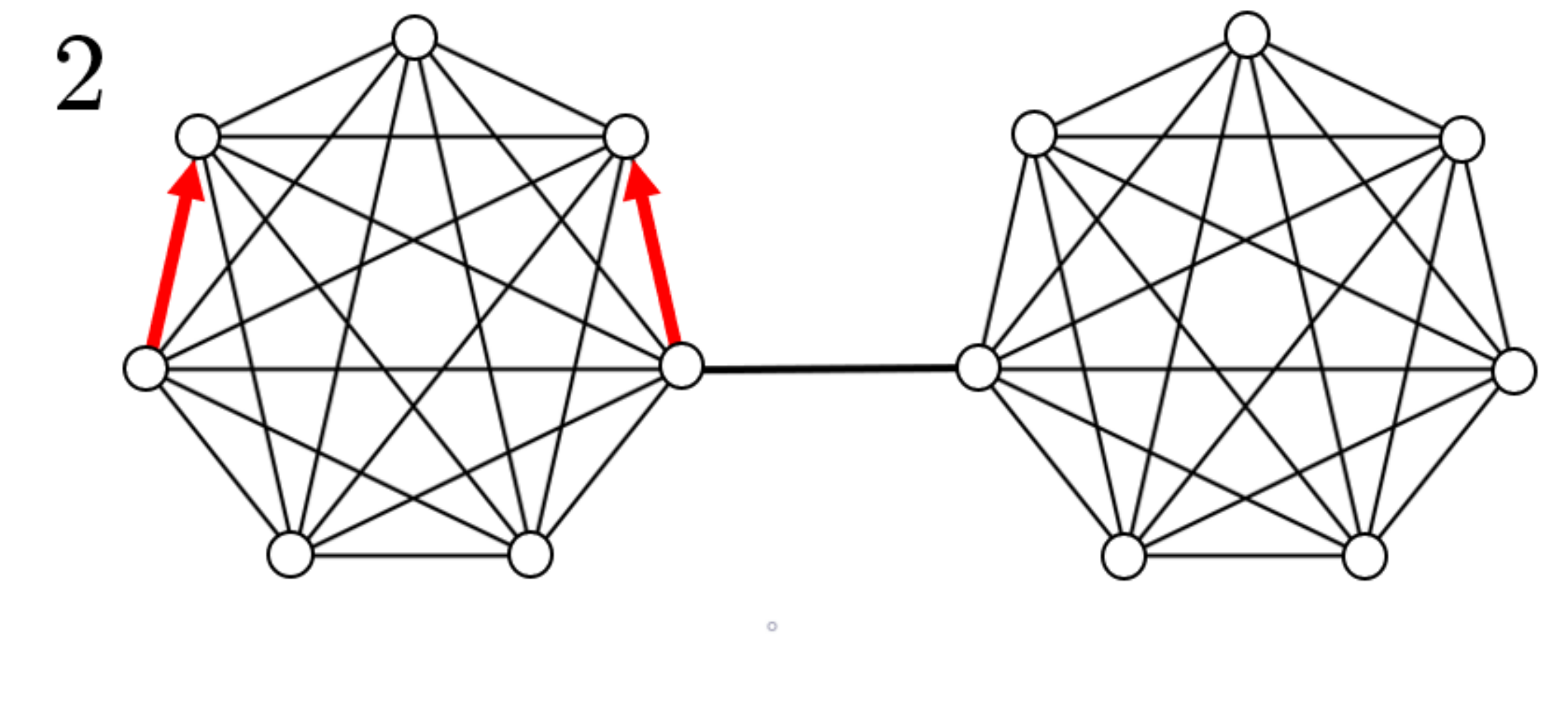}\label{fig:sub2}\quad%
  \includegraphics[width=0.31\textwidth]{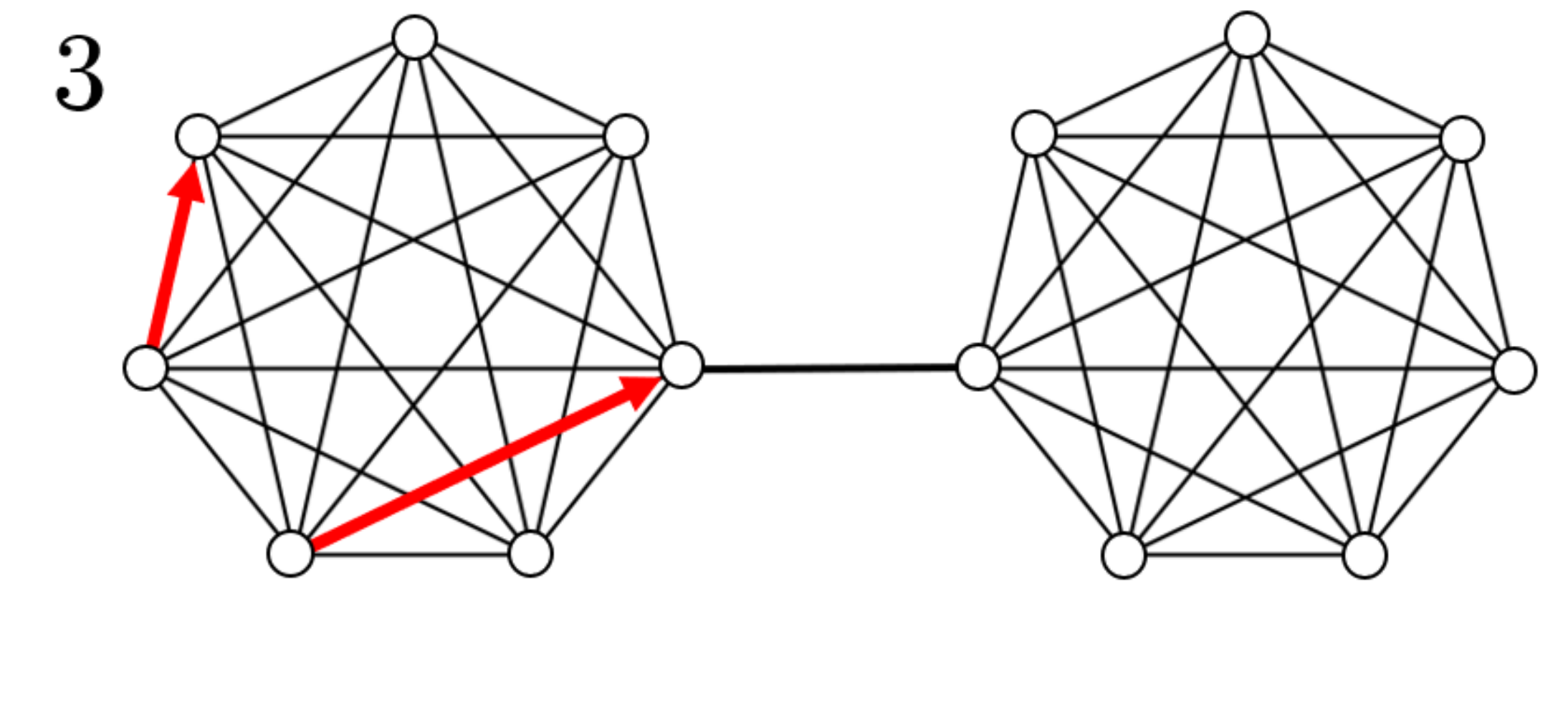}\label{fig:sub3}\quad%
  \includegraphics[width=0.31\textwidth]{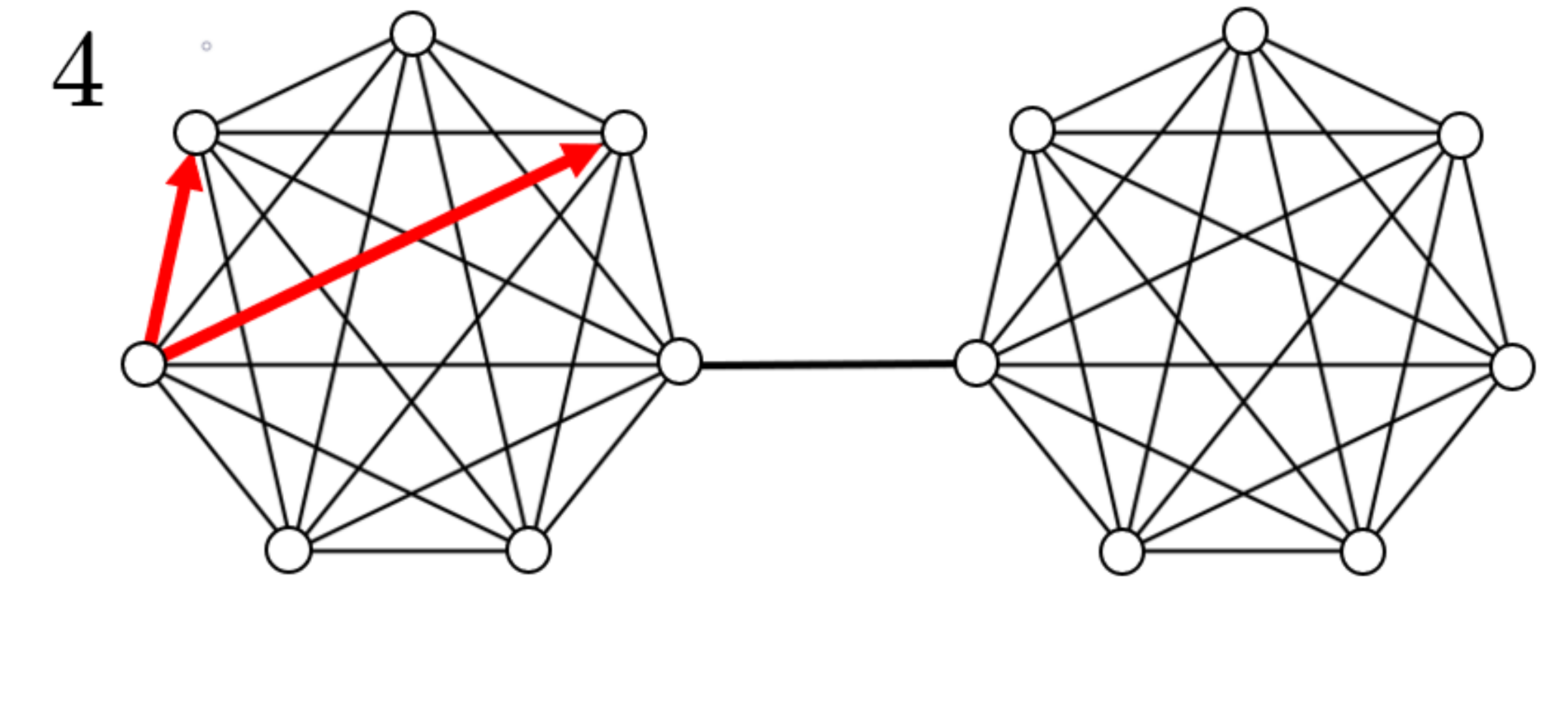}\label{fig:sub4}\quad%
  \includegraphics[width=0.31\textwidth]{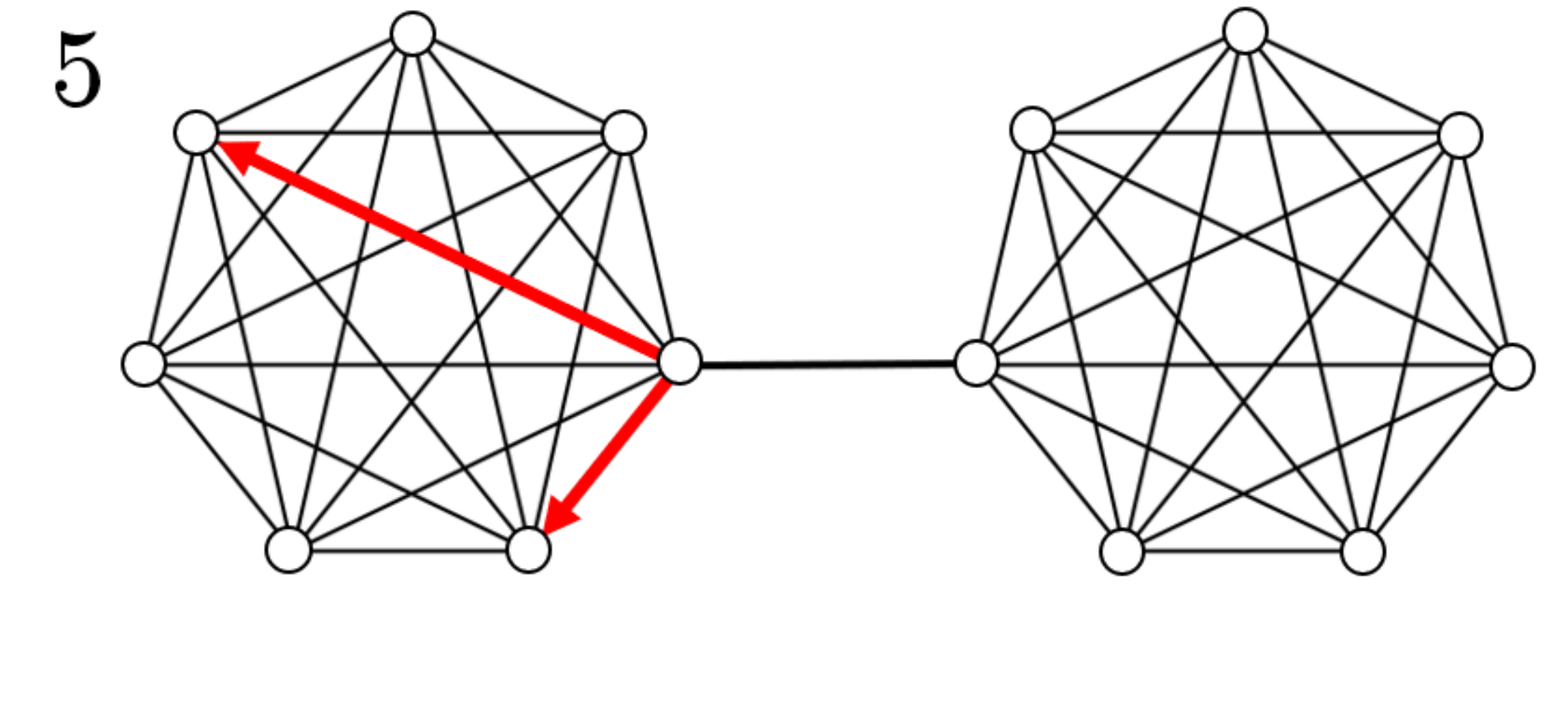}\label{fig:sub5}\quad%
  \includegraphics[width=0.31\textwidth]{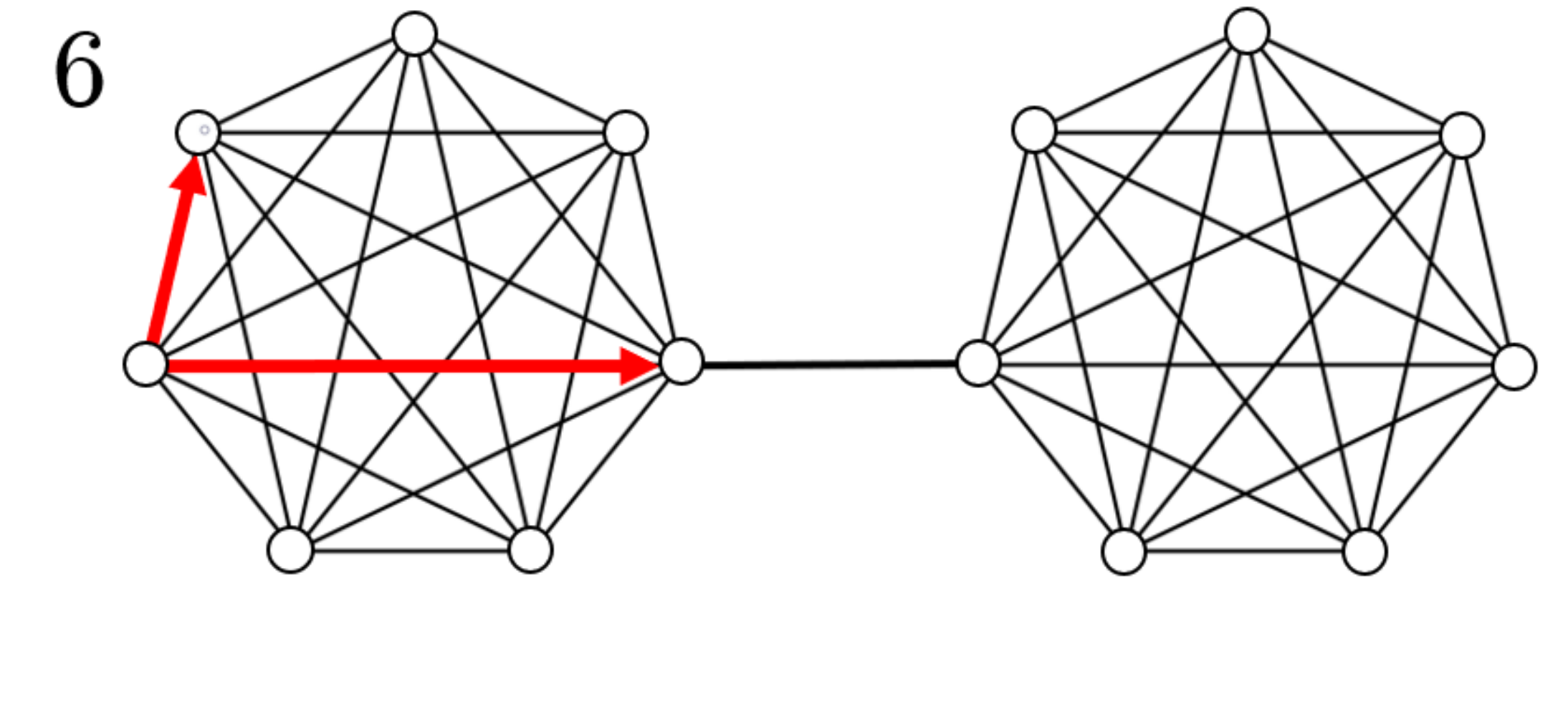}\label{fig:sub6}\quad%
  \includegraphics[width=0.31\textwidth]{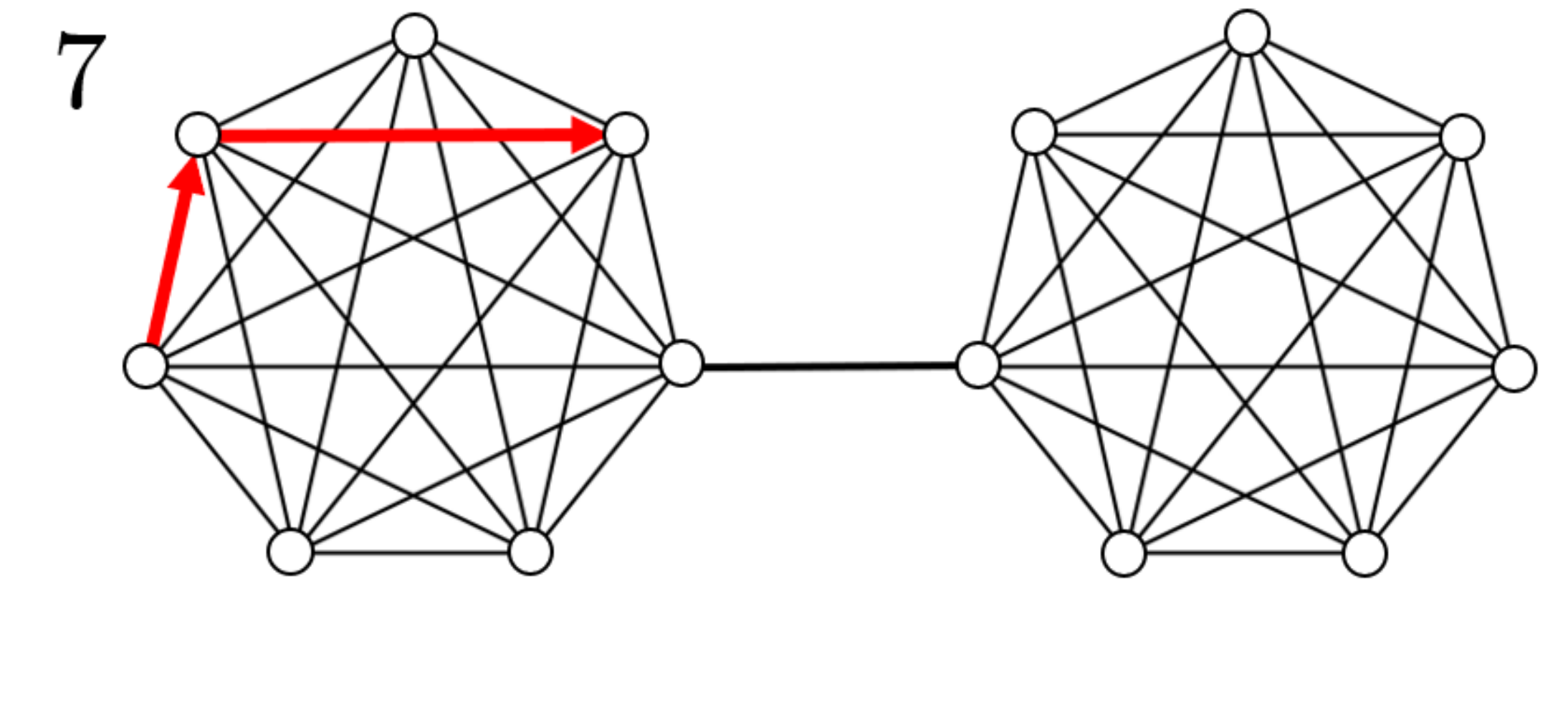}\label{fig:sub7}\quad%
  \includegraphics[width=0.31\textwidth]{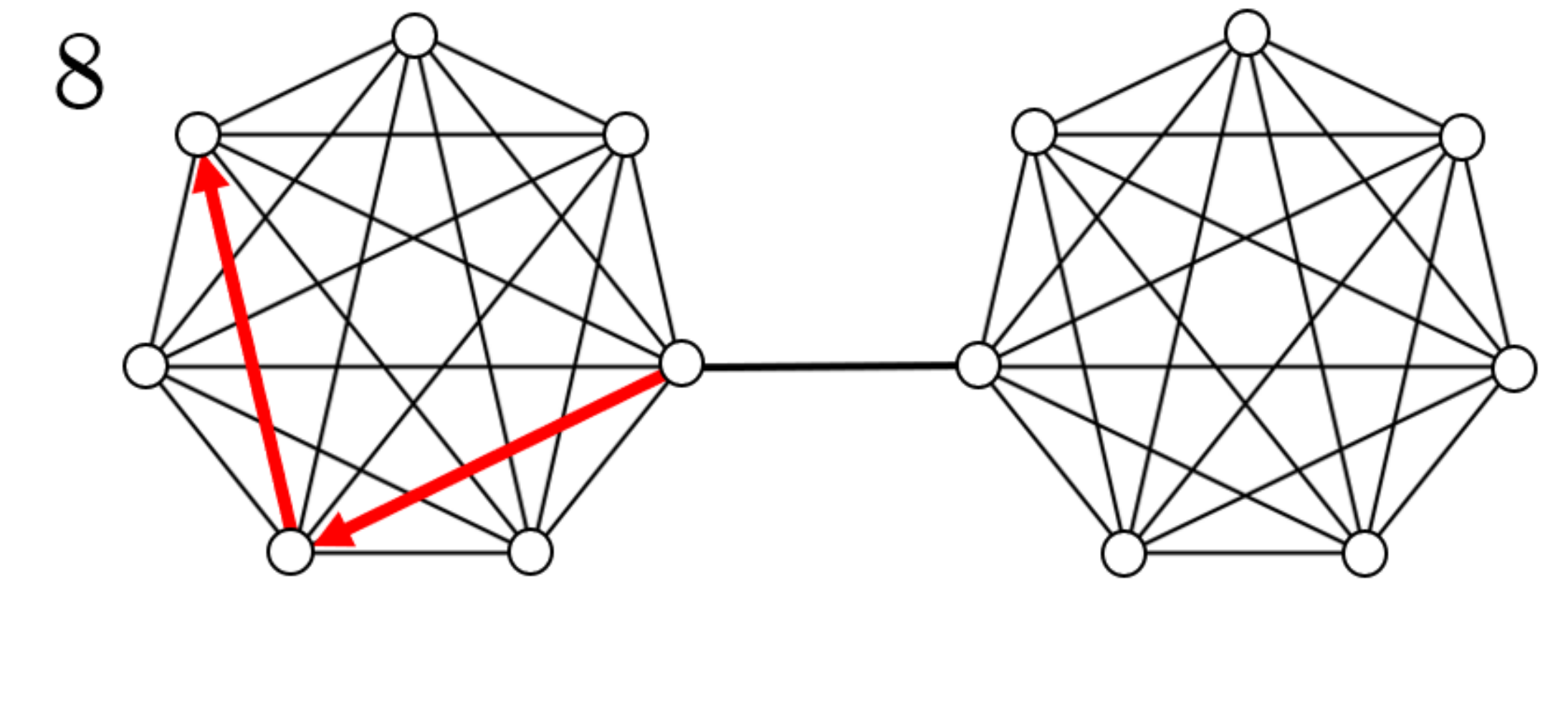}\label{fig:sub8}\quad%
  \includegraphics[width=0.31\textwidth]{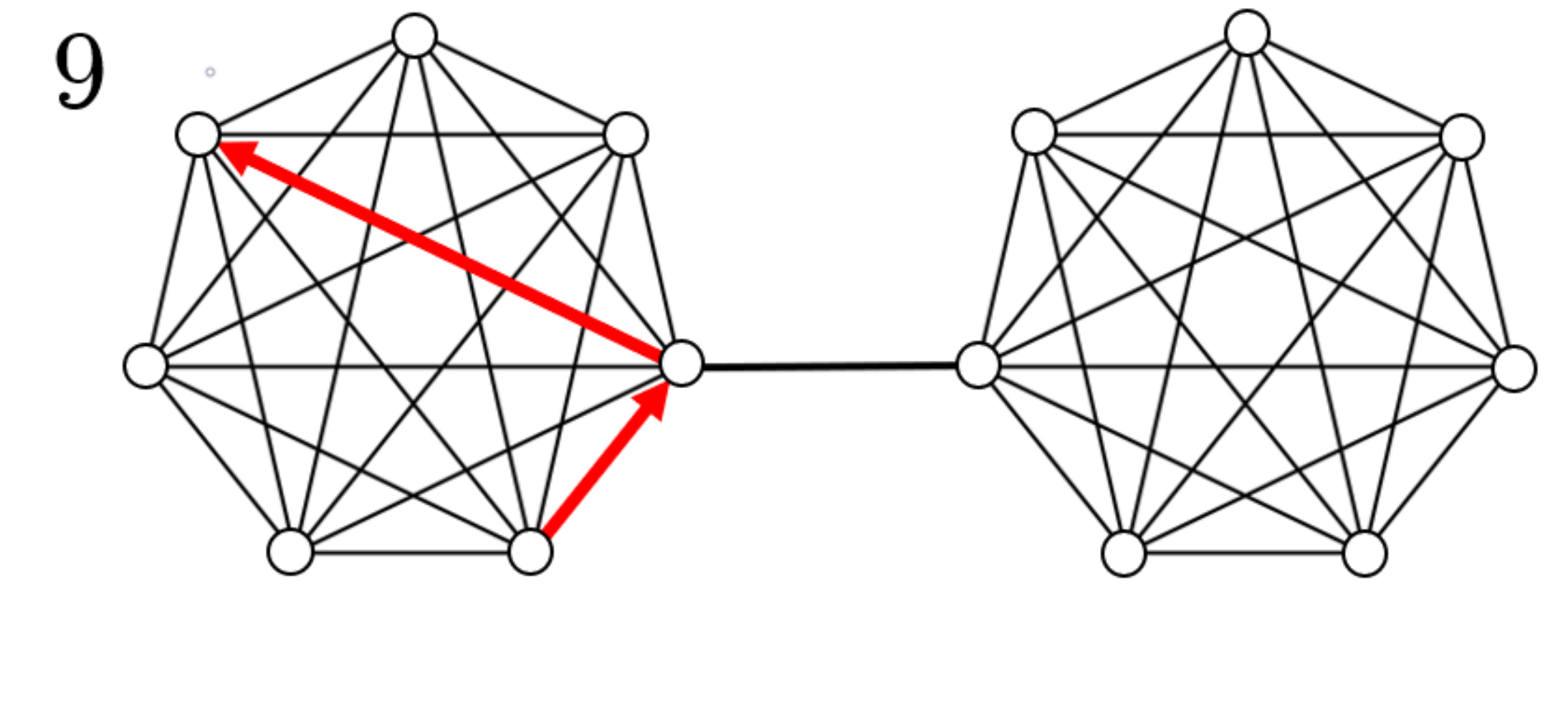}\label{fig:sub9}\quad%
  \includegraphics[width=0.31\textwidth]{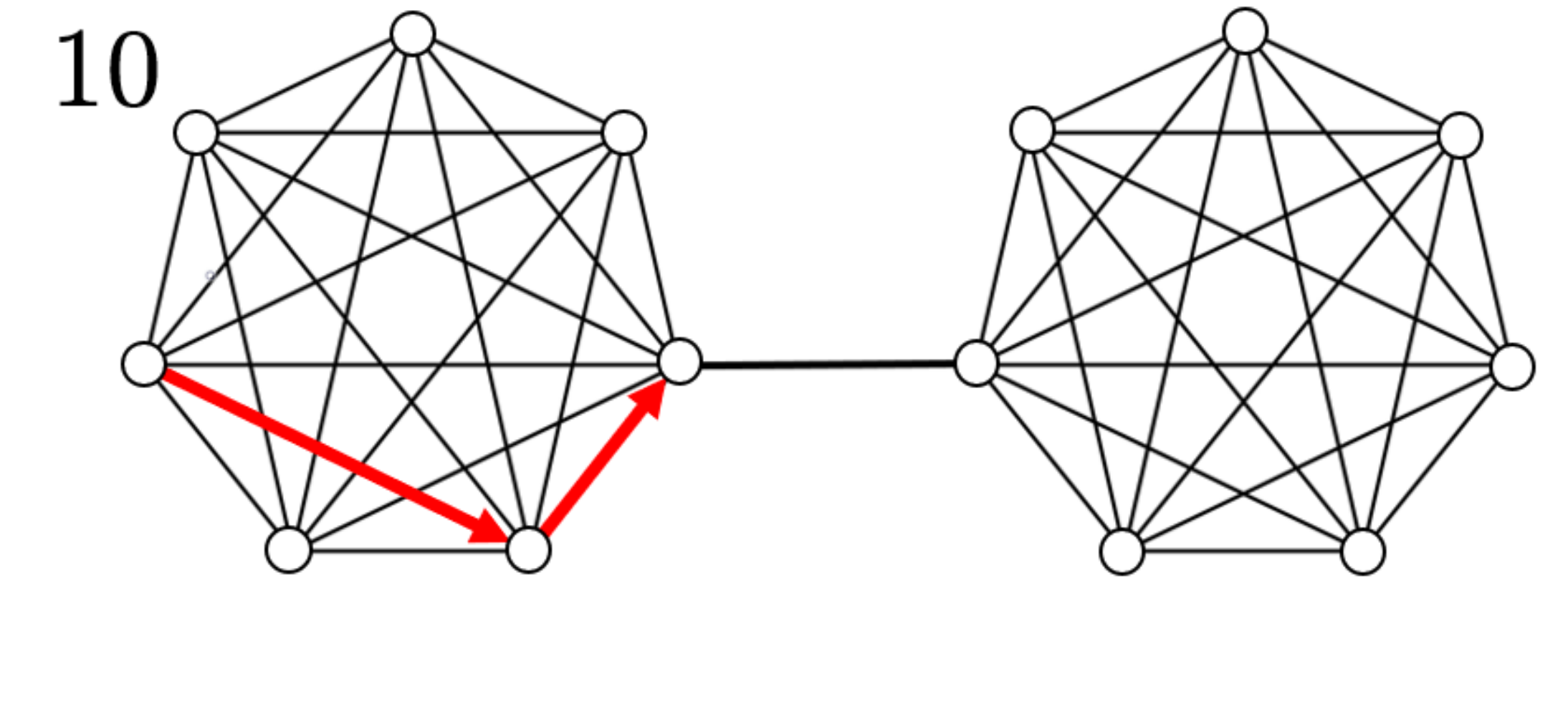}\label{fig:sub10}\quad%
  \includegraphics[width=0.31\textwidth]{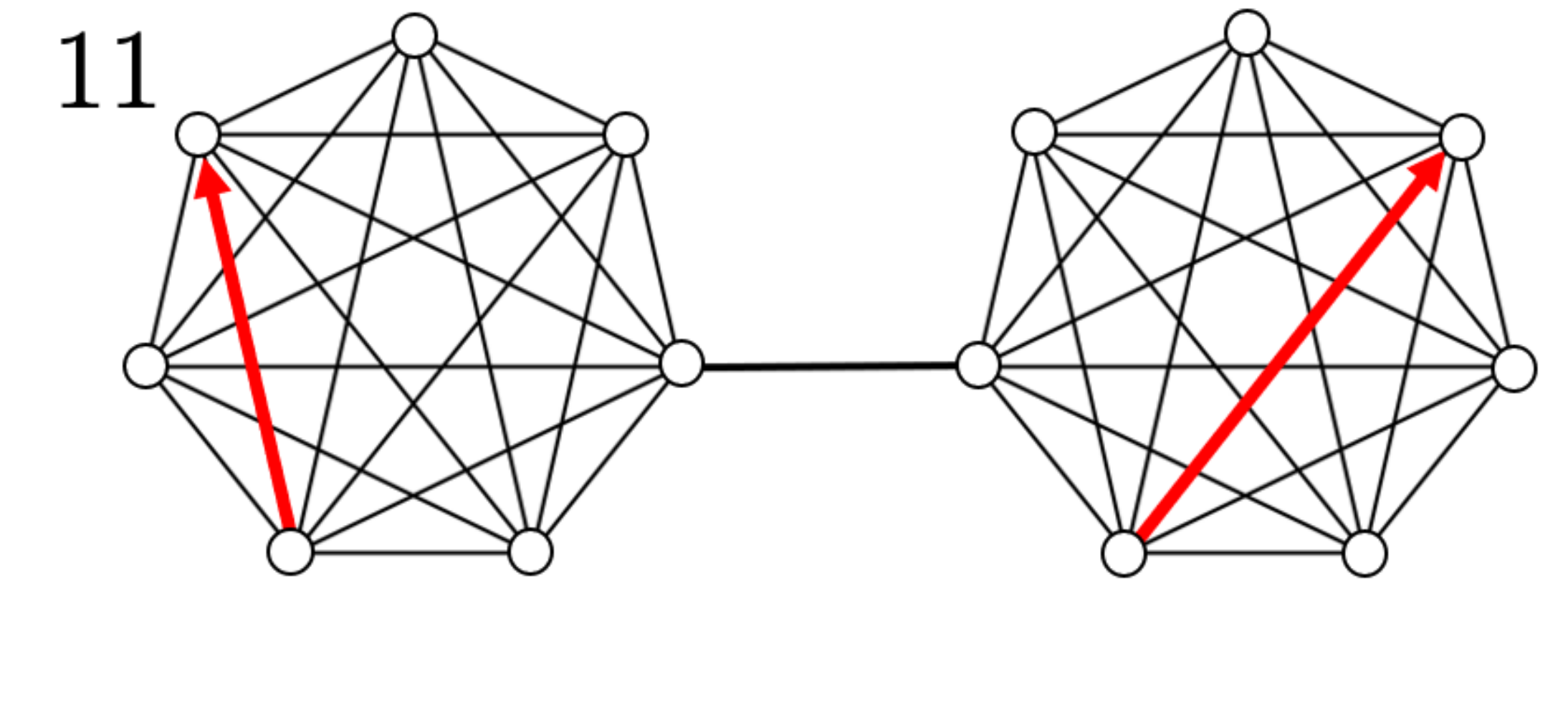}\label{fig:sub11}\quad%
  \includegraphics[width=0.31\textwidth]{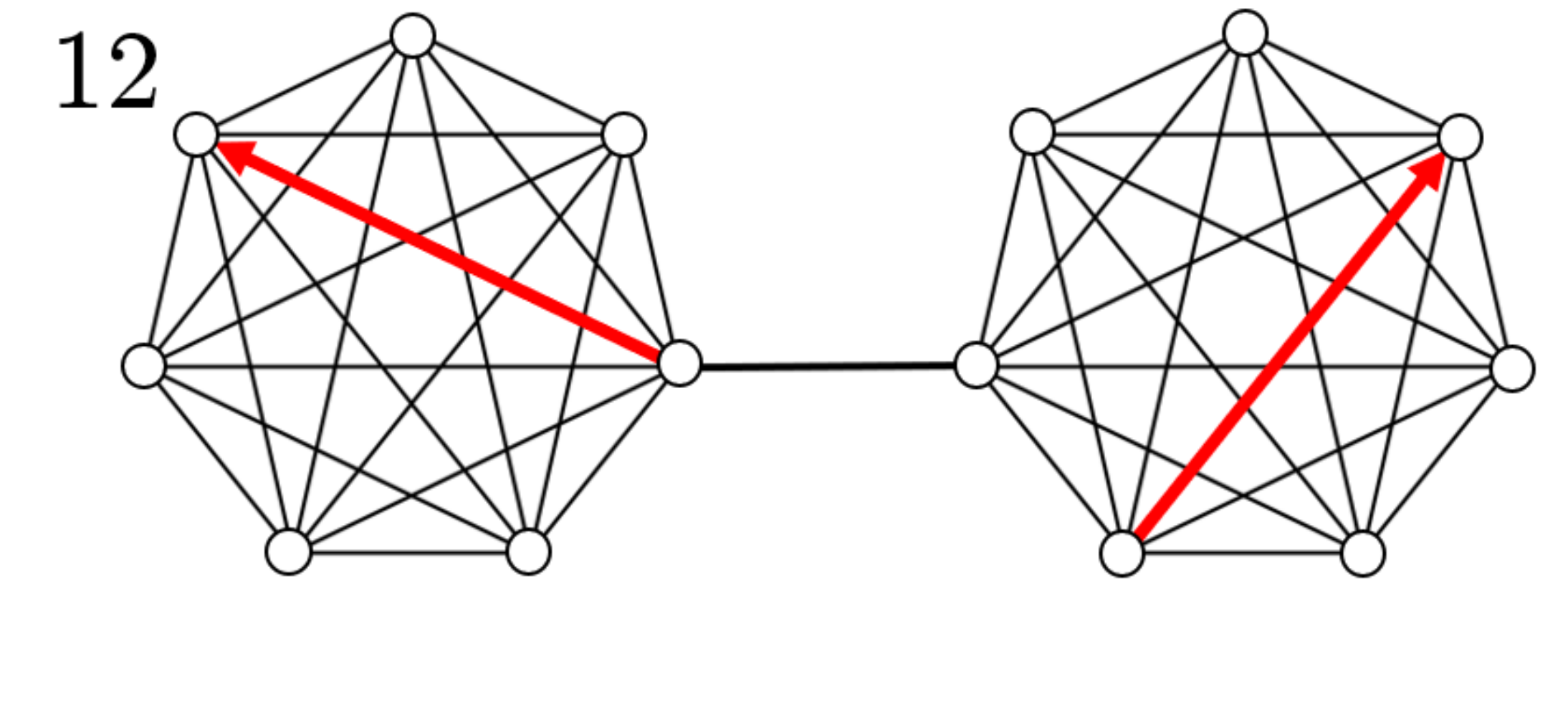}\label{fig:sub12}\quad%
  \includegraphics[width=0.31\textwidth]{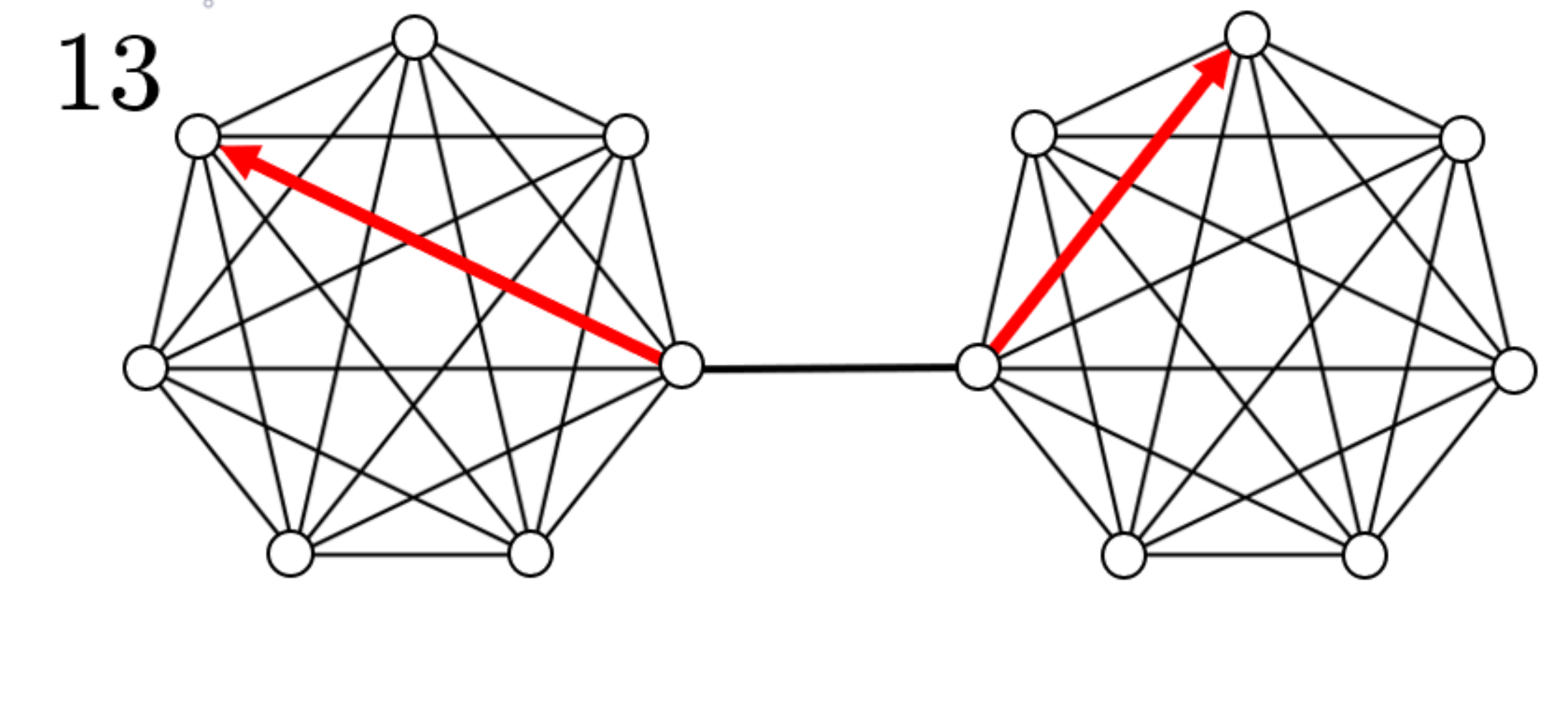}\label{fig:sub13}\quad%
  \includegraphics[width=0.31\textwidth]{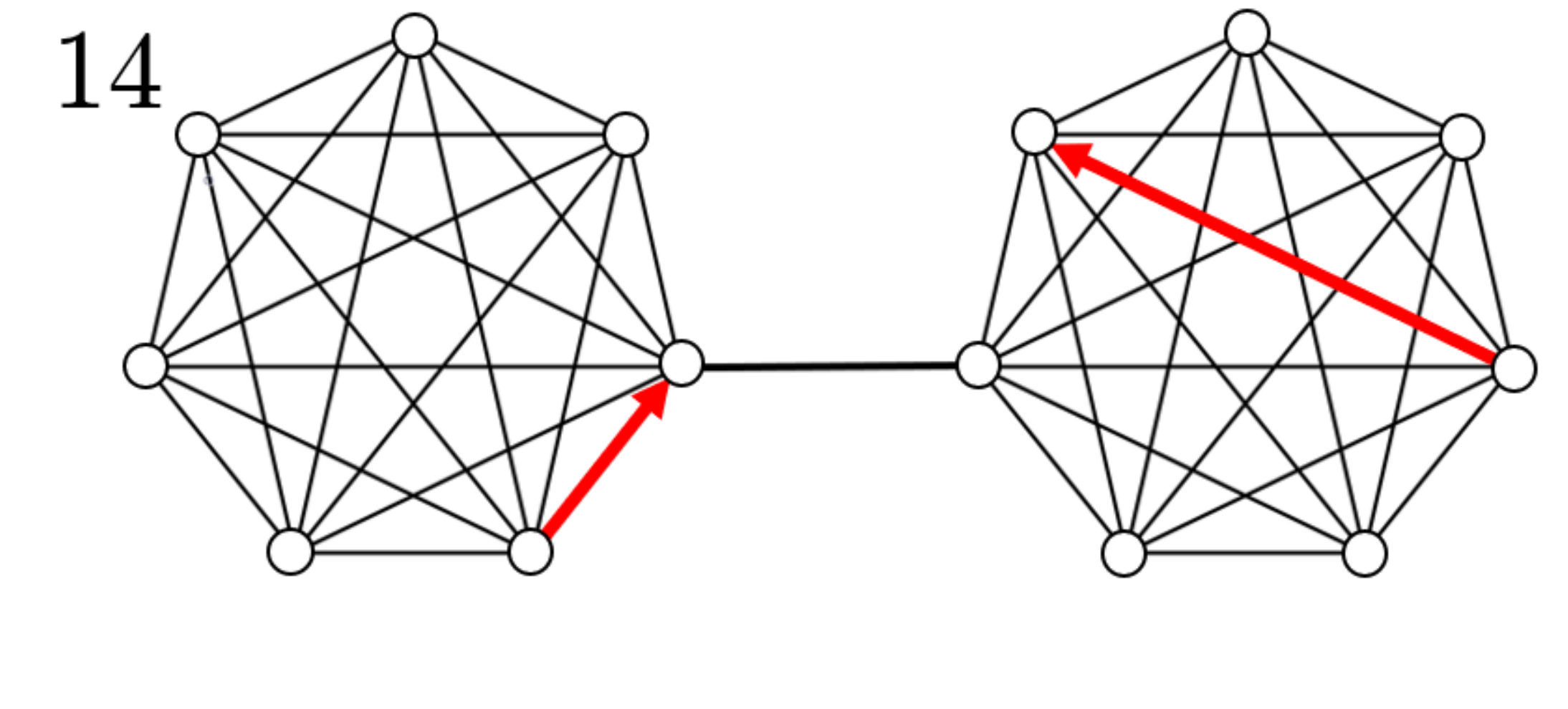}\label{fig:sub14}\quad%
  \includegraphics[width=0.31\textwidth]{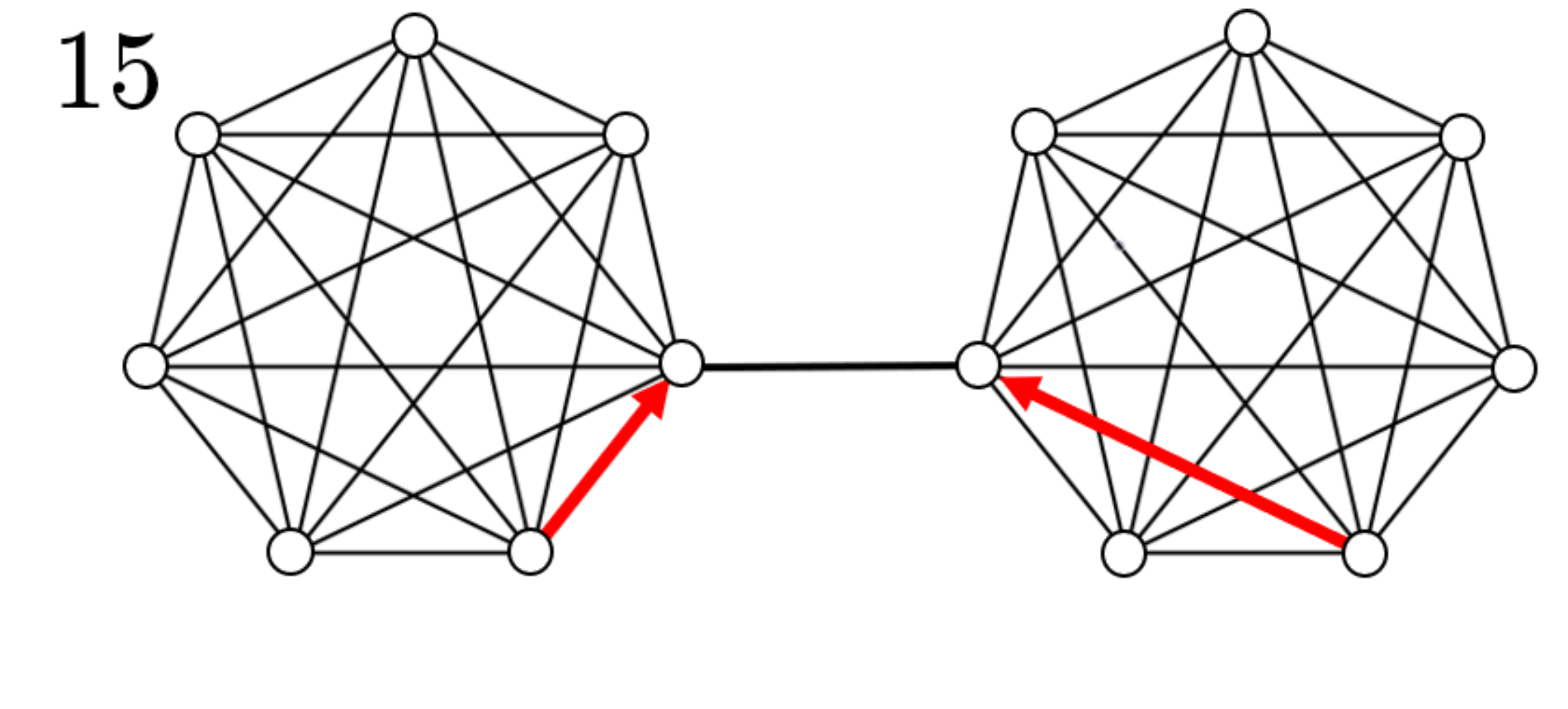}\label{fig:sub15}\quad%
  \includegraphics[width=0.31\textwidth]{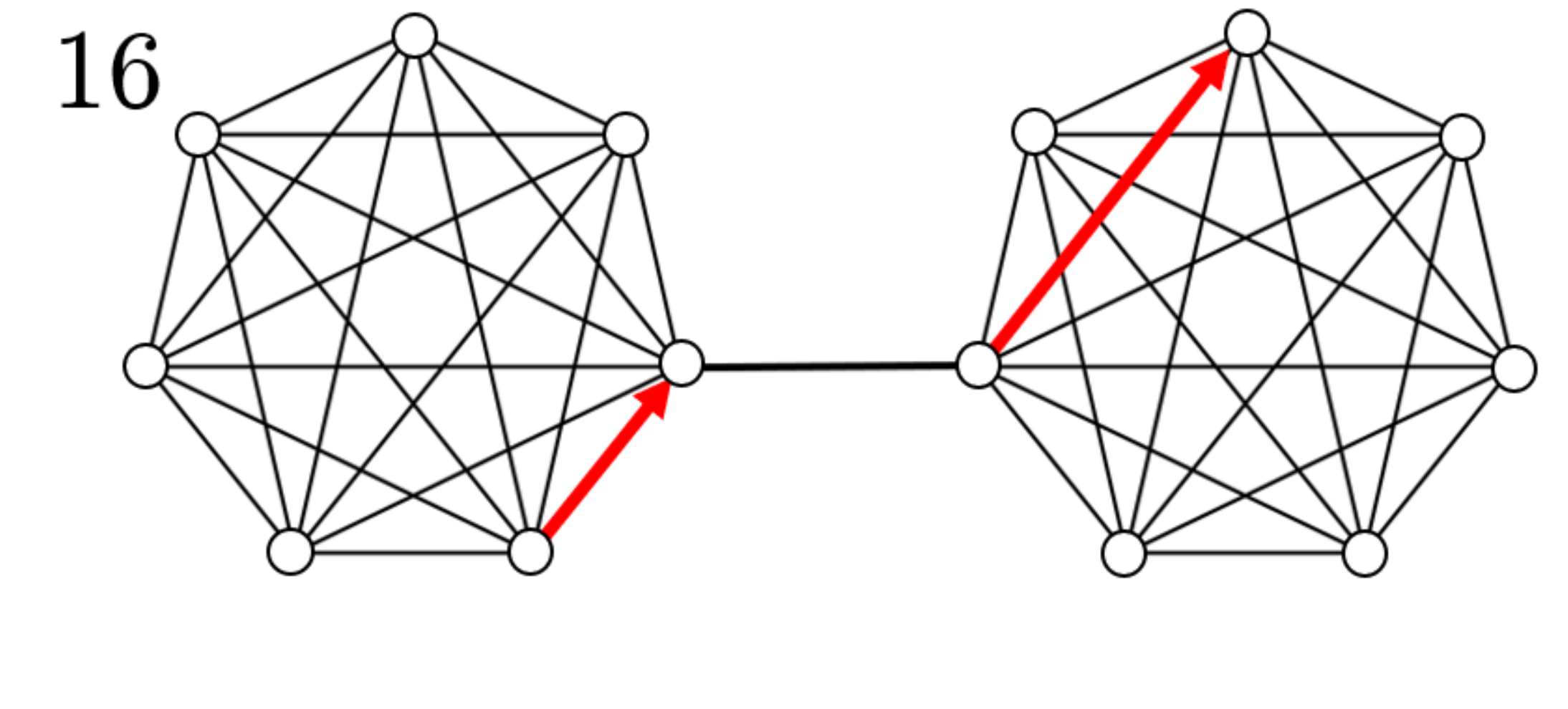}\label{fig:sub16}\quad%
  \includegraphics[width=0.31\textwidth]{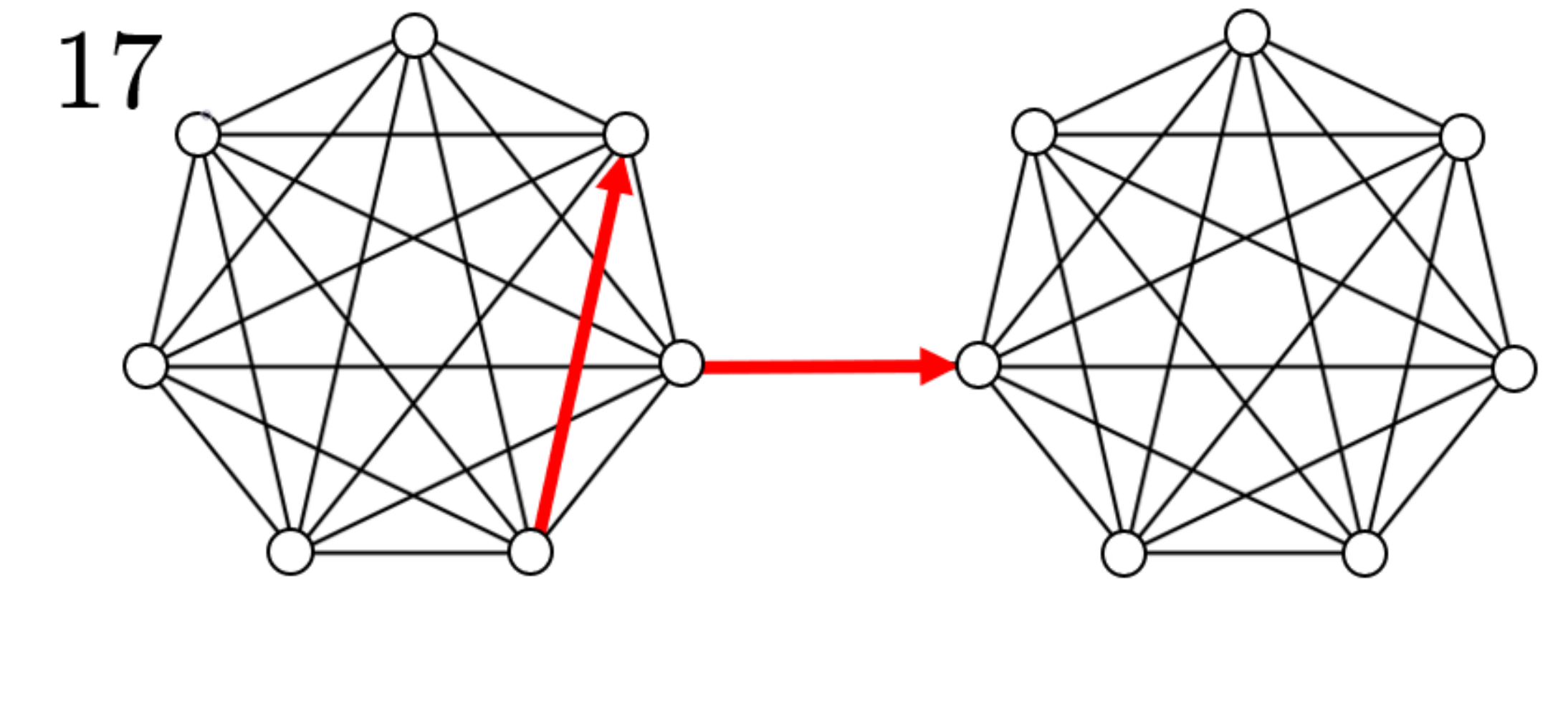}\label{fig:sub17}\quad%
  \includegraphics[width=0.31\textwidth]{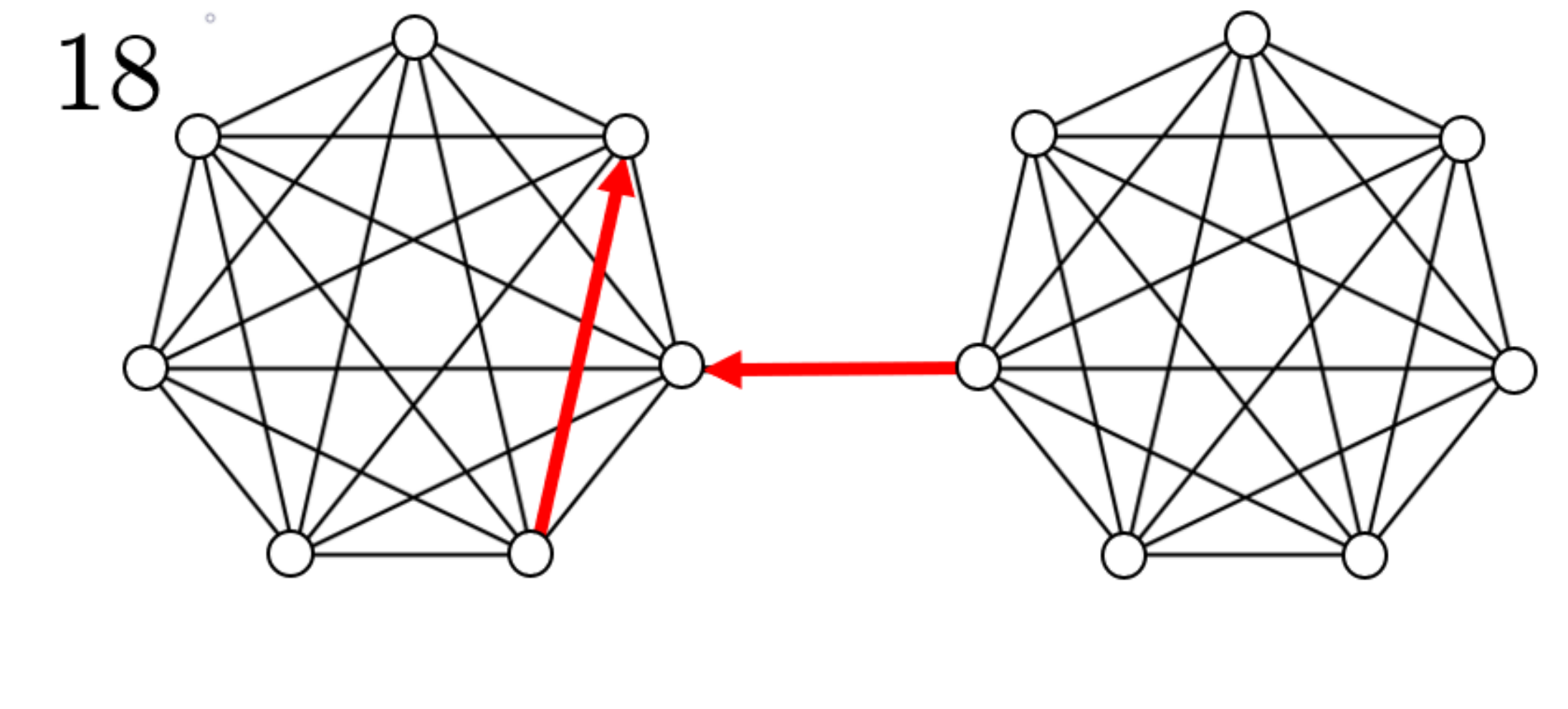}\label{fig:sub18}\quad%
  \includegraphics[width=0.31\textwidth]{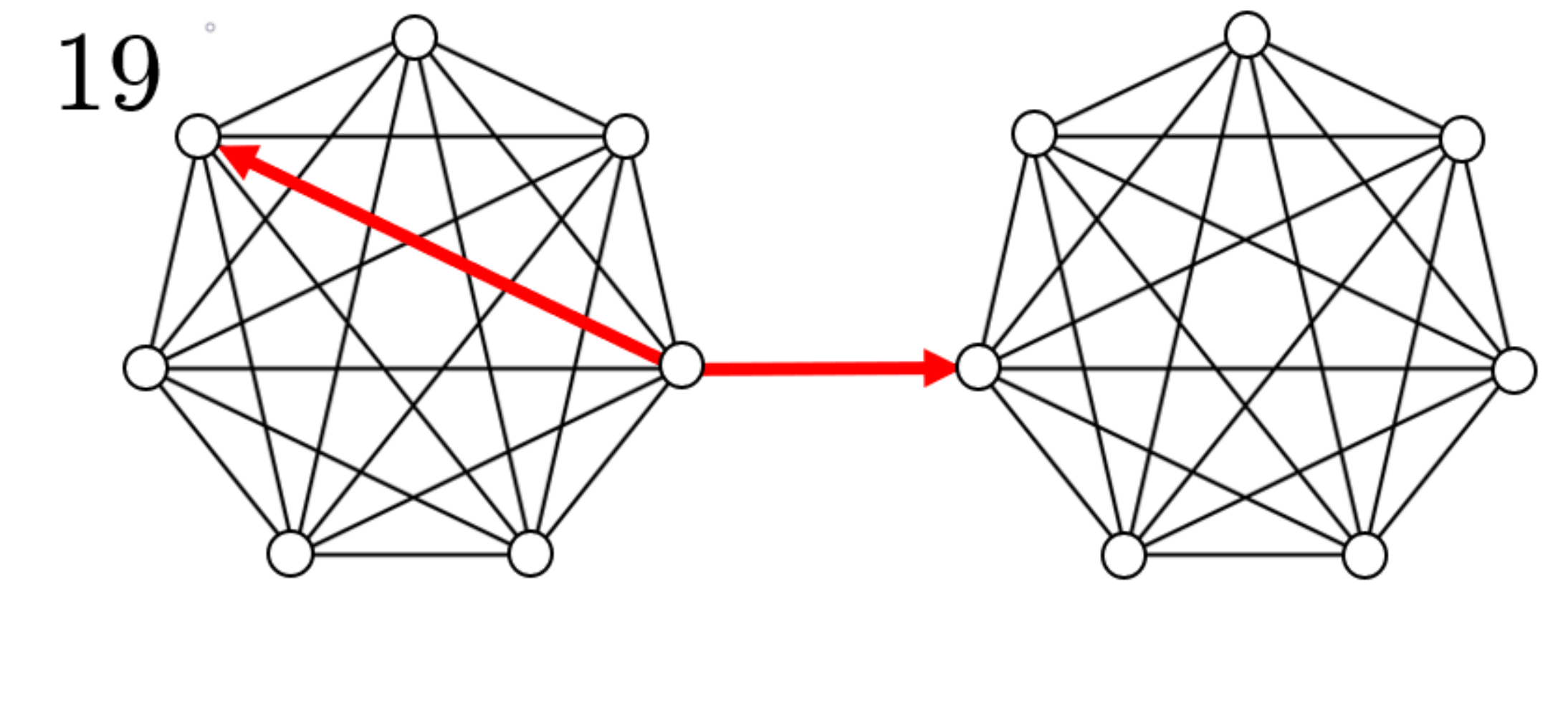}\label{fig:sub19}\quad%
  \includegraphics[width=0.31\textwidth]{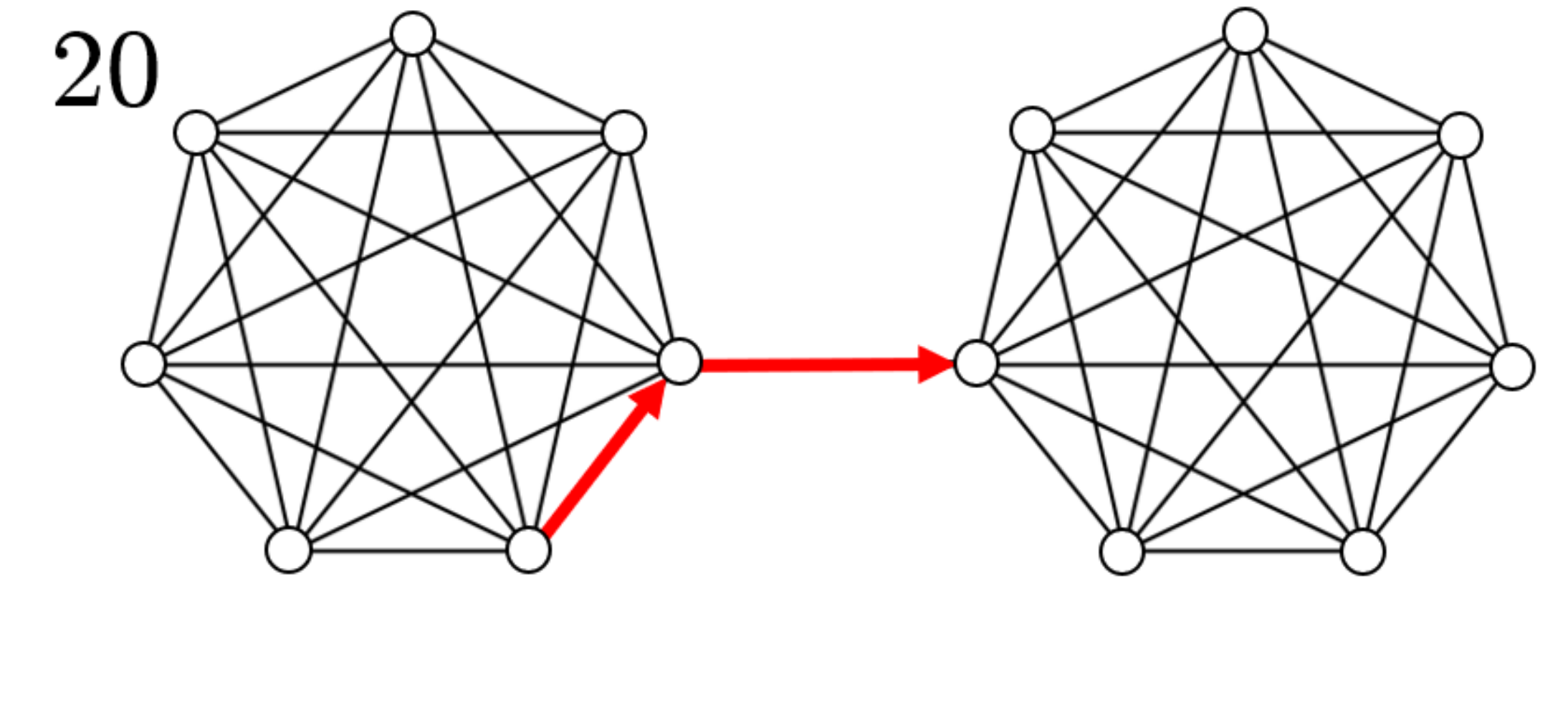}\label{fig:sub20}\quad%
  \includegraphics[width=0.31\textwidth]{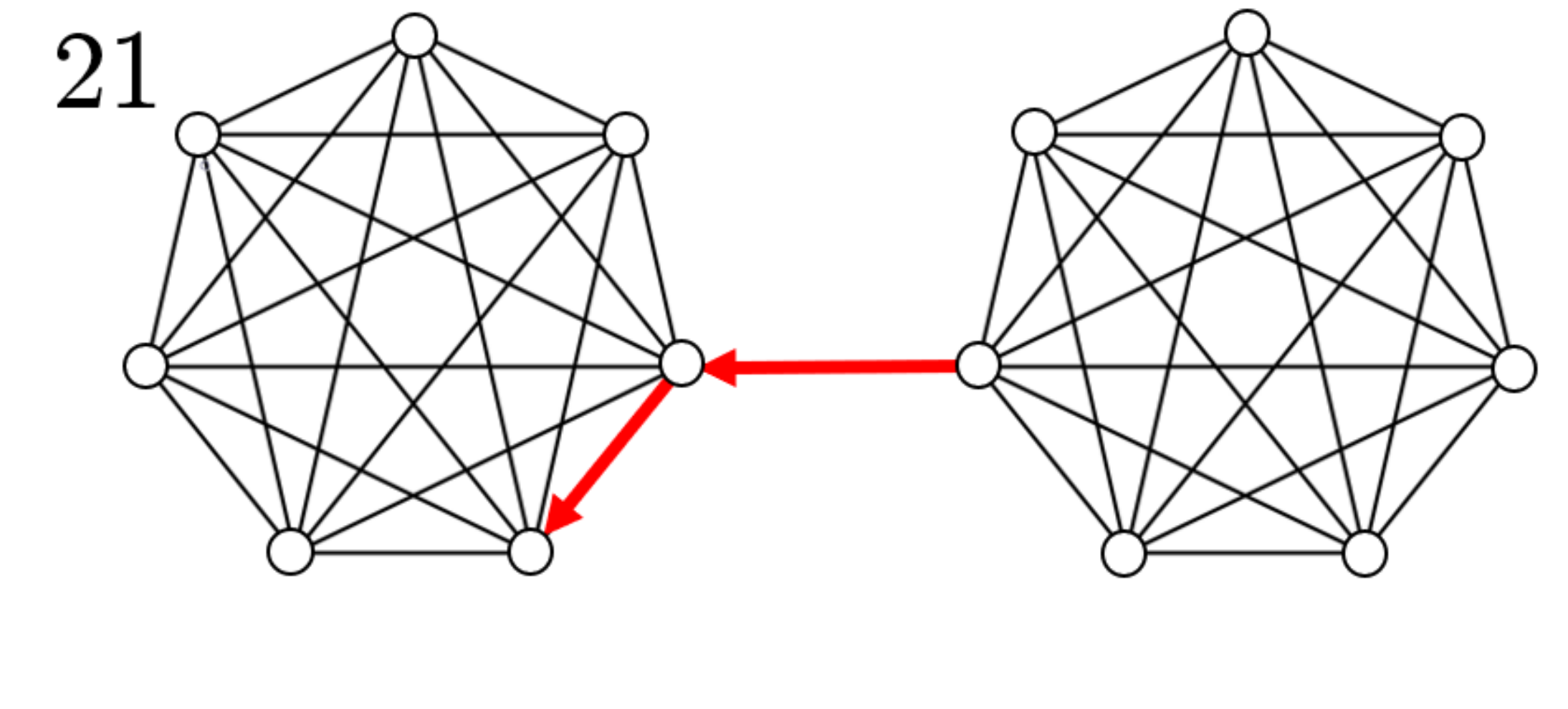}\label{fig:sub21}\quad%
  \caption{Schematic of the 21 states of the coalescencing node2vec random walk. The coalescent state is omitted.}\label{fig7}
  \vspace*{-7pt}
\end{figure}

We consider a weighted network composed of two cliques each of which has $N^\prime = N/2$ nodes; by definition, each pair of nodes in a clique is adjacent to each other. We assume that the edges forming a clique has weight $1$ and that the two cliques are connected by one edge with weight $w$, which we call the bridge (Fig. \ref{fig7}). We refer to the two nodes that are incident to the bridge as portal nodes. Unless $w$ is extremely large, this network is composed of well distinguished two communities such that diffusion needs a long time when $N$ is large. Because the two portal nodes are automorphically equivalent and so are the $N-2$ non-portal nodes, the stationary probability for a single node2vec random walker is given by
\begin{align}
    p_i^*=
    \begin{cases}
    \frac{1}{\mathcal{N}} \left[\frac{N^\prime-1}{w}\cdot\frac{\alpha+(N^\prime-2)\beta+w}{\alpha w+N^\prime-1}\right] & \text{if node}\ i\ \text{is a non-portal node}, \\
    \frac{1}{\mathcal{N}} \left[\frac{N^\prime-1}{w}\cdot\frac{\alpha+(N^\prime-2)\beta+w}{\alpha w+N^\prime-1}+1\right] & \text{if node}\ i\ \text{is a portal node}, \\
    \end{cases}
    \label{eq3.32station}
\end{align}
where $\mathcal{N}=\frac{N(N^\prime-1)}{w}\cdot\frac{\alpha+(N^\prime-2)\beta+w}{\alpha w+N^\prime-1}+2$; we show the derivation of Eq. (\ref{eq3.32station}) 
in the SM. Note that $p_i^* \approx 1/N$ for all nodes when $w=o(1)$.

The state of two coalescing node2vec random walkers is described by the currently visited node and the last visited node of each walker. In every time step, we update the position of one of the two walkers using the link dynamics rule \cite{antal2006evolutionary, sood2008voter}. In other words, we select one of the two walkers with the equal probability (i.e., $1/2$) and then the selected walker makes a single move according to the rule of node2vec. This dynamics repeats until the two walkers meet at the same node to coalesce.

\begin{table}[!t]
    \centering
    \caption{States of a pair of directed edges in two-clique networks. If the two edges are in the same clique and chasing, we assume $e_1(1) = e_2(0)$ without loss of generality within this table. We do so to distinguish between states 8, 9, and 10. Note that this convention does not apply to states 20 and 21. If one of the two edges coincides with the bridge, we assume that edge $e_2$ coincides with the bridge without loss of generality within this table. This convention is to distinguish between states 17 and 18 and between states 20 and 21.}
    \begin{tabular}{ p{0.5cm} p{3.5cm} p{1.5cm} p{8cm}}
    \hline
    State &  & State & Additional condition \\
    \hline
    1 & same clique & disjoint & No edge touches a portal node. \\
    2 & same clique & disjoint & $e_1(0)$ or $e_2(0)$, not both, is a portal node. \\
    3 & same clique & disjoint & $e_1(1)$ or $e_2(1)$, not both, is a portal node. \\
    4 & same clique & divergent & No edge touches a portal node. \\
    5 & same clique & divergent & $e_1(0)=e_2(0)$ is a portal node. \\
    6 & same clique & divergent & $e_1(1)$ or $e_2(1)$, not both, is a portal node. \\
    7 & same clique & chasing & No edge touches a portal node. \\
    8 & same clique & chasing & $e_1(0)$ is a portal node. \\
    9 & same clique & chasing & $e_2(0) (= e_1(1))$ is a portal node. \\
    10 & same clique & chasing & $e_2(1)$ is a portal node. \\
    11 & opposite cliques & disjoint & No edge touches a portal node. \\
    12 & opposite cliques & disjoint & $e_1(0)$ or $e_2(0)$, not both, is a portal node. \\
    13 & opposite cliques & disjoint & Both $e_1(0)$ and $e_2(0)$ are portal nodes. \\
    14 & opposite cliques & disjoint & $e_1(1)$ or $e_2(1)$, not both, is a portal node. \\
    15 & opposite cliques & disjoint & Both $e_1(1)$ and $e_2(1)$ are portal nodes. \\
    16 & opposite cliques & disjoint & $e_1(0)$ and $e_2(1)$, or $e_1(1)$ and $e_2(0)$ are portal nodes. \\
    17 & one edge on bridge & disjoint & Edge $e_1$ and node $e_2(0)$ are in the same clique (so, $e_2(1)$ is in the other clique).\\
    18 & one edge on bridge & disjoint & Edge $e_1$ and node $e_2(1)$ are in the same clique (so, $e_2(0)$ is in the other clique).\\
    19 & one edge on bridge & divergent & $e_1(0)=e_2(0)$ is a portal node. \\
    20 & one edge on bridge & chasing & Edge $e_1$ and node $e_2(0)$ are in the same clique (so, $e_2(1)$ is in the other clique). \\
    21 & one edge on bridge & chasing & Edge $e_1$ and node $e_2(1)$ are in the same clique (so, $e_2(0)$ is in the other clique). \\
    22 & & confluent  \\
    \hline
    \end{tabular}
    \label{tab2}
\end{table}

Specifying the currently visited and last visited nodes for the two walkers is equivalent to specifying two directed edges (while the network is assumed to be undirected). By exploiting the automorphical equivalence of the two portal nodes and that of the $N-2$ non-portal nodes, we only need to distinguish the following types of the pairs of directed edges for specifying the state of the pair of the walkers. The possible states are enumerated in Table 2 and schematically shown in Fig. \ref{fig7}.

A first level of classification of the pair of directed edges is whether they are in the same or different cliques, or on the bridge. Owing to the symmetry, if the two directed edges are contained in the same clique, we do not need to know which of the two cliques contains the two edges. There are ten such states. Alternatively, the two edges may belong to the opposite cliques. There are six such states. As the third and last possibility, one of the two edges may be on the bridge. There are five such states. Note that it is impossible for both edges to be on the bridge because it would mean that the walkers coalesced in a previous time step.

\begin{figure}[!h]
  \includegraphics[width=0.47\textwidth]{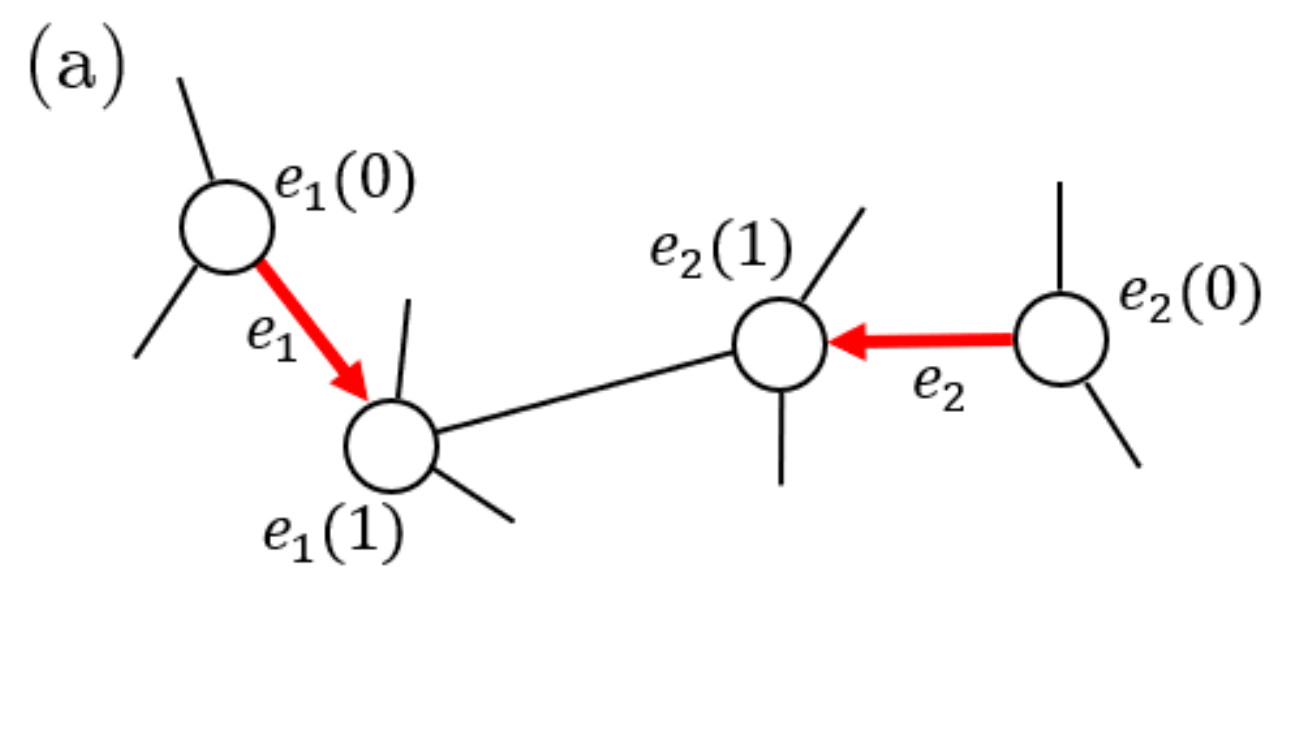}\label{figtotalsep}\quad%
  \includegraphics[width=0.47\textwidth]{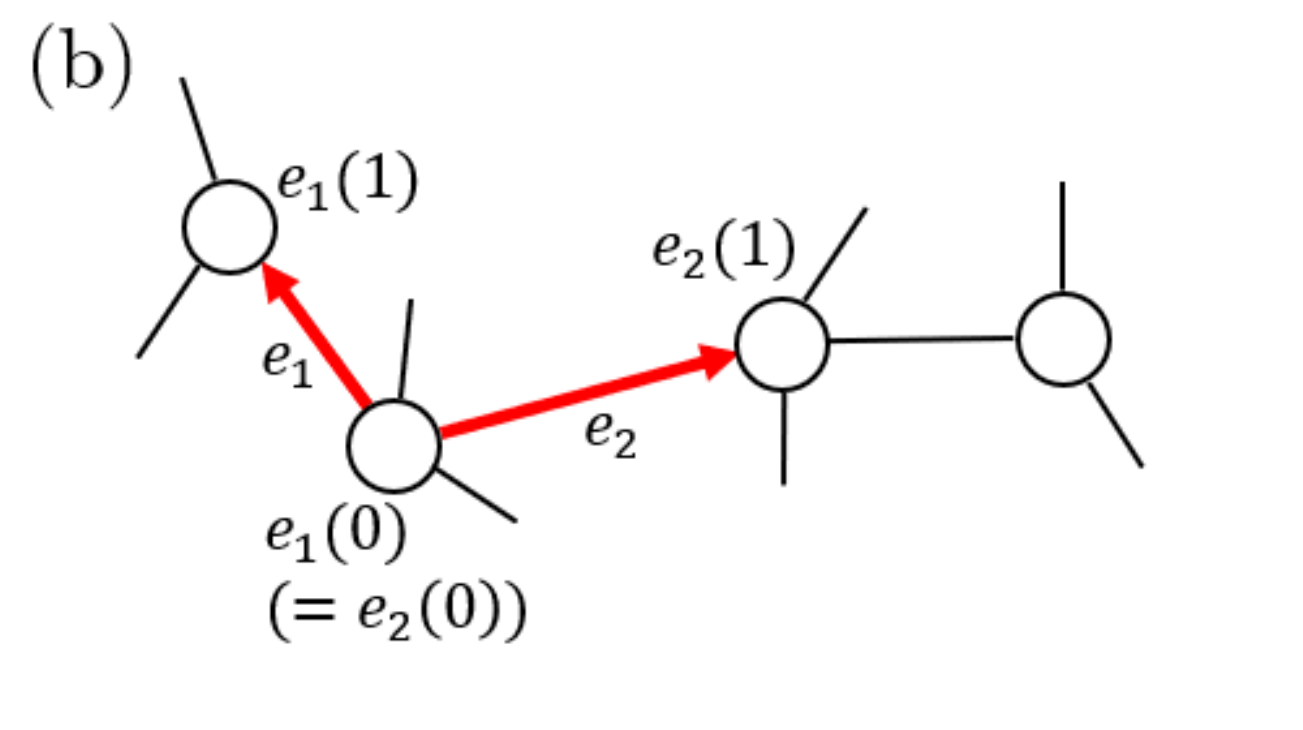}\label{figsep}\quad%
  \includegraphics[width=0.47\textwidth]{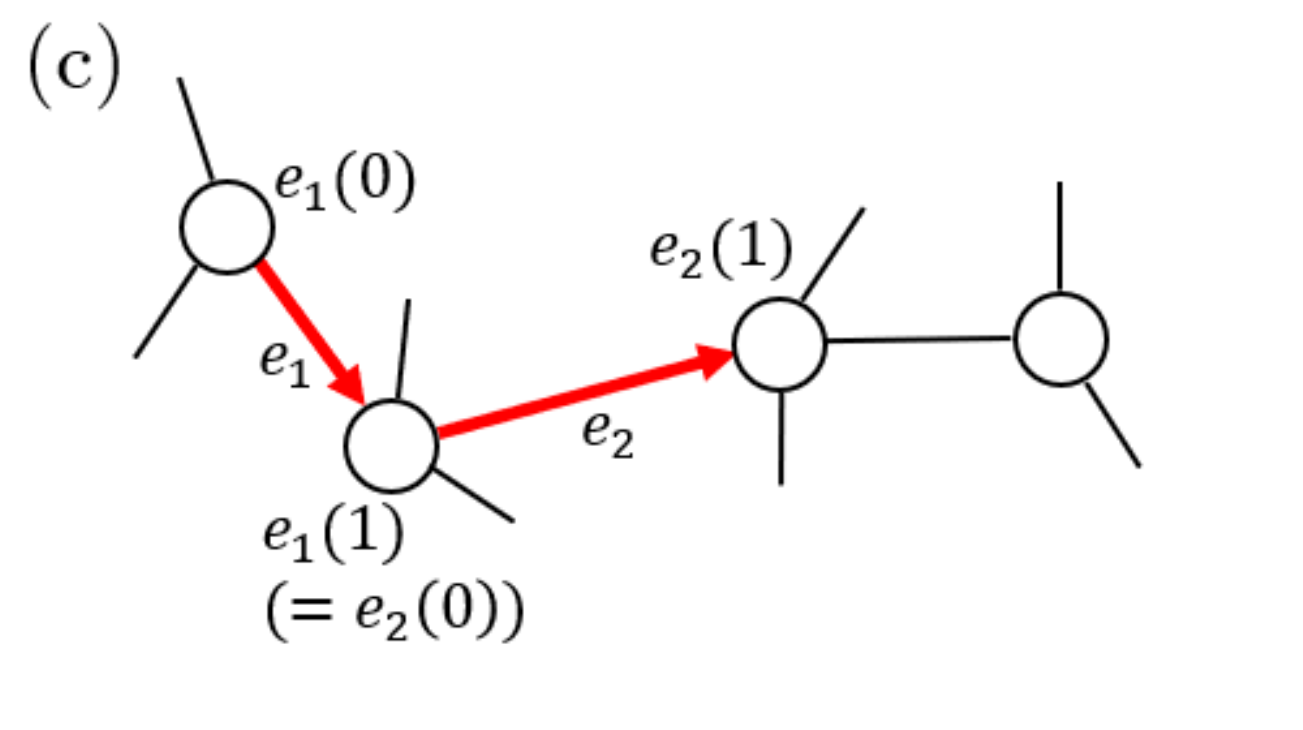}\label{figadj}\quad%
  \includegraphics[width=0.47\textwidth]{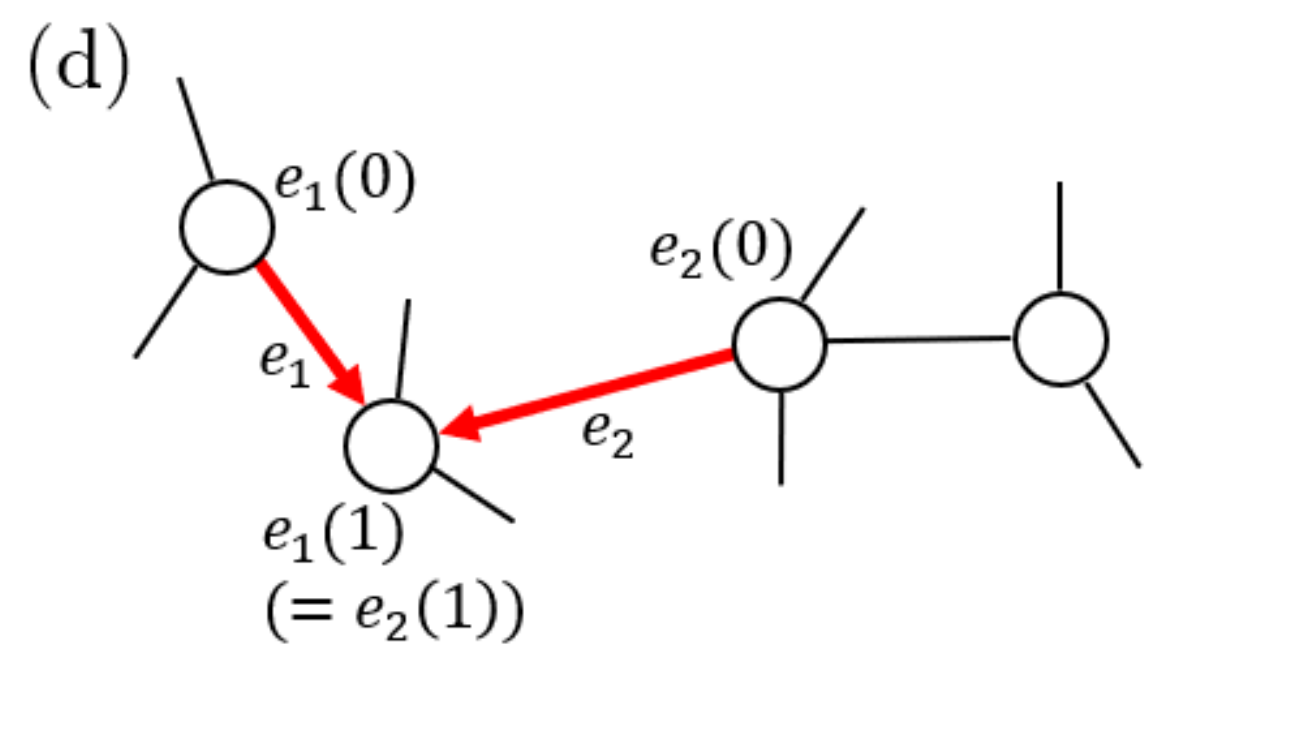}\label{figabsorbing}\quad%
  \caption{Classification of a pair of directed edges in the two-clique network. (a) Disjoint. (b) Divergent. (c) Chasing. (d) Confluent.}\label{figcoalesence}
  \vspace*{-7pt}
\end{figure}

A second level of classification is based on whether or not and how the two directed edges share a node. At this classification level, we distinguish between four configurations, which are schematically shown in Fig. \ref{figcoalesence}. First, we say that two directed edges $e_1$ and $e_2$ are disjoint if they do not share a node, i.e., $e_1(0)\neq e_2(0)$, $e_1(0)\neq e_2(1)$, $e_1(1)\neq e_2(0)$, and $e_1(1)\neq e_2(1)$ (Fig. \ref{figcoalesence}(a)). Second, $e_1$ and $e_2$ are divergent if $e_1(0)= e_2(0)$ and $e_1(1)\neq e_2(1)$ (Fig. \ref{figcoalesence}(b)). Third, the two edges are said to be chasing if $e_1(1)= e_2(0)$ and $e_1(0)\neq e_2(1)$, or $e_1(0)= e_2(1)$ and $e_1(1)\neq e_2(0)$ (Fig. \ref{figcoalesence}(c)). Fourth, if $e_1(1)= e_2(1)$, we say that the two edges are confluent (Fig. \ref{figcoalesence}(d)), which implies the coalescence of the two walkers. 

In some cases, in addition to applying the aforementioned two levels of the classification scheme, one has to distinguish between different states depending on whether or not and how the nodes coincide with the portal node. For example, Table 2 indicates that there are three states for a pair of directed edges that qualify as ``same clique'' (according to the first-level classification) and ``disjoint'' (second-level). The exhaustive classification yields $21$ states excluding the coalescent (i.e., confluent) state. We use the state number from 1 through 21 to inform the row/column index of the transition-probability matrix. We assign state 22 to the coalescent state.

Let $p_i(t)$ be the probability that two walkers are in state $i$ ($i=1,2,\ldots, 21$) at time $t$ and $r(t)$ the probability that the two walkers coalesce at time $t$. Let $\overline{T}^\mathrm{CRW}$ be the $22\times 22$ transition probability matrix derived in the Appendix, and $S$ be the minor of $\overline{T}^\mathrm{CRW}$ that one obtains by removing its last row and column of $\overline{T}^\mathrm{CRW}$ corresponding to the confluent state. Note that $\overline{T}^\mathrm{CRW}_{22,j} = \delta_{j,22}$ where $\delta$ is Kronecker delta. We obtain \cite{masuda2014voter}
\begin{align}
    \boldsymbol{p}(t)=\boldsymbol{p}(0)S^t,
\end{align}
where $\boldsymbol p(t) = (p_1(t), \ldots, p_{21}(t))$, and
\begin{align}
    r(t+1)=\boldsymbol{p}(t) \boldsymbol{v},
\end{align}
where $\boldsymbol{v}=(\overline{T}^\mathrm{CRW}_{1,22}, \ldots, \overline{T}^\mathrm{CRW}_{21,22})^\top$. The mean coalescence time $\langle \tau \rangle$ is given by
\begin{align}
    \langle \tau \rangle 
    & =\displaystyle \sum_{t=1}^\infty t\cdot r(t) \nonumber\\
    & =\boldsymbol{p}(0)A(I-A)^{-2}\boldsymbol{v} \nonumber\\
    & =\boldsymbol{p}(0)A(I-A)^{-1}\boldsymbol{1}^\top.
    \label{eq41}
\end{align}

We consider the two-clique network with $N=200$ nodes (i.e., $N^{\prime}=100$ nodes in each clique) and three initial conditions, i.e., two walkers starting from the same clique, the opposite cliques, or either clique with probability $1/2$ independently for the different walkers. Specifically, we define the initial condition under which the two walkers start from the same clique by $p_j(0)=1/12$ for $j=1,2,3,4,5,6,7,8,9,10,18,21$, and $p_j(0)=0$ otherwise. The initial condition under which the two walkers start from the opposite cliques is defined by $p_j(0)=1/9$ for $j=11,12,13,14,15,16,17,19,20$, and $p_j(0)=0$ otherwise. The initial condition under which the two walkers start from a uniformly randomly selected clique is defined by $p_j(0) = 1/21$ for $j=1, \ldots, 21$.

\begin{figure}[!h]
  \flushleft  % 图片左对齐
  \includegraphics[width=0.31\textwidth]{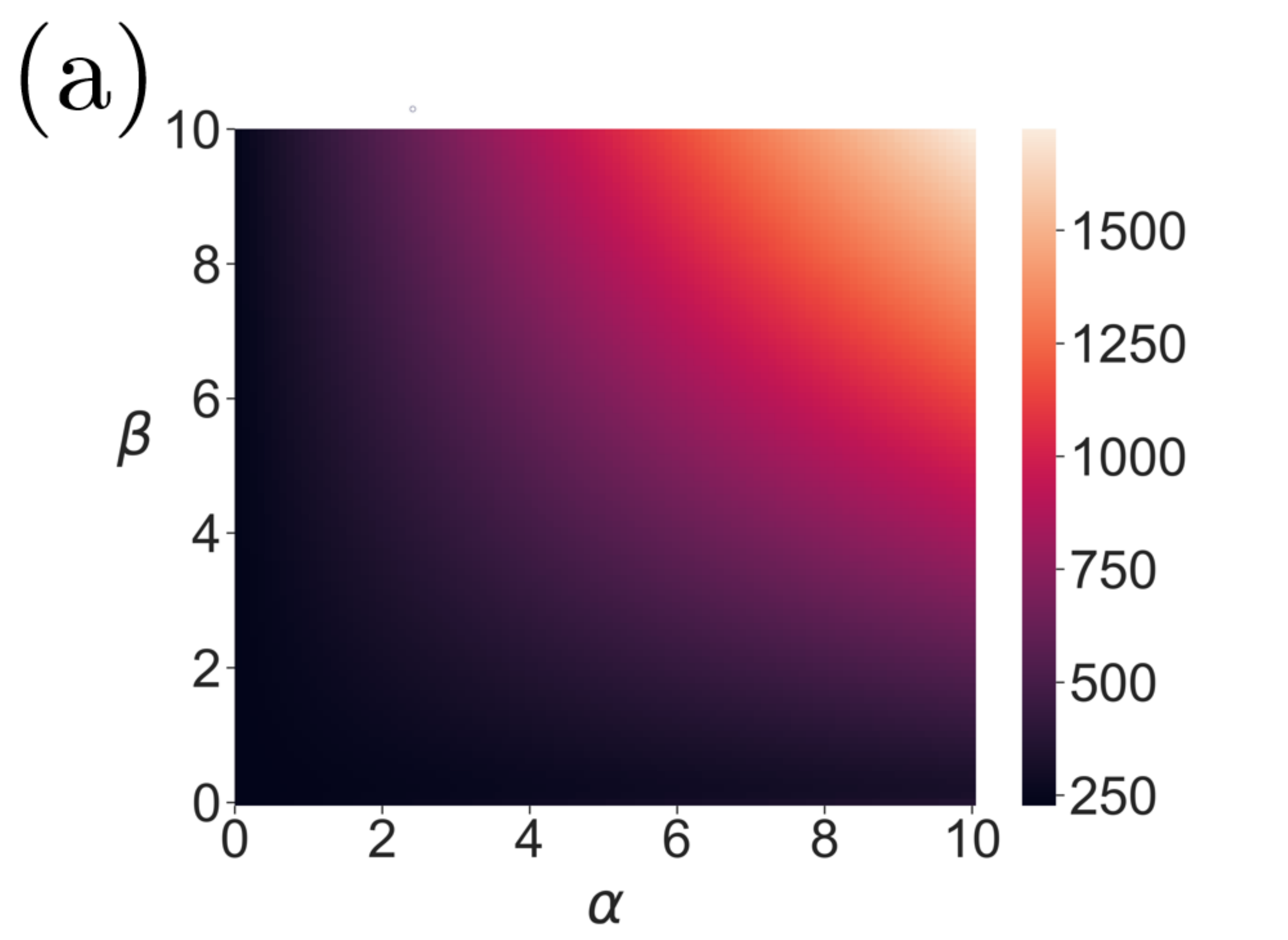}\label{fig9a}\quad%
  \includegraphics[width=0.31\textwidth]{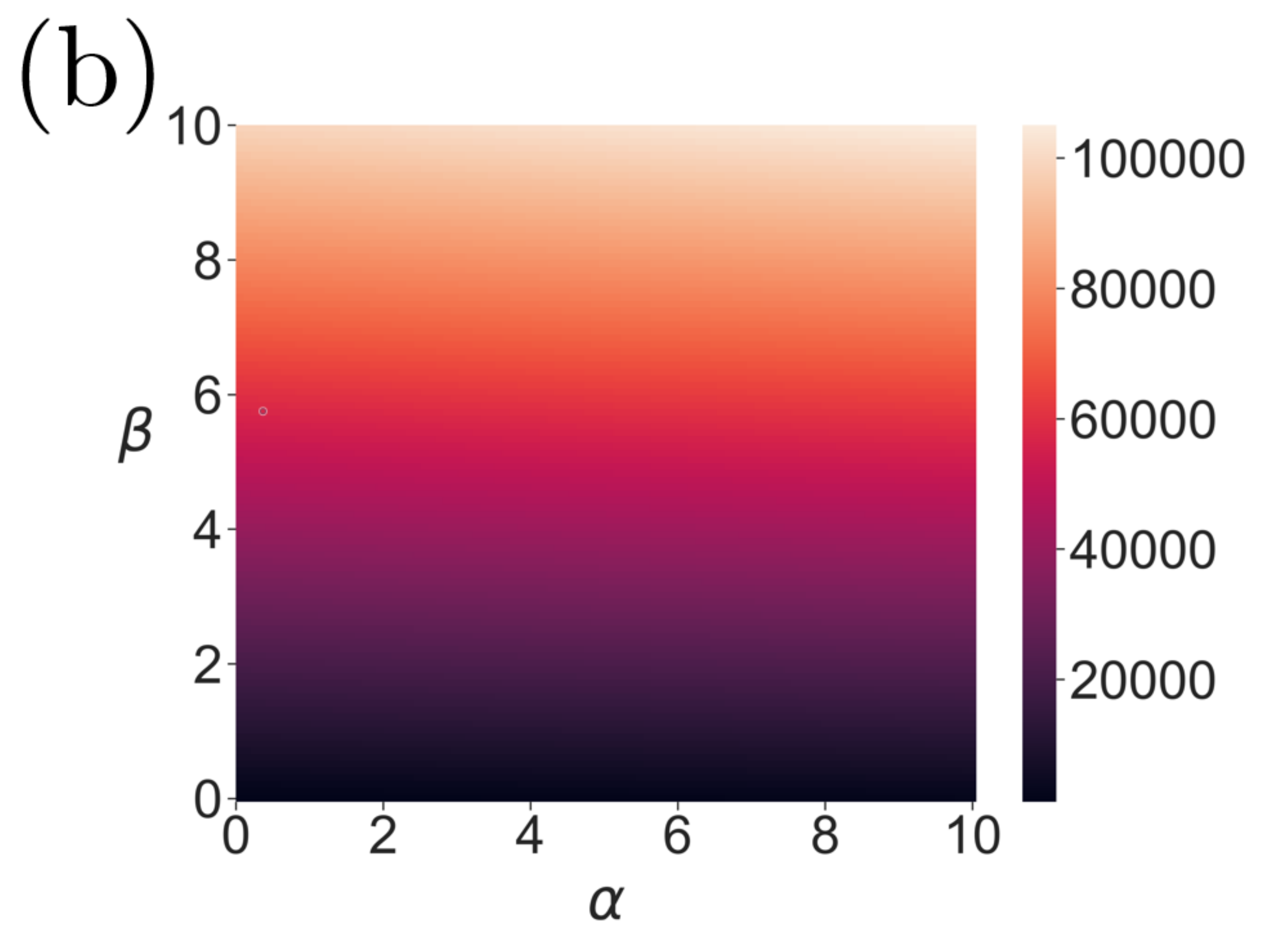}\label{fig9b}\quad%
  \includegraphics[width=0.31\textwidth]{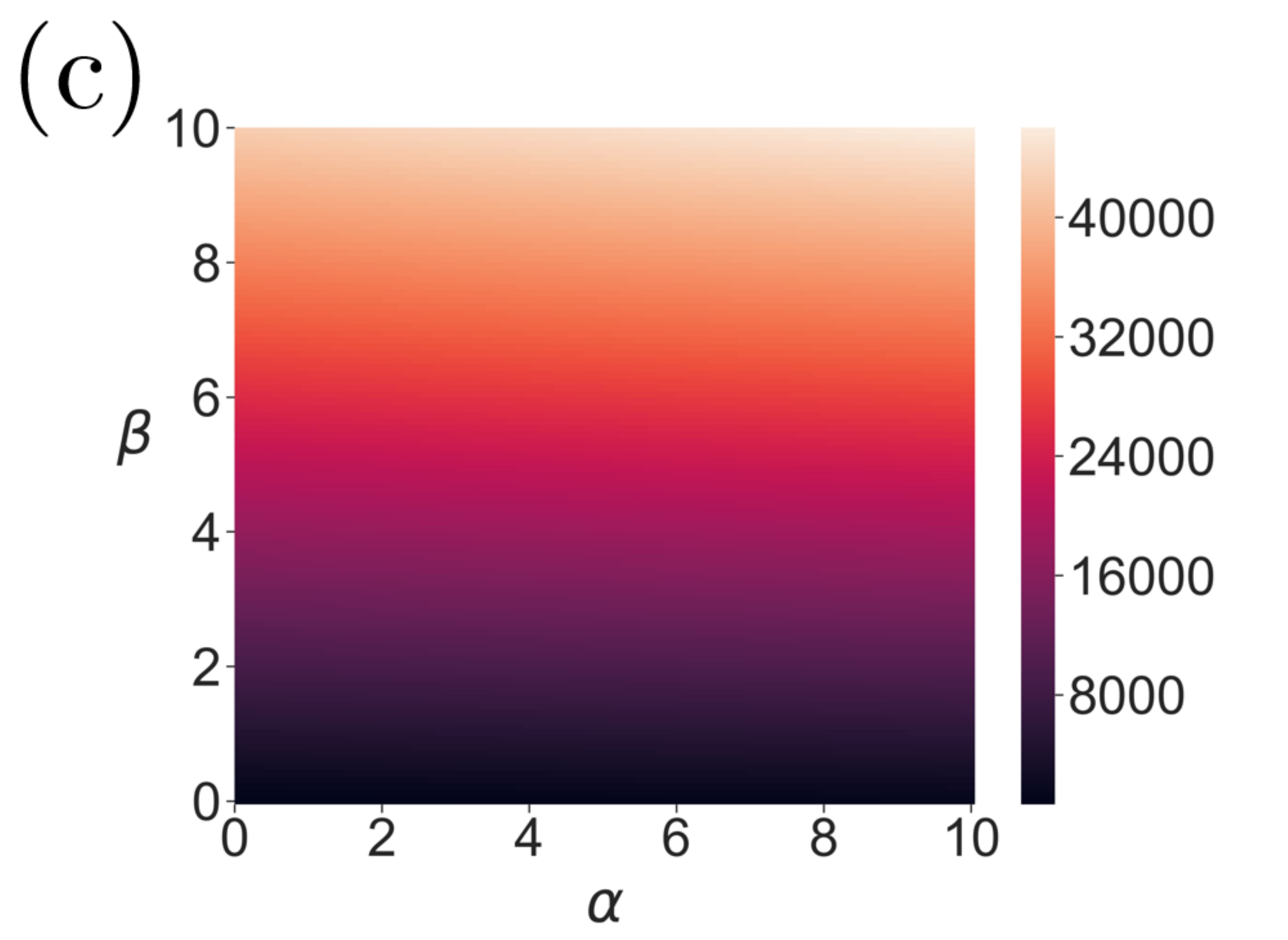}\label{fig9c}\quad%
  \includegraphics[width=0.31\textwidth]{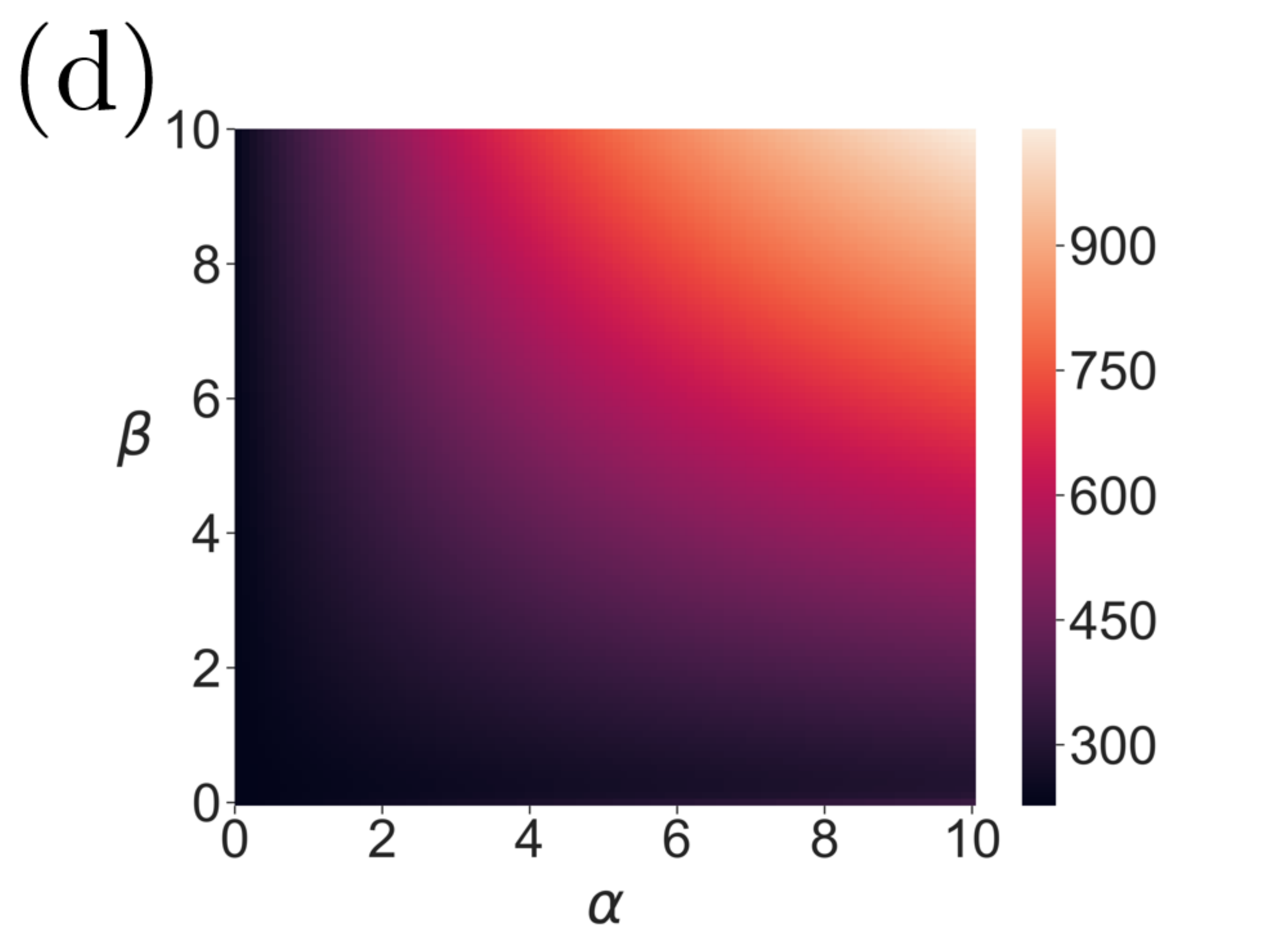}\label{fig9d}\quad%
  \includegraphics[width=0.31\textwidth]{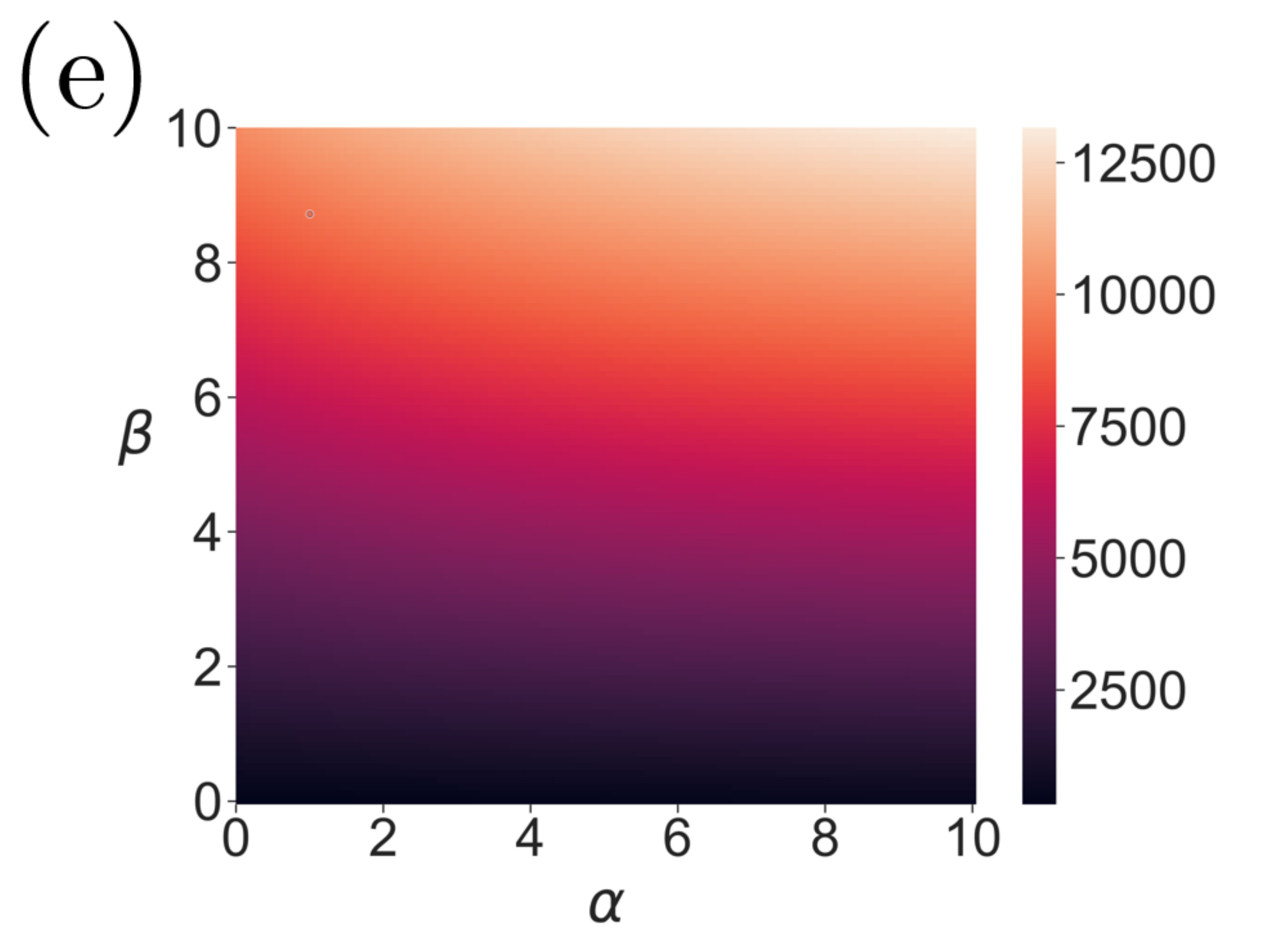}\label{fig9e}\quad%
  \includegraphics[width=0.31\textwidth]{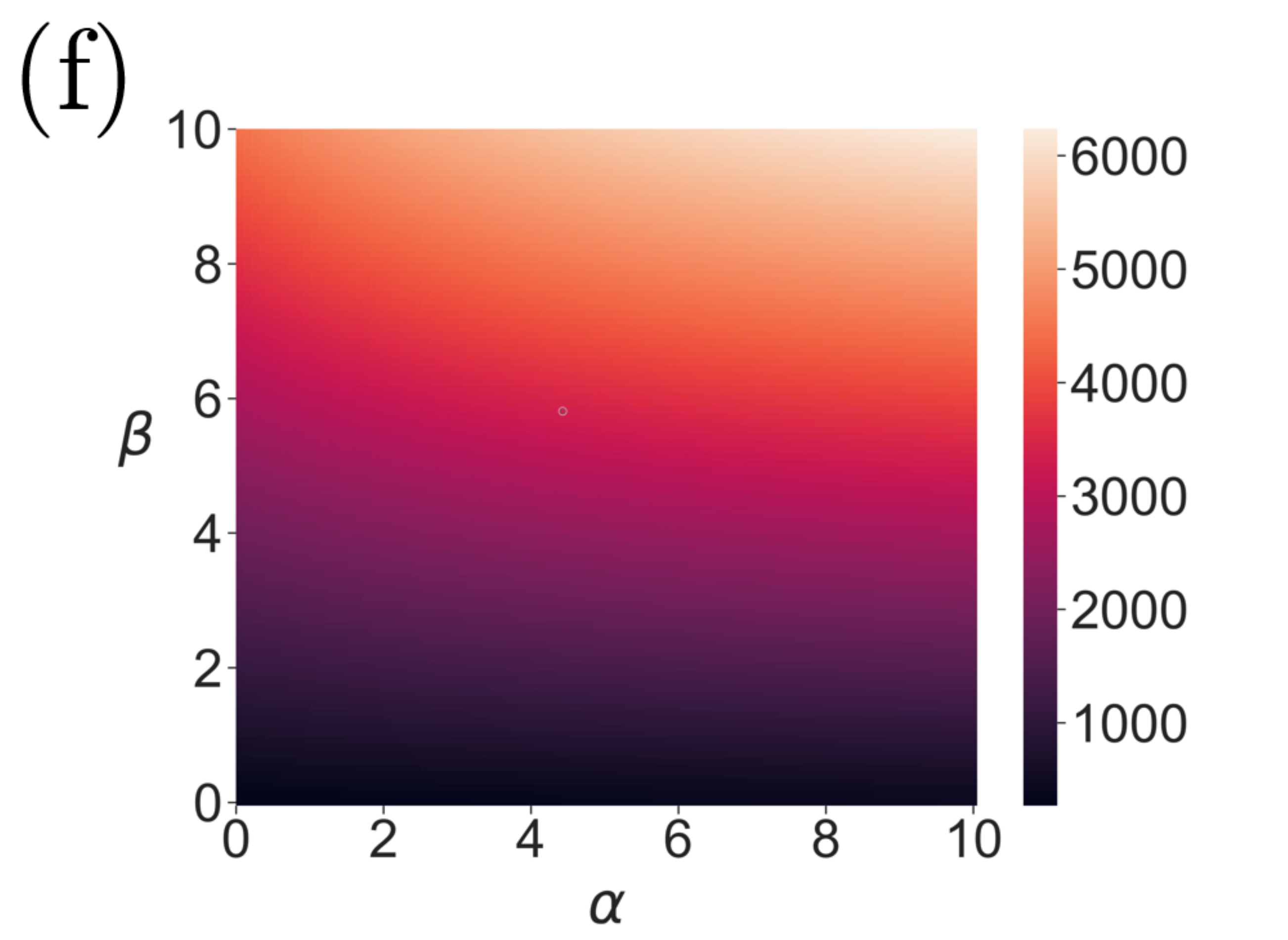}\label{fig9f}\quad%
  \caption{Mean coalescence time of two node2vec random walkers on the two-clique network with $N=200$ nodes. (a)--(c) $w=1$. (d)--(f) $w=10$. The two walkers are initially in the same clique ((a) and (d)), opposite cliques ((b) and (e)), or uniformly randomly selected cliques ((c) and (f)). }\label{fig9}
  \vspace*{-7pt}
\end{figure}

We show the mean coalescence time numerically calculated using Eq. (\ref{eq41}) in Fig. \ref{fig9} for the three initial conditions and two values of $w$ (i.e., $w=1$ and $w=10$). As expected, the mean coalescence time is considerably smaller if the two walkers start in the same clique (Figs. \ref{fig9}(a) and \ref{fig9}(d)) than in the opposite cliques (Figs. \ref{fig9}(b) and \ref{fig9}(e)). The results for the uniformly random initial condition (Figs. \ref{fig9}(c) and \ref{fig9}(f)) are intermediate between the other two initial conditions. Under each initial condition, the mean coalescence time is smaller for $w=1$ (Figs. \ref{fig9}(a)--(c)) than $w=10$ (Figs. \ref{fig9}(d)--(f)) because large $w$ enables the two walkers to move between cliques relatively frequently so that they have more chances to coalesce. 

\section{Discussion}

The node2vec has been recognized as a competitive algorithm of network embedding and also inspiring further network embedding algorithms \cite{cai2018comprehensive, goyal2018graph}. However, theoretical properties of the node2vec random walks, which are considered to affect the performance and applicability of node2vec, have been underexplored. A previous study provided a theoretical foundation of the stationary probability of node2vec random walks \cite{qiu2018network}. In the present study, we have investigated properties of node2vec random walks with a particular focus on diffusion speed. We have shown that diffusion measured in terms of the spectral gap and coalescence time is faster when random walkers are encouraged to explore the network without backtracking or visiting common neighbors of the currently visited node and the last visited node. We have confirmed this conclusion for several empirical and model networks except for some cases in which the avoidance of backtracking or visiting the common neighbors is excessive.  

Node2vec random walks are a second-order Markov process. Second-order Markov processes have been shown to be a promising representation of temporal network data, as opposed to first-order (i.e., memoryless) Markov processes \cite{rosvall2014memory, scholtes2014causality}. For temporal network data, second-order random walks find various applications. Therefore, apart from network embedding for which the node2vec random walks are originally used \cite{grover2016node2vec}, they may also find applications in, for example, community detection, ranking of nodes, network search, and collaborative filtering \cite{masuda2017random, xia2019random}. For example, one may be able to accelerate network search and sampling by setting $\alpha$ and $\beta$ to small values. However, we have pointed out that the stationary probability depends on the parameters of node2vec random walks, i.e., $\alpha$ and $\beta$ assuming $\gamma=1$ (also see Ref. \cite{qiu2018network}). Therefore, applications that depend on the stationary probability have to be carefully considered; one may have to calibrate the dependence of the stationary probability on the $\alpha$ and $\beta$ values to realize such applications.

In the analysis of the spectral gap of model networks (Section 3(b)(\ref{section:watts}) and 3(b)(\ref{section:wattstwolayers})), we analyzed networks whose stationary probability is independent of $\alpha$ and $\beta$ values. To this end, we used vertex-transitive networks, in which all nodes are automorphically equivalent to each other. We avoided the complete graph, which is trivially vertex-transitive, because all the triplets of nodes form a triangle such that the approximate depth-first sampling, which is defined to occur with the probability proportional to $\gamma$, is irrelevant. Both of the vertex-transitive networks that we have employed have a large average path length because they are essentially one-dimensional. This choice allowed us to employ a theorem in Ref.~\cite{tee2007eigenvectors} for conveniently calculating the spectrum of block circulant matrices. However, these networks do not resemble most of the empirical networks that have a small average path length relative to the number of nodes, $N$ \cite{newman2018networks, watts1998collective}. In fact, there are various named vertex-transitive networks, and methods to construct vertex-transitive networks such as Cayley graphs are available in algebraic graph theory \cite{biggs1993algebraic}. Analysis of the diffusion speed in vertex-transitive and small-world networks (i.e., having a small average path length and reasonably many triangles) warrants future work. Analysis of second-order Markov chains with other types of memory also warrants future work.

\section*{Appendix: Transition probabilities for a pair of coalescent node2vec random walkers}

In this section, we list the transition probability for a pair of coalescent random walkers on the two-clique graph. The non-zero elements of the $22 \times 22$ transition probability matrix, $\overline{T}^\mathrm{CRW}$, are enumerated as follows:

{\footnotesize
\begin{align}
\overline{T}^\mathrm{CRW}_{1,1}=\frac{\alpha+(N^\prime-5)\beta}{\alpha+(N^\prime-2)\beta},
\end{align}
\begin{align}
\overline{T}^\mathrm{CRW}_{1,3}=\overline{T}^\mathrm{CRW}_{1,7}=\overline{T}^\mathrm{CRW}_{1,22}=\overline{T}^\mathrm{CRW}_{2,22}=\overline{T}^\mathrm{CRW}_{4,3}=\overline{T}^\mathrm{CRW}_{4,22}=\overline{T}^\mathrm{CRW}_{5,22}=\overline{T}^\mathrm{CRW}_{11,14}=\frac{\beta}{\alpha+(N^\prime-2)\beta},
\end{align}
}
{\footnotesize
\begin{align}
\overline{T}^\mathrm{CRW}_{2,1}=\overline{T}^\mathrm{CRW}_{7,1}=\frac{(N^\prime-4)\beta}{2[\alpha+(N^\prime-2)\beta]},
\end{align}
\begin{align}
\overline{T}^\mathrm{CRW}_{2,2}=\overline{T}^\mathrm{CRW}_{3,3}=\overline{T}^\mathrm{CRW}_{7,4}=\frac{\alpha+(N^\prime-4)\beta}{2[\alpha+(N^\prime-2)\beta]},
\end{align}
\begin{align}
\overline{T}^\mathrm{CRW}_{2,3}=\overline{T}^\mathrm{CRW}_{6,10}=\overline{T}^\mathrm{CRW}_{8,6}=\overline{T}^\mathrm{CRW}_{12,14}=\overline{T}^\mathrm{CRW}_{16,15}=\overline{T}^\mathrm{CRW}_{19,20}=\frac{\alpha}{2[\alpha+(N^\prime-2)\beta]},
\end{align}
\begin{align}
\overline{T}^\mathrm{CRW}_{2,7}&=\overline{T}^\mathrm{CRW}_{2,9}=\overline{T}^\mathrm{CRW}_{3,10}=\overline{T}^\mathrm{CRW}_{7,3}=\overline{T}^\mathrm{CRW}_{7,6}=\overline{T}^\mathrm{CRW}_{7,7} \nonumber \\
&=\overline{T}^\mathrm{CRW}_{8,9}=\overline{T}^\mathrm{CRW}_{9,10}=\overline{T}^\mathrm{CRW}_{12,16}=\overline{T}^\mathrm{CRW}_{14,15}=\overline{T}^\mathrm{CRW}_{17,20}=\frac{\beta}{2[\alpha+(N^\prime-2)\beta]},
\end{align}
}
{\footnotesize
\begin{align}
\overline{T}^\mathrm{CRW}_{3,2}=\frac{\alpha+(N^\prime-4)\beta}{2[\alpha+(N^\prime-2)\beta+w]},
\end{align}
\begin{align}
\overline{T}^\mathrm{CRW}_{3,8}=\overline{T}^\mathrm{CRW}_{10,8}=\frac{\beta}{2[\alpha+(N^\prime-2)\beta+w]}, 
\end{align}
\begin{align}
\overline{T}^\mathrm{CRW}_{3,17}=\overline{T}^\mathrm{CRW}_{6,17}=\overline{T}^\mathrm{CRW}_{9,19}=\overline{T}^\mathrm{CRW}_{10,17}=\overline{T}^\mathrm{CRW}_{14,18}=\overline{T}^\mathrm{CRW}_{16,21}=\frac{w}{2[\alpha+(N^\prime-2)\beta+w]}, 
\end{align}
\begin{align}
\overline{T}^\mathrm{CRW}_{3,22}=\overline{T}^\mathrm{CRW}_{6,22}=\frac{\beta}{2[\alpha+(N^\prime-2)\beta]}+\frac{\beta}{2[\alpha+(N^\prime-2)\beta+w]},
\end{align}
}
{\footnotesize
\begin{align}
\overline{T}^\mathrm{CRW}_{4,1}=\frac{(N^\prime-4)\beta}{\alpha+(N^\prime-2)\beta},
\end{align}
\begin{align}
\overline{T}^\mathrm{CRW}_{4,7}=\overline{T}^\mathrm{CRW}_{5,9}=\overline{T}^\mathrm{CRW}_{13,16}=\frac{\alpha}{\alpha+(N^\prime-2)\beta},
\end{align}
}
{\footnotesize
\begin{align}
\overline{T}^\mathrm{CRW}_{5,2}=\frac{(N^\prime-3)\beta}{\alpha+(N^\prime-2)\beta}, 
\end{align}
}
{\footnotesize
\begin{align}
\overline{T}^\mathrm{CRW}_{6,2}=\overline{T}^\mathrm{CRW}_{10,2}=\frac{(N^\prime-3)\beta}{2[\alpha+(N^\prime-2)\beta+w]},
\end{align}
\begin{align}
\overline{T}^\mathrm{CRW}_{6,3}=\overline{T}^\mathrm{CRW}_{8,2}=\overline{T}^\mathrm{CRW}_{8,4}=\overline{T}^\mathrm{CRW}_{9,3}=\frac{(N^\prime-3)\beta}{2[\alpha+(N^\prime-2)\beta]},
\end{align}
\begin{align}
\overline{T}^\mathrm{CRW}_{6,8}=\frac{\alpha}{2[\alpha+(N^\prime-2)\beta+w]}, 
\end{align}
}
{\footnotesize
\begin{align}
\overline{T}^\mathrm{CRW}_{7,22}=\overline{T}^\mathrm{CRW}_{8,22}=\frac{\alpha+\beta}{2[\alpha+(N^\prime-2)\beta]},
\end{align}
}
{\footnotesize
\begin{align}
\overline{T}^\mathrm{CRW}_{9,5}=\frac{\alpha+(N^\prime-3)\beta}{2[\alpha+(N^\prime-2)\beta+w]},
\end{align}
\begin{align}
\overline{T}^\mathrm{CRW}_{9,22}=\frac{\alpha}{2[\alpha+(N^\prime-2)\beta]}+\frac{\beta}{2[\alpha+(N^\prime-2)\beta+w]},
\end{align}
}
{\footnotesize
\begin{align}
\overline{T}^\mathrm{CRW}_{10,6}=\overline{T}^\mathrm{CRW}_{12,12}=\overline{T}^\mathrm{CRW}_{14,14}=\overline{T}^\mathrm{CRW}_{17,17}=\overline{T}^\mathrm{CRW}_{18,18}=\frac{\alpha+(N^\prime-3)\beta}{2[\alpha+(N^\prime-2)\beta]},
\end{align}
\begin{align}
\overline{T}^\mathrm{CRW}_{10,22}=\frac{\beta}{2[\alpha+(N^\prime-2)\beta]}+\frac{\alpha}{2[\alpha+(N^\prime-2)\beta+w]},
\end{align}
}
{\footnotesize
\begin{align}
\overline{T}^\mathrm{CRW}_{11,11}=\frac{\alpha+(N^\prime-3)\beta}{\alpha+(N^\prime-2)\beta},
\end{align}
}
{\footnotesize
\begin{align}
\overline{T}^\mathrm{CRW}_{12,11}=\overline{T}^\mathrm{CRW}_{16,14}=\overline{T}^\mathrm{CRW}_{19,17}=\overline{T}^\mathrm{CRW}_{21,18}=\frac{(N^\prime-2)\beta}{2[\alpha+(N^\prime-2)\beta]},
\end{align}
}
{\footnotesize
\begin{align}
\overline{T}^\mathrm{CRW}_{13,12}=\frac{(N^\prime-2)\beta}{\alpha+(N^\prime-2)\beta}, 
\end{align}
\begin{align}
\overline{T}^\mathrm{CRW}_{14,12}=\overline{T}^\mathrm{CRW}_{16,13}=\overline{T}^\mathrm{CRW}_{20,19}=\frac{\alpha+(N^\prime-2)\beta}{2[\alpha+(N^\prime-2)\beta+w]},
\end{align}
}
{\footnotesize
\begin{align}
\overline{T}^\mathrm{CRW}_{15,16}=\frac{\alpha+(N^\prime-2)\beta}{\alpha+(N^\prime-2)\beta+w}, 
\end{align}
\begin{align}
\overline{T}^\mathrm{CRW}_{15,22}=\frac{w}{\alpha+(N^\prime-2)\beta+w},
\end{align}
}
{\footnotesize
\begin{align}
\overline{T}^\mathrm{CRW}_{17,18}=\overline{T}^\mathrm{CRW}_{18,17}=\overline{T}^\mathrm{CRW}_{19,21}=\overline{T}^\mathrm{CRW}_{21,19}=\frac{w\alpha}{2[w\alpha+(N^\prime-1)]}, 
\end{align}
\begin{align}
\overline{T}^\mathrm{CRW}_{17,12}=\overline{T}^\mathrm{CRW}_{19,13}=\overline{T}^\mathrm{CRW}_{20,16}=\frac{N^\prime-1}{2[w\alpha+(N^\prime-1)]}, 
\end{align}
}
{\footnotesize
\begin{align}
\overline{T}^\mathrm{CRW}_{18,2}=\frac{N^\prime-3}{2[w\alpha+(N^\prime-1)]}, 
\end{align}
\begin{align}
\overline{T}^\mathrm{CRW}_{18,8}=\frac{1}{2[w\alpha+(N^\prime-1)]}, 
\end{align}
\begin{align}
\overline{T}^\mathrm{CRW}_{18,22}=\frac{\beta}{2[\alpha+(N^\prime-2)\beta]}+\frac{1}{2[w\alpha+(N^\prime-1)]},
\end{align}
}
{\footnotesize
\begin{align}
\overline{T}^\mathrm{CRW}_{20,22}=\frac{w}{2[\alpha+(N^\prime-2)\beta+w]}+\frac{w\alpha}{2[w\alpha+(N^\prime-1)]},
\end{align}
}
{\footnotesize
\begin{align}
\overline{T}^\mathrm{CRW}_{21,5}=\frac{N^\prime-2}{2[w\alpha+(N^\prime-1)]}, 
\end{align}
\begin{align}
\overline{T}^\mathrm{CRW}_{21,22}=\frac{\alpha}{2[\alpha+(N^\prime-2)\beta]}+\frac{1}{2[w\alpha+(N^\prime-1)]},
\end{align}
}
and
{\footnotesize
\begin{align}
    \overline{T}^\mathrm{CRW}_{22,22}=1.
\end{align}
}
All the other elements of $\overline{T}^\mathrm{CRW}$ are equal to $0$.

\section*{Data Accessibility}

The empirical network data sets are open resources and available at \cite{vole, dolphin, enron, jazz, coauthorship, email}. The Python codes used in the present study are available on Github \cite{code}.

\section*{Authors’ Contributions}

N.M. conceived and designed the research; L.M. and N.M. carried out the mathematical analysis; L.M. performed the computational experiment; L.M. and N.M. wrote the paper. Both authors gave final approval for publication and agree to be held accountable for the work performed therein.

\section*{Acknowledgments}

N.M. acknowledges the support provided through AFOSR European Office (FA9550-19-1-7024).

\bibliographystyle{elsarticle-num}
\bibliography{Ref.bib}
\end{document}